\documentclass[prd,aps,floatfix,preprint,12pt]{revtex4}
\usepackage{amsmath}
\usepackage{graphicx,color}
\usepackage{epsfig}
\def\pr{{Phys. Rev.}~}
\def\prl{{ Phys. Rev. Lett.}~}
\def\pl{{ Phys. Lett.}~}
\def\npb{{ Nucl. Phys. B}}

\def\non{\nonumber}

\begin{document}

\vspace*{-3mm}
\begin{flushright}
DESY 07-021\\
March 2007\\
\end{flushright}

\title{Charmless non-leptonic $B_s$ decays to $PP$, $PV$ and $VV$ final
states in the pQCD approach}
\author{Ahmed Ali$^a$, Gustav Kramer$^b$}
\affiliation{
$^a$ Deutsches Elektronen-Synchrotron DESY, 22607 Hamburg, Germany\\
$^b$ II. Institut f\"ur Theoretische Physik, Universit\"at Hamburg,
22761 Hamburg, Germany}
\author{Ying Li, Cai-Dian L\"u, Yue-Long Shen, Wei Wang and Yu-Ming Wang}
\affiliation{ Institute of High Energy Physics, CAS, P.O. Box
918(4), 100049, P.R. China}

\date{\today}

\begin{abstract}
We calculate the CP-averaged branching ratios and CP-violating
asymmetries of a number of two-body charmless hadronic decays  $\overline{B_s^0} \to PP,
PV, VV$ in the perturbative QCD (pQCD) approach to leading order in
$\alpha_s$ (here $P$ and $V$
denote light pseudo-scalar and vector mesons, respectively).
The  mixing-induced $CP$
 violation parameters are also calculated for these decays.
  We also
 predict the polarization fractions of $B_s \to VV$ decays and find that the transverse
 polarizations  are  enhanced in  some penguin dominated decays such as
 $\overline{B_s^0} \to K ^{*}\overline{K^{*}}$, $K ^{*}\rho$.
 Some of the
predictions worked out here can already be confronted with the
recently  available data from the CDF collaboration on the branching
ratios for the decays
$ \overline{B_s^0} \to K^+\pi^-$,  $ \overline{B_s^0} \to K^+K^-$ and the
 CP-asymmetry in the decay $ \overline{B_s^0} \to K^+\pi^-$, and are found
to be in agreement within the current errors. A large number of  predictions for the
branching ratios, CP-asymmetries and vector-meson polarizations in
$\overline{B_s^0}$ decays, presented in this paper and compared with the
already existing results in other theoretical frameworks, will be put to stringent
experimental tests in forthcoming experiments at Fermilab, LHC and
Super B-factories.

\end{abstract}

\maketitle

\section{Introduction}

There has been remarkable progress in the study of exclusive
charmless $\overline{B_d^0} \to h_1 h_2$ and $B^\pm \to h_1 h_2$
decays, where $h_1, h_2$ are light pseudo-scalar and/or vector
mesons. Historically, these decays were calculated in the
so-called naive factorization approach~\cite{bsw}, which was
improved by including some perturbative QCD
contributions~\cite{Ali:1997nh,9804363}. Currently, there are
three popular theoretical approaches to study the dynamics of
these decays, which go under the name
 QCD factorization (QCDF)~\cite{9905312},
perturbative QCD (pQCD)~\cite{0004004}, and soft-collinear
effective theory (SCET)~\cite{0107002}. All three are  based on power
expansion in $1/m_b$, where $m_b$ is the $b$-quark mass.
  Factorization of the hadronic matrix elements $\langle h_1 h_2 |{\cal O}_i
  |B\rangle$,
where ${\cal O}_i$ is typically a four-quark or a magnetic moment type operator,
is shown to exist in the leading power in
$1/m_b$ in a class of decays. In addition, these approaches take into account some
contributions in the decays $B \to h_1 h_2$ not included in
the earlier attempts~\cite{bsw,Ali:1997nh,9804363}, in particular the so-called hard
spectator graphs.

Despite being embedded in the $\Lambda/m_b$ approach, justified by
both the large mass, $m_b=O(5$ GeV), and a large energy release in
the decay, with $E_{h_i}=m_B/2$, these methods differ
significantly from each other in a number of important aspects.
For example, these differences pertain to whether one takes into
account the collinear degrees of freedom only as in QCDF and SCET,
or includes also the transverse momenta implemented using the
Sudakov formalism, as followed in the pQCD method. Also, in pQCD,
the power counting is different from the one in QCDF, which makes
some amplitudes differ significantly in the two approaches. The
other differing feature of pQCD and QCDF is  the scale at which
strong interaction effects, including the Wilson coefficients, are
calculated. In pQCD, this scale is low, typically  of order 1  --
2 GeV. In QCDF, typical scales for the Wilson coefficients are
taken as $O(m_b)$, following arguments based on factorization.
There also exist detailed differences between QCDF and SCET,
despite the fact that dedicated studies in the context of SCET
have allowed to gain a better understanding of the QCDF framework.
These differences, though not inherent, lie in how practically the
calculations are done in the two approaches and involve issues
such as the treatment of the so-called charming penguin
contributions~\cite{Ciuchini:1997hb} to the decays $B \to h_1
h_2$. These are argued to be power-suppressed in QCDF, and left as
phenomenological parameters to be determined by data in SCET.
Likewise, the treatment of the hard spectator contribution in
these two approaches is also different. We recall that a generic
factorization formula~\cite{Beneke:2006wv}
\begin{equation}
\langle h_1 h_2 | {\cal O}_i | B \rangle= \Phi_{h_2}(u) * \left(
T^I(u) F^{Bh_1} (0) + C^{II} (\tau, u) * \Xi^{Bh_1}(\tau,0)
\right)
\end{equation}
involves the QCD form factor $ F^{Bh_1} (0)$ and an unknown, non-local
form factor $\Xi^{Bh_1}(\tau,0)$. In QCDF, this non-local form factor
factorizes into  light-cone distribution amplitudes and a jet function
$J(\tau, \omega, v)$,  when the hard-collinear
scale $\sqrt {m_b \Lambda}$ is integrated out. This interpretation of the
hard spectator contribution was not at hand in the BBNS papers~\cite{9905312},
 but was gained subsequently in the SCET-analysis of the
form factors~\cite{Beneke:2003pa}. Amusingly, this SCET-result is not used
in the SCET-based
phenomenology in $B \to h_1 h_2$ decays, for example in the works of Bauer et
al.~\cite{Bauer:2004tj}, where the use of
perturbation theory at the scale $\sqrt {m_b \Lambda}$ is avoided.
Detailed comparisons of their predictions with the data for the decays of
the $\overline{B_d^0}$ (and its charge conjugate)
and $B^\pm$-mesons have been made in the literature. We also refer to a recent
critique~\cite{Beneke:2006wv} of the underlying theoretical assumptions in the
three methods. Data from the B-factory experiments, BABAR and BELLE, 
as well as the CDF collaboration at the Tevatron, do
provide some discrimination among them. With the advent of the LHC physics
program, and the steadily improving  experimental precision at the existing
facilities, it should be possible to disentangle the underlying dynamics
in hadronic $B$-decays.

The experimental program to study non-leptonic decays $\overline{B_s^0} \to
h_1h_2$ has also started~\cite{Chargeconj}, with first measurements for the
branching ratios $\overline{B_s^0} \to K^+ \pi^- $ and
 $\overline{B_s^0} \to K^+ K^- $
made available recently by the CDF collaboration~\cite{ExpBs,CDFBsKKrecent}
at the proton-antiproton collider Tevatron.  Remarkably,
the first direct CP asymmetry involving the decay
$\overline{B_s^0}\to K^+ \pi^-$  and its CP conjugate mode has also been
 reported by
CDF~\cite{CDFBsKKrecent}, which is found to be large, with $A_{\rm
CP}(\overline{B_s^0}\to K^+ \pi^- ) =(39 \pm 15 \pm 8)\%$. This is
in agreement with the predictions of the pQCD approach, as we also
quantify in this paper. With the ongoing $B$-Physics program at
the Tevatron, but, in particular, with the onset of the LHC
experiments, as well as the Super B-factories being contemplated
for the future, we expect a wealth of data involving the decays
 of the hitherto less studied $\overline{B_s^0}$ meson.
The charmless $\overline{B_s^0} \to h_1 h_2$ decays are important
for the CP asymmetry studies and the determination of the inner
angles of the unitarity triangle, in particular $\gamma$ (or
$\phi_3$), which has not yet been precisely measured. In addition,
a number of charmless decays $\overline{B_s^0} \to h_1 h_2$ can be related to
the $\overline{B_d^0} \to h_1 h_2$  decays using SU(3) (or U-spin) symmetry,
and hence data on these decays can be combined to test the
underlying standard model and search for physics beyond the SM under
less (dynamical) model-dependent conditions.
 Anticipating the
experimental developments,
many studies have been devoted to the interesting charmless
 $\overline{B_s^0} \to h_1 h_2$ decays.
Among others, they include detailed estimates undertaken
in the naive  factorization framework~\cite{NFBs}, the so-called
generalized factorization approach~\cite{GFBs},
QCDF~\cite{QCDFBs1,QCDFBs2,Beneke:2006hg}, pQCD~\cite{PQCDBs}
and SCET~\cite{SCETBs}. There are also many
studies~\cite{Bsothers} undertaken, parameterizing the various
parts of the decay amplitudes using distinct topologies and  the
flavor symmetries to relate the $\overline{B_s^0} \to h_1h_2$ and $\overline{B_d^0} \to
h_1 h_2$ decays.
 Possible  New Physics effects in these decays have also been
 explored~\cite{NewPhysicsBs}.

In the applications of the pQCD approach to $\overline{B_s^0} \to
h_1 h_2$ decays, the currently available works concentrate on
specific decays. However, a comprehensive study of the decays
$\overline{B_s^0} \to h_1 h_2$, which have been undertaken in
QCDF and SCET, to the best of our knowledge, is still
lacking in pQCD. Our aim is to fill in this gap and provide a ready
reference to the existing and forthcoming experiments to compare
their data with the predictions in the pQCD approach. In doing this,
we have included the current information on the CKM matrix elements,
updated some input hadronic parameters and have calculated the decay form
factors in the pQCD approach. Since these form factors have to be provided from
outside in  QCDF and SCET (such as by resorting to QCD sum rules), there is
already a potential source of disagreement among these approaches
on this count. However, we remark that the
estimates presented here for the cases $B_s \to PP, PV, VV$ are rather similar
to the corresponding ones in the  existing literature on the light cone QCD
sum rules.
Thus, theoretical predictions presented in this work reflect the detailed
assumptions about the dynamics endemic to pQCD, such as the effective scales
 which are
generated by the strong interaction aspects of the weak non-leptonic two-body
decays, setting the relative strengths of the various competing amplitudes
in magnitudes and phases. As we work in the leading order (LO) in $\alpha_s$,
there is considerable uncertainty related to the scale-dependence, which we
quantify in the estimates of the branching ratios, CP-asymmetries and
polarization fractions for the decays considered in this paper.
Likewise,  parametric uncertainties in the numerical estimates of these
quantities resulting from other input parameters are worked out.
Together, they quantify the theoretical
imprecision in $\overline{B_s^0} \to h_1 h_2$ decays at the current stage
in the pQCD approach. We have made
detailed comparison of our predictions with the existing literature
on the decays  $\overline{B_s^0} \to h_1 h_2$ and in some
benchmark decay widths and rate asymmetries in the corresponding
$\overline{B_d^0} \to h_1 h_2$ decays. Whenever available, we have
also compared our predictions with the data and found that they
are generally compatible with each other.

We also present numerical results for some selected
ratios of the branching ratios involving the decays
$\overline{B_s^0} \to K^+K^-$, $\overline{B_s^0} \to K^+\pi^-$,
$\overline{B_d^0} \to \pi^+\pi^-$ and $\overline{B_d^0} \to
K^-\pi^+$, which are related by SU(3) and U-spin symmetries. For
comparison with other approaches, we also give in Table IV the
contributions of the various topologies for these decays.
The ratios worked out numerically are:
$R_1\equiv
\frac{BR(\overline{B_s^0} \to K^+K^-)}{BR(\overline{B_d^0} \to
\pi^+\pi^-)}$, $R_2\equiv \frac{BR(\overline{B_s^0}
\to K^+K^-)} {BR(\overline{B_d^0} \to K^-\pi^+)}$, and two more, called
$R_3$ and $\Delta$, defined in Eq. (72) and (73), respectively, which
involve the decays $\overline{B_d^0} \to K^-\pi^+$ and
$\overline{B_s^0} \to K^+\pi^-$ and their charge conjugates.
All these ratios have been measured experimentally and the pQCD-based
estimates presented here are in agreement with the data,
except possibly the ratio $R_1$ which turns out too small.
Whether this reflects an intrinsic limitation of the pQCD approach or
the inadequacy of the LO framework remains to be seen, as complete
next-to-leading order (NLO) calculations of all the relevant pieces of
the $B \to h_1 h_2$ decay matrix elements are still not in place.
However, there are sound theoretical arguments why the ratios $R_2$, $R_3$
and $\Delta$ are protected against higher order QCD
corrections, such as the charge conjugation invariance of the strong
interactions (for $R_3$ and $\Delta$), and the dominance of the decay
amplitudes in the numerator and denominator in the ratio $R_2$ by a single
decay topology. The agreement between the pQCD approach and data in
these quantities is, therefore, both non-trivial and encouraging.

This paper is organized as follows: In section II, we briefly
review the pQCD approach and give the essential input quantities
that enter the pQCD approach, including the operator basis used
subsequently and the numerical values of the Wilson coefficients
together with their scale dependence. The wave function of the
$\overline{B_s^0}$-meson, the distribution amplitudes for the
light pseudo-scalar and vector mesons and the input values of the
various mesonic decay constants are also given here. Section III
contains the calculation of the $\overline{B_s^0} \to PP$ mesons,
making explicit the contributions from the so-called emission and
annihilation diagrams. Numerical results for the charge-conjugated
averages of the decay branching ratios, direct CP-asymmetries, the
time-dependent CP asymmetries $S_f$ and the observables $H_f$ in
the time-dependent decay rates are tabulated in Tables III, V and
VI, respectively. These tables also contain detailed comparisons
of our work with the corresponding numerical results obtained in
the QCDF and SCET approaches, as well as with the available data.
Section IV contains the numerical results for the decays
$\overline{B_s^0} \to PV$. They are presented in Tables VII, VIII
and IX for the charge-conjugated averages of the decay branching
ratios, time-integrated CP-asymmetries, the time-dependent CP
asymmetries $S_f$ and the observables $H_f$ in the time-dependent
decay rates, respectively. We also show the corresponding results
from the QCDF approach in  Tables VII and VIII. Section V is
devoted to a study of the decays $\overline{B_s^0} \to VV$, making
explicit the amplitudes for the longitudinal (L), normal (N) and
transverse (T) polarization components of the vector mesons.
Numerical results for the CP-averaged branching ratios are
presented in Table X, and compared with the corresponding results
from the QCDF approach, updated recently in
Ref.~\cite{Beneke:2006hg}, and available data. Results for the
three  polarization fractions $f_0$, $f_{\|}$ , $f_{\perp}$, the
relative strong phases $\phi_{\|}(rad)$, $\phi_{\perp}(rad)$ and
the CP-asymmetries are displayed in Table XI. Appendix A contains
the various functions that enter the factorization formulae in the
pQCD approach. Appendix B gives the analytic formulae for the
$\overline{B_s^0} \to PP$ decays used in the numerical
calculations, while the details of the formulae for the decays
$\overline{B_s^0} \to PV$ and $\overline{B_s^0} \to VV$ are
relegated to Appendix C and D, respectively.

\section{The Effective Hamiltonian and the input quantities}

\subsection{Notations and Conventions}


We specify the weak effective Hamiltonian \cite{buras}:
 \begin{eqnarray}
 {\cal H}_{eff} &=& \frac{G_{F}}{\sqrt{2}}
     \Bigg\{ V_{ub} V_{uq}^{\ast} \Big[
     C_{1}({\mu}) Q^{u}_{1}({\mu})
  +  C_{2}({\mu}) Q^{u}_{2}({\mu})\Big]
  -V_{tb} V_{tq}^{\ast} \Big[{\sum\limits_{i=3}^{10}} C_{i}({\mu}) Q_{i}({\mu})
  \Big ] \Bigg\} + \mbox{H.c.} ,
 \label{eq:hamiltonian01}
 \end{eqnarray}
where $q=d,s$. The functions $Q_{i}$ ($i=1,...,10$) are the local
four-quark operators:
 \begin{itemize}
 \item  current--current (tree) operators
    \begin{eqnarray}
  Q^{u}_{1}=({\bar{u}}_{\alpha}b_{\beta} )_{V-A}
               ({\bar{q}}_{\beta} u_{\alpha})_{V-A},
    \ \ \ \ \ \ \ \ \
   Q^{u}_{2}=({\bar{u}}_{\alpha}b_{\alpha})_{V-A}
               ({\bar{q}}_{\beta} u_{\beta} )_{V-A},
    \label{eq:operator02}
    \end{eqnarray}
     \item  QCD penguin operators
    \begin{eqnarray}
      Q_{3}=({\bar{q}}_{\alpha}b_{\alpha})_{V-A}\sum\limits_{q^{\prime}}
           ({\bar{q}}^{\prime}_{\beta} q^{\prime}_{\beta} )_{V-A},
    \ \ \ \ \ \ \ \ \
    Q_{4}=({\bar{q}}_{\beta} b_{\alpha})_{V-A}\sum\limits_{q^{\prime}}
           ({\bar{q}}^{\prime}_{\alpha}q^{\prime}_{\beta} )_{V-A},
    \label{eq:operator34} \\
     \!\!\!\! \!\!\!\! \!\!\!\! \!\!\!\! \!\!\!\! \!\!\!\!
    Q_{5}=({\bar{q}}_{\alpha}b_{\alpha})_{V-A}\sum\limits_{q^{\prime}}
           ({\bar{q}}^{\prime}_{\beta} q^{\prime}_{\beta} )_{V+A},
    \ \ \ \ \ \ \ \ \
    Q_{6}=({\bar{q}}_{\beta} b_{\alpha})_{V-A}\sum\limits_{q^{\prime}}
           ({\bar{q}}^{\prime}_{\alpha}q^{\prime}_{\beta} )_{V+A},
    \label{eq:operator56}
    \end{eqnarray}
 \item electro-weak penguin operators
    \begin{eqnarray}
     Q_{7}=\frac{3}{2}({\bar{q}}_{\alpha}b_{\alpha})_{V-A}
           \sum\limits_{q^{\prime}}e_{q^{\prime}}
           ({\bar{q}}^{\prime}_{\beta} q^{\prime}_{\beta} )_{V+A},
    \ \ \ \
    Q_{8}=\frac{3}{2}({\bar{q}}_{\beta} b_{\alpha})_{V-A}
           \sum\limits_{q^{\prime}}e_{q^{\prime}}
           ({\bar{q}}^{\prime}_{\alpha}q^{\prime}_{\beta} )_{V+A},
    \label{eq:operator78} \\
     Q_{9}=\frac{3}{2}({\bar{q}}_{\alpha}b_{\alpha})_{V-A}
           \sum\limits_{q^{\prime}}e_{q^{\prime}}
           ({\bar{q}}^{\prime}_{\beta} q^{\prime}_{\beta} )_{V-A},
    \ \ \ \
    Q_{10}=\frac{3}{2}({\bar{q}}_{\beta} b_{\alpha})_{V-A}
           \sum\limits_{q^{\prime}}e_{q^{\prime}}
           ({\bar{q}}^{\prime}_{\alpha}q^{\prime}_{\beta} )_{V-A},
    \label{eq:operator9x}
    \end{eqnarray}
 \end{itemize}
where $\alpha$ and $\beta$ are the color indices and $q^\prime$ are
the active quarks at the scale $m_b$, i.e.
 $q^\prime=(u,d,s,c,b)$.
The left handed current is defined as $({\bar{q}}^{\prime}_{\alpha}
q^{\prime}_{\beta} )_{V-A}= {\bar{q}}^{\prime}_{\alpha} \gamma_\nu
(1-\gamma_5) q^{\prime}_{\beta}  $ and the right handed current
$({\bar{q}}^{\prime}_{\alpha} q^{\prime}_{\beta} )_{V+A}=
{\bar{q}}^{\prime}_{\alpha} \gamma_\nu (1+\gamma_5)
q^{\prime}_{\beta}  $.
 The combinations $a_i$ of Wilson coefficients are
defined as usual \cite{9804363}:
\begin{eqnarray}
a_1= C_2+C_1/3, & a_3= C_3+C_4/3,~a_5= C_5+C_6/3,~a_7=
C_7+C_8/3,~a_9= C_9+C_{10}/3,\nonumber \\
 a_2= C_1+C_2/3, & a_4=
C_4+C_3/3,~a_6= C_6+C_5/3,~a_8= C_8+C_7/3,~a_{10}= C_{10}+C_{9}/3.
\end{eqnarray}
Since we work in the leading order of perturbative QCD
($O(\alpha_s )$), it is consistent to use the leading order Wilson
coefficients. The scale $\mu$ characterizes the typical scale for
the hard scattering, and for four different values of this scale
the results of the Wilson coefficients are listed in
Table~\ref{WCRG}. It is seen that the coefficient $a_2$ and the
penguin operator Wilson coefficients $a_3$, $a_5$ have a large
dependence on the scale which will give large uncertainties to
those channels highly dependent on them, such as the
color-suppressed QCD penguin dominant processes. This situation is
typical of leading order estimates; the reduction in the
scale-dependence requires the calculation of next-to-leading order
results which are beyond the theoretical accuracy to which we are
working.

\begin{table}[tb]
 \caption{Numerical values of the combinations of Wilson coefficients defined
   in the text at different scales ($\mu$).} \label{WCRG}
\begin{center}
  \begin{tabular}{c|c|c|c|c}
  \hline\hline
  $\mu$ (GeV)                 & $2.5$       &  $2.0$       & $1.5$      & $1.0$    \\  \hline
   $a_1$                      & $1.1$       &  $1.1$       & $1.1$      & $1.1$    \\  \hline
   $a_2$($\times 10^{-2}$)    & $1.1$       &  $-2.8$      & $-8.7$     & $-19.4$  \\  \hline
   $a_3$($\times 10^{-3}$)    & $6.2$       &  $7.5$       & $9.7$      & $14.4$   \\  \hline
   $a_4$($\times 10^{-3}$)    & $-32.0$     &  $-35.8$     & $-41.4$    & $-51.3$  \\  \hline
   $a_5$($\times 10^{-3}$)    & $-5.6$      &  $-7.4$      & $-10.5$    & $-17.6$  \\  \hline
   $a_6$($\times 10^{-3}$)    & $-46.8$     &  $-54.9$     & $-68.2$    & $-95.6$  \\  \hline
   $a_7$($\times 10^{-4}$)    & $12.6$      &  $12.7$      & $13.2$     & $14.0$   \\  \hline
   $a_8$($\times 10^{-4}$)    & $9.6$       &  $10.6$      & $12.2$     & $15.7$   \\  \hline
   $a_9$($\times 10^{-4}$)    & $-84.3$     &  $-85.4$     & $-87.2$    & $-91.0$  \\  \hline
   $a_{10}$($\times 10^{-4}$) & $-0.87$     &  $2.3$       & $7.0$      & $15.5$   \\  \hline
    \hline
    \end{tabular}
\end{center}
\end{table}

\begin{figure}
\begin{center}
\includegraphics[scale=0.6]{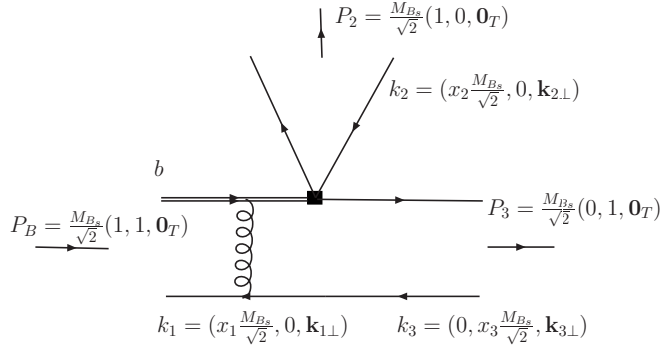}
\caption{The notation in our calculation} \label{notation}
\end{center}
\end{figure}

We consider the $\bar B_s^0\to M_2M_3$ decay. In the emission type
diagrams with the spectator quark $\bar s$
from the initial state $\bar B_s$ recombining to form the meson $M_3$, we
denote  the emitted meson as $M_2$ while the recoiling meson is
$M_3$. The factorization formula is then denoted as $F_{\bar B_s
\to M_3}$. It is convenient to define two light-like vectors: $n$
and $v$. These two vectors satisfy  $n^2=v^2=0$ and $n\cdot v=1$.
The meson $M_2$ is moving along the direction of $n=(1,0,
{\bf{0}}_T)$ and $M_3$ is on $v=(0,1,{\bf{0}}_T)$, we use $x_i$ to
denote the momentum fraction of the anti-quark in each meson,
$k_{i\perp}$ to denote the transverse momentum of the anti-quark:
\begin{eqnarray}
P_B&=&\frac{M_{B_s}}{\sqrt 2}(1,1,{\bf{0}}_T),\;\;
P_2=\frac{M_{B_s}}{\sqrt 2}(1,0,
{\bf{0}}_T),\;\;P_3=\frac{M_{B_s}}{\sqrt
2}(0,1, {\bf{0}}_T),\nonumber\\
k_1&=&(x_1\frac{M_{B_s}}{\sqrt 2},0, {\bf
k}_{1\perp}),\;\;k_2=(x_2\frac{M_{B_s}}{\sqrt 2},0, {\bf
k}_{2\perp}),\;\;k_3=(0,x_3\frac{M_{B_s}}{\sqrt 2},{\bf
k}_{3\perp}),
\end{eqnarray}
which are shown in Figure~\ref{notation}.

\subsection{Wave Functions of the $B_s$ Meson}

In order to calculate the analytic formulas of the decay amplitude, we
use the light cone wave functions $\Phi_{M,\alpha\beta}$ decomposed
in terms of the spin structure. In general, $\Phi_{M,\alpha\beta}$ with
Dirac indices $\alpha,\beta$ can be decomposed into 16 independent
components, $1_{\alpha\beta}$, $\gamma^\mu_{\alpha\beta}$,
$\sigma^{\mu\nu}_{\alpha\beta}$,
$(\gamma^\mu\gamma_5)_{\alpha\beta}$, $\gamma_{5\alpha\beta}$. If
the considered meson $M$ is the $B_s$ meson, a heavy pseudo-scalar
meson,  the $B_s$ meson light-cone matrix element can be decomposed
as \cite{grozin,qiao}
\begin{eqnarray}
&&\int d^4ze^{ik_1\cdot z}
\langle 0|\bar{b}_\alpha(0)s_\beta(z)| {B_s}(P_{B_s})\rangle \nonumber\\
&=&\frac{i}{\sqrt{6}}\left\{(\not\! P_{B_s}+M_{B_s})\gamma_5
\left[\phi_{B_s} ({ k_1})-\frac{\not n -\not v }{\sqrt{2}}
\bar{\phi}_{B_s}({ k_1})\right]\right\}_{\beta\alpha}. \label{aa1}
\end{eqnarray}
From the above equation, one can see that there are two Lorentz
structures in the $B_s$ meson distribution amplitudes. They obey
the following normalization conditions
\begin{equation}
 \int\frac{d^4 k_1}{(2\pi)^4}\phi_{B_s}({
k_1})=\frac{f_{B_s}}{2\sqrt{6}}, ~~~\int \frac{d^4
k_1}{(2\pi)^4}\bar{\phi}_{B_s}({ k_1})=0.
\end{equation}

In general, one should consider these two Lorentz structures in the
calculations of $B_s$ meson decays. However, it is found that
the contribution of $\bar{\phi}_{B_s}$ is numerically small
\cite{ly}, thus its contribution can be
 neglected. With this approximation,  we only retain the first term in the
square bracket from  the full Lorentz
structure in Eq.~(\ref{aa1})
\begin{equation}
 \Phi_{B_s}= \frac{i}{\sqrt{6}} (\not \! P_{B_s} +M_{B_s}) \gamma_5
\phi_{B_s} ({ k_1}). \label{bmeson}
\end{equation}
   In the next section, we
will see that the hard part is always independent of one of the
$k_1^+$ and/or $k_1^-$, if we make  the approximations shown in the next
section. The ${B_s}$ meson wave function is then a function of the
variables $k_1^-$ (or $k_1^+$) and $k_1^\perp$ only,
\begin{equation}
  \phi_{B_s} (k_1^-,
k_1^\perp)=\int \frac{d k_1^+}{2\pi} \phi_{B_s} (k_1^+, k_1^-,
k_1^\perp)~. 
\label{int}
\end{equation}
Then,  the $B_s$ meson's wave function in the $b$-space can be
expressed by
\begin{equation}
 \Phi_{B_s}(x,b) = \frac{i}{\sqrt{6}}
\left[ \not \! P_{B_s} \gamma_5 + M_{B_s} \gamma_5 \right]
\phi_{B_s}(x,b),
\end{equation}
where $b$ is the conjugate space coordinate of the transverse momentum
$k^\perp$.

\begin{figure}
 \centerline{
\psfig{file=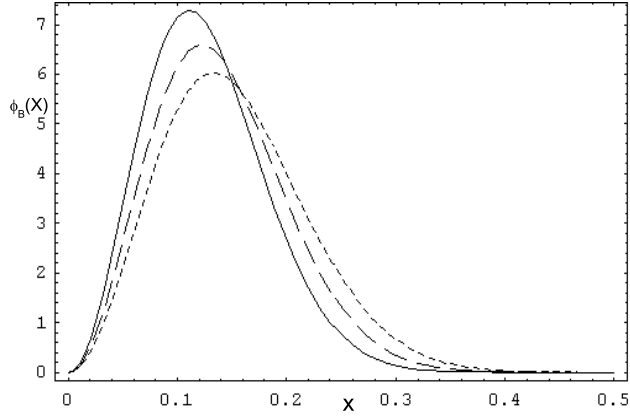,width=9.0cm,angle=0}} \caption{ $B_s$
meson distribution amplitudes. The solid-, dashed-, and
tiny-dashed- lines correspond to $\omega_B=0.45~\mathrm{GeV}$,
$0.5~\mathrm{GeV}$,  and $0.55~\mathrm{GeV}$.}\label{bswave}
 \end{figure}

In this study, we use the model function similar to that of the
$B$ meson which is
\begin{equation}
\phi_{B_s}(x,b) = N_{B_s} x^2(1-x)^2 \exp \left[ -\frac{M_{B_s}^2\
x^2}{2 \omega_b^2} -\frac{1}{2} (\omega_b b)^2 \right],\label{waveb}
\end{equation}
with $N_{B_s}$ the normalization factor. In recent years, a lot of studies have been
performed for the $B_d^0$ and $B^\pm$ decays  in the
pQCD approach \cite{0004004}. The parameter
$\omega_b=0.40~\mathrm{GeV}$  has been fixed there using the rich
experimental data on the $B_d^0$ and $B^\pm$ mesons. In the SU(3) limit, this
parameter should be
the same in $B_s$ decays. Considering  a small SU(3) breaking,
the $s$ quark momentum fraction here should be  a little
larger than that of the $u$ or $d$ quark in the lighter $B$ mesons, since  the
  $s$ quark is heavier than the $u$ or $d$ quark. The shape
of the distribution amplitude is shown in Fig.\ref{bswave} for
$\omega_B=0.45~\mathrm{GeV}$, $0.5~\mathrm{GeV}$, and
$0.55~\mathrm{GeV}$. It is easy to see that the larger $\omega_b$ gives a
larger momentum fraction to the $s$ quark.  We will use
$\omega_b=0.50\pm 0.05~\mathrm{GeV}$ in this paper for the $B_s$ decays.

\subsection{Distribution Amplitudes of Light Pseudo-scalar Mesons}

The decay constant $f_P$ of the pseudo-scalar meson is defined by the
matrix element of the axial current:
\begin{eqnarray}
\langle 0|\bar q_1\gamma_\mu\gamma_5 q_2|P(P)\rangle=if_PP_\mu.
\end{eqnarray}
The pseudo-scalar  decay constants are shown in table~\ref{fpi}, taken from
the Particle Data Group~\cite{PDG}. The vector meson longitudinal decay constants are
extracted from the data on $\tau^- \to (\rho^-,K^{*-}) \nu_\tau$~\cite{PDG}
and the transverse decay constants are taken from QCD sum
rules~\cite{LCSRBZ,adoptedvectorwf}. The input values given in
table~\ref{fpi} are very similar to the ones used in~\cite{QCDFBs1}.

 \begin{table} \caption{Input values of the decay constants  of the pseudo-scalar and vector
mesons (in MeV)~\cite{PDG,Li:2006jv}}
\begin{tabular}{cccccccccc}
\hline\hline
 $f_\pi $ & $f_K $ & $f_\rho $ & $ f_\rho^T $ & $ f_\omega $ & $ f_\omega^T $
 & $ f_{K^*} $ & $ f_{K^*}^T $ & $f_\phi $ & $
f_\phi^T $  \\
131 & 160& $ 209\pm 2$& $ 165\pm 9 $&
 $ 195\pm 3$&
 $ 145\pm 10$&
$ 217\pm 5$&
 $185\pm 10$&
 $ 231\pm 4$&
 $ 200\pm 10$\\
\hline \hline
\end{tabular}\label{fpi}
 \end{table}

The light-cone distribution amplitudes are defined by the matrix
elements of the non-local light-ray operators at the light-like
separation $z_\mu$ with $z^2=0$, and sandwiched between the vacuum
and the meson state. The two-particle light-cone distribution
amplitudes of an outgoing  pseudo-scalar meson $P$, up to twist-3
accuracy, are defined by \cite{PseudoscalarWV}:
\begin{eqnarray}
\langle P(P)|{\bar q}_2(z)\gamma_\mu \gamma_5q_1(0)|0\rangle&=&
-i{f_P} P_{\mu}\int_0^1 dx e^{i x P\cdot z}\phi_2(x),
\label{paf}\\
\langle P(P)|{\bar q}_2(z)\gamma_5q_1(0)|0\rangle&=& -i{f_P}
m_{0}\int_0^1 dx e^{i x P\cdot z}\phi_3^P(x)\;,
\label{psf}\\
\langle P(P)|{\bar q}_2(z)\sigma_{\mu\nu} \gamma_5q_1(0)|0\rangle&=&
\frac{i}{6}{f_P} m_{0} (P_{\mu}z_\nu-P_{\nu}z_\mu) \int_0^1 dx e^{i
x P\cdot z}\phi_3^\sigma(x)\;. \label{pt}
\end{eqnarray}
where we have omitted the Wilson line connecting the two
space-time points. $\phi_2(x)$, $\phi_3^P(x)$ and
$\phi_3^\sigma(x)$ have unit normalization. $M_P$ is the mass of
the pseudo-scalar meson, $m_0$ is the chiral scale parameter which
is defined using the meson mass and the quark masses as $m_0=
\frac{M_P^2}{m_{q_1}+m_{q_2}}$: $m_0^\pi=1.4$ GeV and $m_0^K=1.9$
GeV, $x$ is the momentum fraction associated with the quark $q_2$.
It is easy to observe that the contribution from $\phi_2(x)$,
independent of the mass, is twist-2. The contributions from
$\phi_3^P(x)$ and $\phi_3^\sigma(x)$, proportional to
$r=m_{0}/M_{B_s}$, are twist-3.

The above definitions can be collected as
\begin{eqnarray}
\langle P(P)|{\bar q}_{2\beta}(z)q_{1\alpha}(0)|0\rangle &=&
-\frac{i}{\sqrt{6}}\int_0^1 dx e^{ixP\cdot z} \left[\gamma_5\not\!
P\phi^A(x) +m_0\gamma_5\phi^P(x) -m_0\sigma^{\mu\nu}\gamma_5
P_{\mu}z_{\nu} \frac{\phi^{\sigma}(x)}{6}\right]_{\alpha\beta}  \nonumber\\
&=&-\frac{i}{\sqrt{6}} \int_0^1dx e^{ixP\cdot z}\left[\gamma_5\not
\! P\phi^A(x) + \gamma_5m_0\phi^P(x) +m_0\gamma_5(\not \! n\not
\!v-1) \phi^T(x)\right]_{\alpha\beta}\;,\nonumber\\
&&
 \label{fpd}
\end{eqnarray}
with the redefinitions of the distribution amplitudes:
\begin{eqnarray} \phi^A(x)=\frac{f_P}{2\sqrt 6}
\phi_2(x),\,\,\phi^P(x)=\frac{f_P}{2\sqrt 6}\phi_3^P(x)
,\;\;\phi^\sigma(x)=\frac{f_P}{2\sqrt 6}\phi_3^\sigma(x),
\end{eqnarray} and we have performed the integration by parts for the
third terms and $\phi^T(x)=\frac{1}{6}\frac{d}{dx}\phi^\sigma(x)$.

In the pQCD approach, the only non-perturbative inputs are the
meson decay constants and meson light cone distribution
amplitudes, and they are both channel independent. The meson decay
constants are either measured through the leptonic decays of the
mesons and the semileptonic decays of the $\tau$-lepton or
calculated from the measured ones using broken SU(3) symmetry.
Therefore there is not much uncertainty in them. The wave
functions depend on the factorization scale and also the
factorization scheme. In principle, they should be determined by
experiment. Although there is no direct experimental measurement
for the moments yet, the non-leptonic $B^0$ and $B^\pm$ decays
already give much information on them \cite{0004004,li2003}. Since
the pQCD approach gives very good results for these decays,
especially the direct CP asymmetries in $B^0\to \pi^+\pi^-$ and
$B^0 \to K^+\pi^-$ decays \cite{direct}, we will use the well
constrained  light cone distribution amplitudes of the mesons in
these papers~\cite{PseudoscalarWV} (see \cite{Pseudoscalar-Ball}
for a summary and update of the LCDAs):
\begin{eqnarray}
 \phi_{\pi}^A(x) &=& \frac{3f_{\pi}}{\sqrt{6}} x(1-x)[ 1 +0.44C_2^{3/2}(t)], \\
 \phi_{\pi}^P(x) &=& \frac{f_{\pi}}{2\sqrt{6}}[1 +0.43C_2^{1/2}(t) ], \\
 \phi_{\pi}^T(x) &=& -\frac{f_{\pi}}{2\sqrt{6}}[C_1^{1/2} (t)+0.55 C_3^{1/2} (t) ]
 ,\\
 \phi_{K}^A(x) &=& \frac{3f_{K}}{\sqrt{6}}x(1-x)[1+0.17C_1^{3/2}(t)+0.2C_2^{3/2}(t)], \\
 \phi_{K}^P(x) &=& \frac{f_{K}}{2\sqrt{6}} [1+0.24C_2^{1/2}(t)], \\
 \phi_{K}^T(x)
 &=&-\frac{f_{K}}{2\sqrt{6}}[C_1^{1/2} (t)+0.35 C_3^{1/2} (t)] ,
\end{eqnarray}
with Gegenbauer polynomials defined as:
\begin{equation}
\begin{array}{ll}
 C^{1/2}_{1}(t)=t ,&C^{3/2}_{1}(t)=3t  \\
 C_2^{1/2}(t)=\frac{1}{2} (3t^2-1),& C_2^{3/2} (t)=\frac{3}{2}
(5t^2-1),
 \\
C_3^{1/2} (t) = \frac{1}{2} t (5t^2 -3)~,
\end{array}
\end{equation}
and $t=2x-1$. In the LCDAs for $\phi_\pi^A(x)$, $\phi_\pi^P(x)$ and
$\phi_K^P(x)$, we have dropped the terms proportional to  $C_4^{1/2,3/2}$,
and take into account only the first two terms in their expansion, consistently
with the rest of the LCDAs.  
 These distribution amplitudes are very close to the
previous QCD sum rule results~\cite{PseudoscalarWV}. In recent
years, there have been continuing updates of the light cone
distribution amplitudes  \cite{PseudoscalarUpdate}. However, the
changes in the coefficients of the Gegenbauer polynomials do
not affect our results significantly, as shown in the next section.
We also point out that
  the default value of the scale at which the Gegenbauer coefficients
are given above is 1 GeV. 
  However, the scale of the perturbative calculation in the pQCD
approach where these LCDAs enter is typically 2 GeV.  The LCDAs can be scaled
  up, as the required anomalous
dimensions are known. We have done this and find that the resulting numerical
differences are small. Strictly speaking, these differences are part of the
NLO corrections. With the current theoretical accuracy, they can be absorbed
in the  uncertainties on the input Gegenbauer coefficients in the numerical calculations.

As for the mixing of $\eta$ and $\eta^\prime$, we use the
quark flavor basis proposed by Feldmann, Kroll and Stech \cite{FKmixing},
i.e. these two mesons are made of $\bar nn=(\bar uu+\bar dd)/\sqrt
2$ and $\bar ss$:
\begin{equation}
   \left( \begin{array}{c}
    |\eta\rangle \\ |\eta^\prime\rangle
   \end{array} \right)
   = U(\phi)
   \left( \begin{array}{c}
    |\eta_{n}\rangle \\ |\eta_s\rangle
   \end{array} \right),
\end{equation}
with the matrix,
\begin{equation}
U(\phi)=\left( \begin{array}{cc}
    \cos\phi & ~-\sin\phi \\
    \sin\phi & \phantom{~-}\cos\phi
   \end{array} \right)\;,
\end{equation}
where the mixing angle $\phi=39.3^\circ\pm1.0^\circ$. In principle,
this mixing mechanism is equivalent to the singlet and octet
formalism, as discussed in~\cite{FKmixing}, and  the advantage
here is that explicitly only two decay constants are needed:
\begin{eqnarray}
   \langle 0|\bar n\gamma^\mu\gamma_5 n|\eta_n(P)\rangle
   &=& \frac{i}{\sqrt2}\,f_n\,P^\mu \;,\nonumber \\
   \langle 0|\bar s\gamma^\mu\gamma_5 s|\eta_s(P)\rangle
   &=& i f_s\,P^\mu \;.\label{deffq}
\end{eqnarray}

We assume that the distribution amplitudes of $\bar{n}n$ and $\bar ss$ are
 the same
as the distribution amplitudes of $\pi$, except for the different decay
constants and the chiral scale parameters. We use~\cite{FKmixing}
\begin{eqnarray}
 f_n=(1.07\pm0.02)f_{\pi}=139.1\pm 2.6~{\rm MeV}, \ \ \
 f_s=(1.34\pm0.06)f_\pi=174.2 \pm 7.8~{\rm MeV},
 \end{eqnarray}
as the averaged results from the experimental data. The chiral
enhancement factors are chosen as
 \begin{eqnarray}
  m_0^{\bar {n} n}=\frac{1}{2m_n}[m^2_\eta\cos^2\phi+
  m_{\eta'}^2\sin^2\phi-\frac{\sqrt 2f_s}{f_n}(m_{\eta'}^2-m_\eta^2)\cos\phi\sin\phi],\\
  m_0^{\bar{s} s}=\frac{1}{2m_s}[m^2_{\eta'}\cos^2\phi+
  m_{\eta}^2\sin^2\phi-\frac{f_n}{\sqrt
  2f_s}(m_{\eta'}^2-m_\eta^2)\cos\phi\sin\phi].
\end{eqnarray}

There are gluonic contributions  which have been investigated in
\cite{GluonicCKL}, with the result that these parts do not change the numerical
results significantly. So, we will not consider this kind of contribution in
this work.

\subsection{Distribution Amplitudes of Light Vector Mesons}

We choose the vector meson momentum $P$ with $P^2=M_V^2$, which is
mainly in the plus direction. The polarization vectors $\epsilon$,
satisfying $P\cdot \epsilon=0$, include one longitudinal
polarization vector $\epsilon_L$ and two transverse polarization
vectors $\epsilon_T$. Following a similar procedure as for the
pseudo-scalar mesons, we can derive the vector meson distribution
amplitudes up to twist-3 \cite{rho}:
\begin{eqnarray}
\langle V(P,\epsilon^*_L)|\bar q_{2\beta}(z) q_{1\alpha}
(0)|0\rangle &=&\frac{1}{\sqrt{6}}\int_0^1 dx e^{ixP\cdot z}
\left[M_V\not\! \epsilon^*_L \phi_V(x) +\not\! \epsilon^*_L\not\!
P \phi_{V}^{t}(x) +M_V \phi_V^s(x)\right]_{\alpha\beta},
\label{lpf}\\
\ \ \ \
 \langle V(P,\epsilon^*_T)|\bar q_{2\beta}(z) q_{1\alpha}
(0)|0\rangle &=&\frac{1}{\sqrt{6}}\int_0^1 dx e^{ixP\cdot z}
\left[ M_V\not\! \epsilon^*_T\phi_V^v(x)+ \not\!\epsilon^*_T\not\!
P\phi_V^T(x)\right.
\nonumber\\
& & \left.+M_V
i\epsilon_{\mu\nu\rho\sigma}\gamma_5\gamma^\mu\epsilon_T^{*\nu}
n^\rho v^\sigma \phi_V^a(x)\right ]_{\alpha\beta}\;, \label{spf}
\end{eqnarray}
for longitudinal polarization and transverse polarization,
respectively. Here  $x$ is the momentum fraction associated with the
$q_2$ quark. We adopt the convention $\epsilon^{0123}=1$ for the
Levi-Civita tensor $\epsilon^{\mu\nu\alpha\beta}$.

The twist-2 distribution amplitudes for a longitudinally polarized
vector meson can be parameterized as:
\begin{eqnarray}
\phi_\rho (x)&=&\frac{3f_\rho}{\sqrt{6}} x (1-x)\left[1+
a_{2\rho}^{||}C_2^{3/2} (t) \right]\;,\label{phirho}\\
\phi_\omega(x)&=&\frac{3f_\omega}{\sqrt{6}} x (1-x)\left[1+
a_{2\omega}^{||}C_2^{3/2} (t)\right]\;,\\
\phi_{K^*}(x)&=&\frac{3f_{K^*}}{\sqrt{6}} x
(1-x)\left[1+a_{1K^*}^{||}C^{3/2}_1(t)+
a_{2K^*}^{||}C_2^{3/2} (t)\right]\;,\\
\phi_{\phi}(x)&=&\frac{3f_{\phi}}{\sqrt{6}} x (1-x)\left[1+
a_{2\phi}^{||}C_2^{3/2} (t)\right]\;.\label{phiphi}
\end{eqnarray}
 Here $f_{V}$ is the decay constant of
the vector meson with longitudinal  polarization, whose values are
shown in table \ref{fpi}. The Gegenbauer moments have been studied
extensively in the literatures \cite{rho,previousvectorwf}, here
we adopt the following values from the recent
updates~\cite{adoptedvectorwf,LCSRBZ,twist3}:
\begin{eqnarray}
a_{1K^*}^{||}=0.03\pm0.02,\;\;
a_{2\rho}^{||}=a_{2\omega}^{||}=0.15\pm0.07,\;\;
a_{2K^*}^{||}=0.11\pm0.09,\;\;a_{2\phi}^{||}=0.18\pm0.08,
\end{eqnarray}
 and we use the asymptotic form
\cite{PolarizationLi}:
\begin{eqnarray}
\phi^t_V(x) = \frac{3f^T_V}{2\sqrt 6}t^2,
  \hspace*{0.5cm} \phi^s_V(x)=\frac{3f_V^T}{2\sqrt 6} (-t)~.
\end{eqnarray}

The twist-2 transversely polarized distribution amplitudes
$\phi_V^T$ have a similar form as the longitudinally polarized ones
in eq.(\ref{phirho}-\ref{phiphi}), with the moments
\cite{adoptedvectorwf,LCSRBZ}:
\begin{eqnarray}
a_{1K^*}^\perp=0.04\pm0.03,\;\;
a_{2\rho}^{\perp}=a_{2\omega}^{\perp}=0.14\pm0.06,\;\;
a_{2K^*}^{\perp}=0.10\pm0.08,\;\;a_{2\phi}^{\perp}=0.14\pm0.07.
\end{eqnarray}
 The asymptotic form of the twist-3 distribution amplitudes
$\phi^v_V$ and $\phi_V^a$ are
\begin{eqnarray}
\phi_V^v(x)&=&\frac{3f_V}{8\sqrt6}(1+t^2),\;\;\; \ \ \
 \phi_V^a(x)=\frac{3f_V}{4\sqrt6}(-t).
\end{eqnarray}
The above choices of vector meson distribution amplitudes can essentially
explain the measured $B\to K^* \phi$, $B\to K^*\rho$ and $B\to
\rho \rho$ polarization fractions
\cite{kphi,krho,rhorho,PolarizationLi}, together with the right
branching ratios.

\subsection{A Brief Review of the pQCD Approach}

The basic idea of the pQCD approach is that it takes into account the
transverse momentum of the valence quarks in the hadrons which results in the Sudakov
factor in the decay amplitude.  As an example, taking the first diagram in
Fig.~\ref{emission}, the emitted particle $M_2$ in the decay can
be factored out (in terms of the appropriate vacuum to $M_2$ transition matrix
element) and the rest of the amplitude can be expressed as the convolution
of the wave functions $\phi_{B_s}$, $\phi_{M_3}$ and the hard
scattering kernel $T_H$, integrated over the longitudinal and the
transverse momenta. Thus, 
\begin{eqnarray}
{\cal M}\propto \int^1_0dx_1dx_3\int
 \frac{d^2{\vec{k}}_{1T}}{(2\pi)^2}\frac{d^2{\vec{k}}_{3T}}{(2\pi)^2}
\phi_B(x_1,{\vec{k}}_{1T},p_1,t)
T_H(x_1,x_3,{\vec{k}}_{1T},{\vec{k}}_{3T},t)
\phi_V(x_3,{\vec{k}}_{3T},p_3,t)~.
\end{eqnarray}
It is convenient to calculate the decay amplitude in 
coordinate space. Through the Fourier transformation, the above
equation can be expressed as:
\begin{eqnarray}
{\cal M}\propto\int^1_0dx_1dx_3\int
{d^2{\vec{b}}_{1}}{d^2{\vec{b}}_{3}}{\phi}_B(x_1,{\vec{b}}_{1},p_1,t)
T_H(x_1,x_3,{\vec{b}}_{1},{\vec{b}}_{3},t){\phi}_V(x_3,{\vec{b}}_{3},p_3,t)~.
\end{eqnarray}
Loop effects can, in principle, be taken into account in the above expression.
In general, individual higher order diagrams suffer
from two types of infrared divergences: soft and collinear. Soft
divergence arise from the region of a loop momentum where all it's
 components in the light-cone coordinate vanish:
\begin{eqnarray}
l^{\mu}=(l^+,l^-,\vec{l}_T)=(\Lambda,\Lambda,\vec{\Lambda}).
\end{eqnarray}
Collinear divergence originates from the gluon momentum region
which is parallel to the massless quark momentum,
\begin{eqnarray}
l^{\mu}=(l^+,l^-,\vec{l}_T) \sim
(m_B,{\Lambda}^2/m_B,\vec{\Lambda}).
\end{eqnarray}
In both cases, the loop integration corresponds to ${\int d^4
l/l^4 \sim \log{\Lambda}}$, so logarithmic divergences are
generated. It has been shown order by order in perturbation theory
that these divergences can be separated from the hard perturbative kernel and
absorbed into the meson wave functions using the eikonal approximation
\cite{Li:1994iu} . One also encounters double logarithm divergences when
soft and collinear momenta overlap. These large double logarithm
can be re-summed into the Sudakov factor and the explicit form is
given in Appendix \ref{PQCDfunctions}.

Loop corrections to the weak decay vertex
also give rise to double logarithms. For example, the first
diagram in Fig.~\ref{emission}  gives an amplitude proportional to
$1/(x_3^2 x_1)$. In the threshold region with $x_3\to 0$,
additional collinear divergences are associated with the internal
quark. The QCD loop corrections to the weak vertex can
produce the double logarithm $\alpha_s\ln^2 x_3$ and the re-summation
of this type of double logarithms leads to the Sudakov factor
$S_t(x_3)$. Similarly, the re-summation of $\alpha_s\ln^2 x_1$ due to
loop corrections in the other diagram lead to the Sudakov factor
$S_t(x_1)$. These double logarithm can also be factored out from
the hard part and grouped into the quark jet function \cite{L3}.
This type of factor decreases faster than any power of ${x}$
as ${x\rightarrow 0}$, so it removes the endpoint singularity. For
simplicity, this factor has been parameterized in a form which is
independent of the  decay channels, twist and flavors~\cite{L4}.

Combining all the elements together,  the typical
factorization formula in the pQCD approach reads as:
\begin{eqnarray}
{\cal M}&\propto&\int^1_0dx_1dx_3\int
 {d^2{\vec{b}}_{1}}{d^2{\vec{b}}_{3}}{\phi}_B(x_1,{\vec{b}}_{1},p_1,t)\nonumber\\
&&\times
T_H(x_1,x_3,{\vec{b}}_{1},{\vec{b}}_{3},t){\phi}_{M_3}(x_3,{\vec{b}}_{3},p_3,t)
S_t(x_3)\exp[-S_B(t)-S_3(t)]~.
\end{eqnarray}
The threshold Sudakov function $S_t(x_3)$ and the Sudakov exponents $S_B(t)$,
$S_2(t)$ and $S_3(t)$ are given in Appendix A. Again, strictly speaking, the
Sudakov improvements result from higher order contributions. However, they are
included traditionally in the pQCD approach, though in most applications,
 the perturbative function
 $T_H(x_1,x_3,{\vec{b}}_{1},{\vec{b}}_{3},t){\phi}_{M_3}(x_3,{\vec{b}}_{3},p_3,t)$
is calculated only in the leading order in $\alpha_s$~. 

\section{Calculation of  the $B_s\to PP$ decay amplitudes in the PQCD approach}

In the following we will give the general factorization formulae for
$\bar B_s\to PP$ decays, and the pQCD functions can be found in
Appendix \ref{PQCDfunctions}. We will use $LL$ to denote the
contribution from $(V-A)(V-A)$ operators, $LR$ to denote the
contribution from $(V-A)(V+A)$ operators and $SP$ to denote the
contribution from $(S-P)(S+P)$ operators which result from the Fierz
transformation of the $(V-A)(V+A)$ operators.

\subsection{Emission Diagram}

\begin{figure}
\vspace{-0.0cm}
\begin{center}
\psfig{file=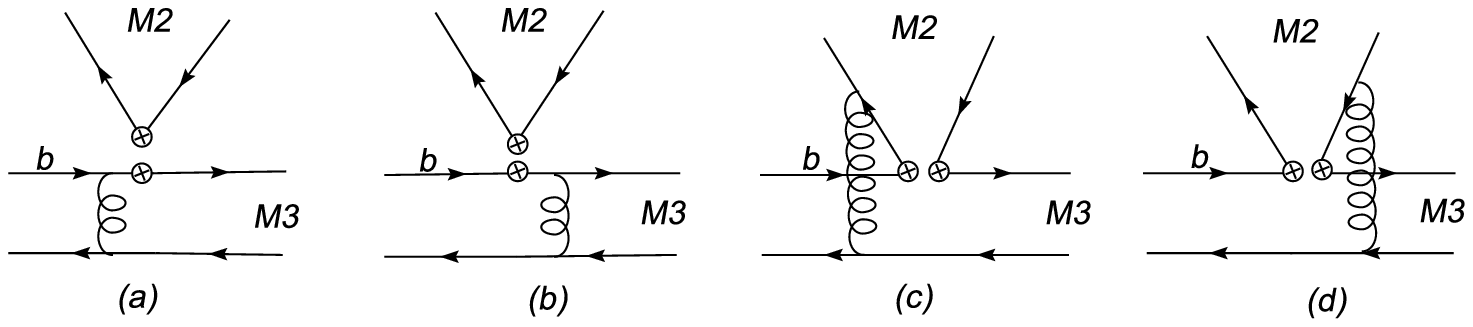,width=12.0cm,angle=0}
\end{center}
\vspace{-0.2cm} \caption{{  The Feynman diagrams for emission
contribution, with possible four-quark operator insertions
}}\label{emission}
\end{figure}

The emission diagrams are depicted in Figure~\ref{emission}. The
first two diagrams are called factorizable. They will give
the $B\to M_3$ decay form factor, if we factor out the corresponding
Wilson coefficients $a_i$.
\begin{itemize}
\item $(V-A)(V-A)$ operators:
  \begin{eqnarray}
  f_{M_2} F^{LL}_{B_s\to M_3} (a_i)&=&8\pi
  C_FM_{B_s}^4f_{M_2}\int^1_0dx_1dx_3\int^\infty_0b_1db_1b_3db_3
\phi_{B_s}(x_1,b_1)
  \Big\{a_i(t_a) E_e(t_a)
  \nonumber\\
  &&\times \Big[(1+x_3)\phi_3^A(x_3)+r_3(1-2x_3)(\phi_3^P(x_3)+\phi_3^T(x_3))
  \Big]h_e(x_1,x_3,b_1,b_3)
 \nonumber\\ && \;\;+2r_3\phi_3^P(x_3)a_i(t_a^\prime) E_e(t_a^\prime)h_e(x_3,x_1,b_3,b_1)
 \Big\},\label{ppefll}
  \end{eqnarray}

\item $(V-A)(V+A)$ operators:
\begin{eqnarray}
  F^{LR}_{B_s\to M_3}(a_i)&=&-F^{LL}_{B_s\to M_3}(a_i),\label{ppeflr}
\end{eqnarray}

\item $(S-P)(S+P)$ operators:
  \begin{eqnarray}
 f_{M_2} F^{SP}_{B_s\to M_3}(a_i)&=& 16\pi r_2
  C_FM_{B_s}^4f_{M_2}\int^1_0dx_1dx_3\int^\infty_0b_1db_1b_3db_3
\phi_{B_s}(x_1,b_1)
  \Big\{a_i(t_a)E_e(t_a)\nonumber\\
  &&\times\Big[\phi_3^A(x_3)+r_3(2+x_3)\phi_3^P(x_3)-r_3x_3\phi_3^T(x_3)\Big]
  h_e(x_1,x_3,b_1,b_3)\nonumber\\
  &&\;\;\;+2
  r_3\phi_3^P(x_3)a_i(t^\prime_a)E_e(t_a^\prime)h_e(x_3,x_1,b_3,b_1)\Big\},\label{ppefsp}
  \end{eqnarray}
\end{itemize}
with $C_F=4/3$ and $a_i$ the corresponding Wilson coefficients for
specific channels. In the above functions, $r_{i}=m_{0i}/m_{B_s}$,
where $m_{0i}$ is the chiral scale parameter. The functions $E_i$, the
definitions of the factorization scales $t_i$ and the hard functions $h_i$ are given in
Appendix~\ref{PQCDfunctions}.

The last two diagrams in Fig.\ref{emission} (c)(d) are the
non-factorizable diagrams, whose contributions are
\begin{itemize}
\item $(V-A)(V-A)$ operators:
\begin{eqnarray} M_{B_s\to M_3}^{LL}(a_i)&=&32\pi
C_FM_{B_s}^4/\sqrt{6}\int^1_0dx_1dx_2dx_3\int^\infty_0b_1db_1b_2db_2
\phi_{B_s}(x_1,b_1)\phi_2^A(x_2)
\nonumber\\
&&\times
\Big\{\Big[(1-x_2)\phi_3^A(x_3)-r_3x_3(\phi_3^P(x_3)-\phi_3^T(x_3))\Big]
a_i(t_b)E_e^\prime(t_b)\nonumber\\
&&~\times h_n(x_1,1-x_2,x_3,b_1,b_2)+h_n(x_1,x_2,x_3,b_1,b_2)\nonumber\\
 &&\;\;\times\Big[-(x_2+x_3)\phi_3^A(x_3)+r_3x_3(\phi_3^P(x_3)+\phi_3^T(x_3))\Big]
 a_i(t_b^\prime) E_e^\prime(t_b^\prime)\Big\},\label{ppenll}
 \end{eqnarray}

\item $(V-A)(V+A)$ operators:
\begin{eqnarray} M_{B_s\to M_3}^{LR}( a_i)&=&32\pi C_FM_{B_s}^4r_2/\sqrt{6}
      \int^1_0dx_1dx_2dx_3\int^\infty_0b_1db_1b_2db_2\phi_{B_s}(x_1,b_1)\nonumber\\
     &&\times \Big\{h_n(x_1,1-x_2,x_3,b_1,b_2)\Big[(1-x_2)\phi_3^A(x_3)
     \left(\phi_2^P(x_2)+\phi_2^T(x_2)\right)\nonumber\\
     &&\;\;+r_3x_3\left(\phi_2^P(x_2)-\phi_2^T(x_2)\right)
     \left(\phi_3^P(x_3)+\phi_3^T(x_3)\right)\nonumber\\
     &&\;\;+(1-x_2)r_3\left(\phi_2^P(x_2)+\phi_2^T(x_2)\right)\left(\phi_3^P(x_3)
      -\phi_3^T(x_3)\right)\Big]a_i(t_b)
         E_e^\prime(t_b) \nonumber\\
       &&\;\;-h_n(x_1,x_2,x_3,b_1,b_2)\Big[x_2\phi_3^A(x_3)(\phi_2^P(x_2)-\phi_2^T(x_2))\nonumber\\
       &&\;\;+r_3x_2(\phi_2^P(x_2)-\phi_2^T(x_2))(\phi_3^P(x_3)-\phi_3^T(x_3))\nonumber\\
      &&\;\;+r_3x_3(\phi_2^P(x_2)+\phi_2^T(x_2))(\phi_3^P(x_3)+\phi_3^T(x_3))\Big]a_i(t^\prime_b)
       E_e^\prime(t_b^\prime)\Big\},\label{ppenlr}
 \end{eqnarray}

\item $(S-P)(S+P)$ operators:
\begin{eqnarray} M^{SP}_{B_s\to M_3}( a_i) &=&32\pi C_F
M_{B_s}^4/\sqrt{6}\int^1_0dx_1dx_2dx_3\int^\infty_0b_1db_1b_2db_2
\phi_{B_s}(x_1,b_1)\phi_2^A(x_2)
\nonumber\\
&&\times\Big\{
\Big[(x_2-x_3-1)\phi_3^A(x_3)+r_3x_3(\phi_3^P(x_3)+\phi_3^T(x_3))\Big]\nonumber\\
&&\times
a_i(t_b)E_e^\prime(t_b)h_n(x_1,1-x_2,x_3,b_1,b_2)+a_i(t_b^\prime)
E^\prime_e(t_b^\prime)\nonumber\\
&&\times
 \Big[x_2\phi_3^A(x_3)+r_3x_3(\phi_3^T(x_3)-\phi_3^P(x_3))\Big]h_n(x_1,x_2,x_3,b_1,b_2)\Big\}.
 \label{ppensp}
\end{eqnarray}
\end{itemize}
From these formulas we can see that there are cancellations between
the two diagrams of Fig.\ref{emission} (c) and (d). If the Wilson
coefficients are the same, the non-factorizable contributions
(proportional to the small $x_i$) are power suppressed compared to the
factorizable emission diagram contributions in
eq.(\ref{ppefll}-\ref{ppefsp}).

\subsection{Annihilation type Diagrams}

\begin{figure}
\vspace{-0.0cm}
\begin{center}
\psfig{file=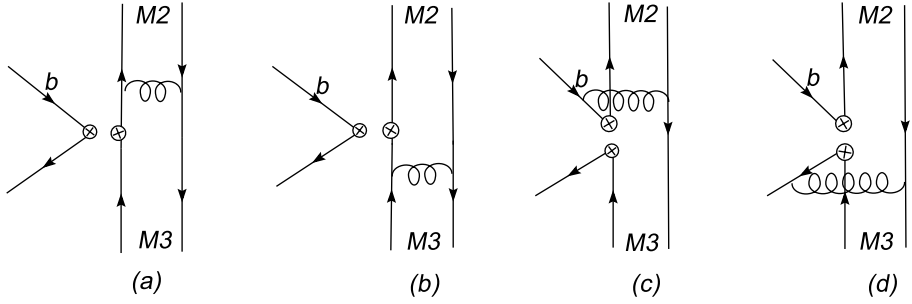,width=12.0cm,angle=0}
\end{center}
\vspace{-0.2cm} \caption{{  The Feynman diagrams for annihilation
contribution, with possible four-quark operator
insertions}}\label{annihilation}
\end{figure}

We group all the W annihilation and W-exchange, space-like penguin
 and time-like penguin annihilation diagrams together and refer to
them as the annihilation type
diagrams. In Fig.\ref{annihilation}, the first two diagrams are the
factorizable annihilation diagrams, whose contributions are
\begin{itemize}
\item $(V-A)(V-A)$ operators:
\begin{eqnarray}
f_{B_s} F_{ann}^{LL}( a_i)&=&8\pi
C_FM_{B_s}^4f_{B_s}\int^1_0dx_2dx_3\int^\infty_0b_2db_2b_3db_3\Big\{a_i(t_c)
E_a(t_c)
\nonumber\\
&&
\times\Big[(x_3-1)\phi_2^A(x_2)\phi_3^A(x_3)-4r_2r_3\phi_2^P(x_2)\phi_3^P(x_3)\nonumber
\\
&&+2r_2r_3x_3\phi_2^P(x_2)(\phi_3^P(x_3)-\phi_3^T(x_3))\Big]h_a(x_2,1-x_3,b_2,b_3)\nonumber
\\
&&+\Big[x_2\phi_2^A(x_2)
\phi_3^A(x_3)+2r_2r_3(\phi_2^P(x_2)-\phi_2^T(x_2))\phi_3^P(x_3)\nonumber\\
&&+2r_2r_3x_2(\phi_2^P(x_2)+\phi_2^T(x_2))\phi_3^P(x_3)\Big]
a_i(t_c^\prime)
E_a(t_c^\prime)h_a(1-x_3,x_2,b_3,b_2)\Big\}.\label{ppafll}
 \end{eqnarray}
There is a big cancellation between the two factorizable diagrams
Fig.\ref{annihilation}(a) and (b), such that they are highly power
suppressed, which agrees with the long time argument that the
annihilation contributions are negligible. Especially, if the two
final state mesons are identical, the formula of eq.(\ref{ppafll})
gives exactly zero.

 \item
$(V-A)(V+A)$ operators:
\begin{eqnarray}
F_{ann}^{LR}( a_i)=F_{ann}^{LL}(a_i),\label{ppaflr}
\end{eqnarray}

 \item $(S-P)(S+P)$ operators:
 \begin{eqnarray}
 f_{B_s} F_{ann}^{SP}(a_i)&=&16\pi
  C_FM_{B_s}^4f_{B_s}\int^1_0dx_2dx_3\int^\infty_0b_2db_2b_3db_3
  \Big\{\Big[2r_2\phi_2^P(x_2)\phi_3^A(x_3)\nonumber\\
  &&\;\;+(1-x_3)r_3\phi_2^A(x_2)(\phi_3^P(x_3)
  +\phi_3^T(x_3))\Big]
 a_i(t_c) E_a(t_c)h_a(x_2,1-x_3,b_2,b_3)\nonumber\\
  &&\;\;+\Big[2r_3\phi_2^A(x_2)\phi_3^P(x_3)+r_2x_2(\phi_2^P(x_2)-\phi_2^T(x_2))\phi_3^A(x_3)
  \Big]\nonumber\\
  &&\;\;\times
  a_i(t_c^\prime)E_a(t_c^\prime)h_a(1-x_3,x_2,b_3,b_2)\Big\}.\label{ppafsp}
\end{eqnarray}
\end{itemize}
It is interesting to see that the two diagrams
Fig.\ref{annihilation} (a) and (b) give constructive contributions
here. Furthermore, they are not power suppressed as the $(V-A)(V-A)$
operator contribution  in eq.(\ref{ppafll}) (proportional to the
small $x_i$), but proportional to $2r_2$ or $2r_3$. This gives the
chirally enhanced contributions in the annihilation type diagrams.
The operator $O_6$-induced  space-like penguin contributions produce a large
strong phase, which is essential to explain the large direct CP
asymmetry in the $B^0\to \pi^+\pi^-$ and $B^0\to K^+\pi^-$ decays
\cite{direct}.

The last two diagrams in Fig.\ref{annihilation} are the
non-factorizable annihilation diagrams, whose contributions are
\begin{itemize}
\item $(V-A)(V-A)$ operators:
\begin{eqnarray}
 M_{ann}^{LL}( a_i)&=&32\pi C_FM_{B_s}^4/\sqrt
 {6}\int^1_0dx_1dx_2dx_3\int^\infty_0b_1db_2b_2db_2\phi_{B_s}(x_1,b_1)\nonumber\\
 &&\times \Big\{h_{na}(x_1,x_2,x_3,b_1,b_2)\Big[-x_2\phi_2^A(x_2)\phi_3^A(x_3)-4r_2r_3
 \phi_2^P(x_2)\phi_3^P(x_3)\nonumber\\
 &&\;\;\;+r_2r_3(1-x_2)(\phi_2^P(x_2)+\phi_2^T(x_2))(\phi_3^P(x_3)-\phi_3^T(x_3))
 \nonumber\\
 &&\;\;+r_2r_3x_3(\phi_2^P(x_2)-\phi_2^T(x_2))(\phi_3^P(x_3)+\phi_3^T(x_3))\Big]a_i(t_d)
 E_a^\prime(t_d)\nonumber\\
 &&\;\;+h_{na}^\prime(x_1,x_2,x_3,b_1,b_2)\Big[(1-x_3)\phi_2^A(x_2)\phi_3^A(x_3)
 \nonumber\\
 &&\;\;+(1-x_3)r_2r_3(\phi_2^P(x_2)+\phi_2^T(x_2))(\phi_3^P(x_3)-\phi_3^T(x_3))
 \nonumber\\
 &&\;\;+x_2r_2r_3(\phi_2^P(x_2)-\phi_2^T(x_2))(\phi_3^P(x_3)+\phi_3^T(x_3))\Big]
 a_i(t_d^\prime)
 E_a^\prime(t_d^\prime)\Big\},\label{ppanll}
 \end{eqnarray}

 \item $(V-A)(V+A)$ operators:
 \begin{eqnarray}
 M_{ann}^{LR}(M_2,M_3, a_i)&=&32\pi C_FM_{B_s}^4/\sqrt
 {6}\int^1_0dx_1dx_2dx_3\int^\infty b_1db_1b_2db_2\phi_{B_s}(x_1,b_1)\nonumber\\
 &&\;\;\times\Big\{h_{na}(x_1,x_2,x_3,b_1,b_2)\Big[r_2(2-x_2)(\phi_2^P(x_2)+\phi_2^T(x_2))
 \phi_3^A(x_3)\nonumber\\
 &&\;\;-r_3(1+x_3)\phi_2^A(x_2)(\phi_3^P(x_3)-\phi_3^T(x_3))\Big]a_i(t_d)E_a^\prime(t_d)
 \nonumber\\
 &&\;\;+h_{na}^\prime
 (x_1,x_2,x_3,b_1,b_2)\Big[r_2x_2\left(\phi_2^P(x_2)+\phi_2^T(x_2)\right)\phi_3^A(x_3)
 \nonumber\\
 &&\;\;+r_3(x_3-1)\phi_2^A(x_2)
 (\phi_3^P(x_3)-\phi_3^T(x_3))\Big]  a_i(t_d^\prime)E_a^\prime(t_d^\prime)
 \Big\},\label{ppanlr}
 \end{eqnarray}

 \item $(S-P)(S+P)$ operators:
 \begin{eqnarray}
 M_{ann}^{SP}( a_i)&=&32\pi C_F M_{B_s}^4/\sqrt {6}\int^1_0dx_1dx_2dx_3\int^\infty_0b_1db_1b_2db_2
 \phi_{B_s}(x_1,b_1)\nonumber\\
 &&\times \Big\{a_i(t_d)E_a^\prime(t_d)h_{na}(x_1,x_2,x_3,b_1,b_2)\Big[(x_3-1)
 \phi_2^A(x_2)\phi_3^A(x_3)\nonumber\\
 &&\;\; -4r_2r_3\phi_2^P(x_2)\phi_3^P(x_3)+r_2r_3x_3(\phi_2^P(x_2)+\phi_2^T(x_2))
 (\phi^P_3(x_3)-\phi_3^T(x_3))\nonumber\\
 &&\;\;+r_2r_3(1-x_2)(\phi_2^P(x_2)-\phi_2^T(x_2))(\phi^P_3(x_3)+\phi_3^T(x_3))\Big]
 \nonumber\\
 &&\;\;+a_i(t_d^\prime)
 E_a^\prime(t_d^\prime)h_{na}^\prime(x_1,x_2,x_3,b_1,b_2)
  \Big[x_2\phi_2^A(x_2)\phi_3^A(x_3)\nonumber
 \\
 &&\;\;+x_2r_2r_3(\phi_2^P(x_2)+\phi_2^T(x_2))
 (\phi_3^P(x_3)-\phi_3^T(x_3)))\nonumber\\
 &&\;\;+r_2r_3(1-x_3)(\phi_2^P(x_2)-\phi_2^T(x_2))(\phi_3^P(x_3)+\phi_3^T(x_3))\Big]\Big\}.
 \label{ppansp}
 \end{eqnarray}
\end{itemize}
They are all power suppressed compared to the factorizable
emission diagrams.

\subsection{Results for $B_s\to PP$ decays}

First we give the numerical results in the pQCD approach for the form
factors at maximal recoil. For the  form factors, we obtain:
\begin{eqnarray}
F_{0}^{B\to\pi}&=& 0.23^{+0.05+0.00}_{-0.04-0.00},\;\;F_{0}^{B\to
K}=0.28^{+0.06+0.00}_{-0.05-0.00},\\
  F_{0}^{B_s\to K}&=&0.24^{+0.05+0.00}_{-0.04-0.01},
\end{eqnarray}
where $f_B=0.19 \pm 0.02\mbox{ GeV}$, $\omega_B=0.40\mbox{ GeV}$
(for the $B^\pm$ and $B_d^0$ mesons) and $f_{B_s}=0.23 \pm 0.02$
GeV, $\omega_{B_s}=0.50 \pm 0.05$ GeV (for the $B_s^0$ meson) have
been used. They quantify the SU(3)-symmetry breaking effects in
the form factors in the pQCD approach. The input values for $f_B$
and $f_{B_s}$ are in agreement with the unquenched lattice
results~\cite{Gray:2005ad} $f_B=0.216\pm\ 0.022 \mbox{ GeV}$ and
$f_{B_s}=0.259 \pm\ 0.032 \mbox{  GeV}$, and with the results from
the QCD sum rules~\cite{Penin:2001ux,Jamin:2001fw}. We also
mention in passing that a recent calculation of the distribution
amplitudes for the pion and kaon with the QCD sum
rules~\cite{NewPWV}, which includes a new logarithmic divergent
term in the twist-3 distribution amplitudes, leads to very large
SU(3)-breaking effects in the form factors, calculating the form
factors in the standard pQCD approach~\cite{LMS}. This has also
been observed in \cite{Li:2006jv}. Whether this feature also
emerges in the light-cone sum rule approach is not clear to us.

\begin{table}[tb]
 \caption{The $CP$-averaged branching ratios ($\times
 10^{-6}$) of $B_s\to PP$ decays obtained in the pQCD approach
(This work); the errors for these entries correspond to the
uncertainties in the input hadronic quantities, from the scale-dependence,
and
the CKM matrix elements, respectively. We have also listed  the
current experimental  measurements and upper limits (90\% C.L.)
wherever available~\cite{CDFBsKKrecent}. For comparison, we also
cite the theoretical estimates of the branching ratios in the
QCD factorization framework \cite{QCDFBs1},  and in SCET
\cite{SCETBs}, quoting two estimates in the latter case for some decays.}
 \label{BRPP}
\begin{center}
{\footnotesize
 \begin{tabular}{c|c|c|c|c|c}
  \hline\hline
        {Modes}    & Class  &  QCDF & SCET&   This work & Exp.
          \\  \hline
   ${\overline{B}}^{0}_{s}{\to}K^{+}\pi^-$     & $T$
    & $10.2^{+4.5+3.8+0.7+0.8}_{-3.9-3.2-1.2-0.7}$
                                                          & $4.9\pm1.2\pm1.3\pm0.3$
                                                          & $7.6^{+3.2+0.7+0.5}_{-2.3-0.7-0.5}$
                                                          & $5.0\pm 0.75\pm 1.0$ \\
   ${\overline{B}}^{0}_{s}{\to}K^{0}{\pi}^{0}$ & $ C$
     & $0.49^{+0.28+0.22+0.40+0.33}_{-0.24-0.14-0.14-0.17}$
                                                          & $0.76\pm0.26\pm0.27\pm0.17$
                                                          & $0.16^{+0.05+0.10+0.02}_{-0.04-0.05-0.01}$
                                                          &          \\
   $\overline B^0_s\to K^+ K^-$                & $P$
   & $22.7^{+3.5+12.7+2.0+24.1}_{-3.2-~ 8.4-2.0-~9.1}$
                                                          & $18.2\pm6.7\pm1.1\pm0.5$
                                                          & $13.6^{+4.2+7.5+0.7}_{-3.2-4.1-0.2}$
                                                          & $24.4\pm 1.4\pm 4.6$  \\
   $\overline B^0_s\to K^0\overline K^0$       & $P$
     & $24.7^{+2.5+13.7+2.6+25.6}_{-2.4-~9.2-2.9-~9.8} $
                                                          & $17.7\pm6.6\pm0.5\pm0.6$
                                                          & $15.6^{+5.0+8.3+0.0}_{-3.8-4.7-0.0}$
                                                          &          \\
   $\overline B^0_s\to\pi^0\eta$         & $P_{EW}$
                                                            & $0.075^{+0.013+0.030+0.008+0.010}_{-0.012-0.025-0.010-0.007} $
                                                          & $0.014\pm0.004\pm0.005\pm0.004$
                                                          & $0.05^{+0.02+0.01+0.00}_{-0.02-0.01-0.00}$&
                                                          $<1000$\\
                                               &          &
                                                          & $0.016\pm0.0007\pm0.005\pm0.006$
                                                          & \\
   $\overline B^0_s\to\pi^0\eta^\prime$  & $P_{EW}$
      & $0.11^{+0.02+0.04+0.01+0.01}_{-0.02-0.04-0.01-0.01}$
                                                          & $0.006\pm0.003\pm0.002^{+0.064}_{-0.006}$
                                                          & $0.11^{+0.05+0.02+0.00}_{-0.03-0.01-0.00}$
                                                          & \\
                                               &          &
                                                          & $0.038\pm0.013\pm0.016^{+0.260}_{-0.036}$
                                                          &  \\
   $\overline B^0_s\to K^0\eta$              & $ C$
     & $0.34^{+0.19+0.64+0.21+0.16}_{-0.16-0.27-0.07-0.08}$
                                                          & $0.80\pm0.48\pm0.29\pm0.18$
                                                          & $0.11^{+0.05+0.06+0.01}_{-0.03-0.03-0.01}$
                                                          & \\
                                               &          &
                                                          & $0.59\pm0.34\pm0.24\pm0.15$
                                                          &\\
   $\overline B^0_s\to K^0\eta^\prime$       & $ C$
      & $2.0^{+0.3+1.5+0.6+1.5}_{-0.3-1.1-0.3-0.6}$
                                                          & $4.5\pm1.5\pm0.4\pm0.5$
                                                          & $0.72^{+0.20+0.28+0.11}_{-0.16-0.17-0.05}$
                                                          &\\
                                               &          &
                                                          & $3.9\pm1.3\pm0.5\pm0.4$
                                                          & \\
   $\overline B^0_s\to\eta\eta$                & $P$
     & $15.6^{+1.6+9.9+2.2+13.5}_{-1.5-6.8-2.5-~5.5}$
                                                          & $7.1\pm6.4\pm0.2\pm0.8$
                                                          & $8.0^{+2.6+4.7+0.0}_{-1.9-2.5-0.0}$
                                                          & $<1500$\\
                                               &          &
                                                          & $6.4\pm6.3\pm0.1\pm0.7$
                                                          & \\
   $\overline B^0_s\to\eta\eta^\prime$         & $P$
     & $54.0^{+5.5+32.4+8.3+40.5}_{-5.2-22.4-6.4-16.7}$
                                                          & $24.0\pm13.6\pm1.4\pm2.7$
                                                          & $21.0^{+6.0+10.0+0.0}_{-4.6-5.6-0.0}$
                                                          &\\
                                               &          &
                                                          & $23.8\pm13.2\pm1.6\pm2.9$&\\
   $\overline B^0_s\to\eta^\prime\eta^\prime$  & $P$
     & $41.7^{+4.2+26.3+15.2+36.6}_{-4.0-17.2-~8.5-15.4}$
                                                          & $44.3\pm19.7\pm2.3\pm17.1$
                                                          & $14.0^{+3.2+6.2+0.0}_{-2.7-3.9-0.0}$
                                                          & \\
                                               &          &
                                                          & $49.4\pm20.6\pm8.4\pm16.2$
                                                          & \\\hline
   $\overline B^0_s\to\pi^+\pi^-$              & ann
     & $0.024^{+0.003+0.025+0.000+0.163}_{-0.003-0.012-0.000-0.021}$
                                                          & ---
                                                          & $0.57^{+0.16+0.09+0.01}_{-0.13-0.10-0.00}$
                                                          & $<1.36$  \\
   $\overline B^0_s\to\pi^0\pi^0$              & ann
     & $0.012^{+0.001+0.013+0.000+0.082}_{-0.001-0.006-0.000-0.011}$
                                                          & ---
                                                          & $0.28^{+0.08+0.04+0.01}_{-0.07-0.05-0.00}$
                                                          & $<210$\\
 \hline\hline\end{tabular}
 }
\end{center}
 \end{table}

For the CKM matrix elements, we adopt the updated results  from
\cite{CKMfitter:2006} and drop the (small) errors on $V_{ud}$, $V_{us}$, $V_{ts}$
and $V_{tb}$:
\begin{equation}
 \begin{array}{lll}
 |V_{ud}|=0.974,   &|V_{us}|=0.226,    &|V_{ub}|=(3.68^{+0.11}_{-0.08})\times 10^{-3},
\\
  |V_{td}|=(8.20^{+0.59}_{-0.27})\times 10^{-3}, &|V_{ts}|=40.96\times 10^{-3},   &|V_{tb}|=1.0,\\
 \alpha={(99^{+4}_{-9.4})}^\circ,&   \gamma=(59.0^{+9.7}_{-3.7})^\circ,&
  \mbox{arg}[-V_{ts}V^*_{tb}]={1.0}^\circ.
\end{array}
\end{equation}
The CKM factors mostly   give an overall factor to the branching
ratios. However, the CKM angles do give large uncertainties to the
branching ratios of some decays and to all the non-zero CP asymmetries.
 We will discuss their effects separately.

The formulas of the decay amplitudes  for the various channels in terms of
 above topologies are given in Appendix \ref{bspp}. The
$CP$-averaged branching ratios  of $B_s\to PP$ decays are listed in
Table~\ref{BRPP}. The dominant topologies contributing to these decays
are also indicated through the symbols $T$ (tree), $P$ (penguin), $P_{\rm EW}$
(electroweak penguins), $C$ (color-suppressed tree), and ann (annihilation).
 The first error in these entries arises from the input
 hadronic parameters, which is dominated by the $B_s$-meson decay constant
(taken as $f_{B_s}=0.23 \pm 0.02$ GeV) and the $B_s$ meson wave
function shape parameter ( taken as $\omega_b= 0.50\pm0.05$ GeV).
The second error is from the hard scale $t$, defined in Eqs. (A1)
-- (A8) in Appendix A, which we vary from $0.75t$ to $1.25t$, and
from $\Lambda^{(5)}_{QCD}=0.25\pm 0.05$ GeV. The scale-dependent
uncertainty can be reduced only if the next-to-leading order
contributions in the pQCD approach are known. A part of this
perturbative improvement coming from the Wilson coefficients in
the NLO approximation can be implemented already. However, the
complete NLO corrections to the hard spectator kernels are still
missing.
 The third error is the combined uncertainty in the $CKM$
matrix elements and the angles of the unitarity triangle.

 For comparison, we also reproduce verbatim
 the corresponding numerical results evaluated in the framework of the QCD
 factorization
(QCDF) \cite{QCDFBs1}, and the ones obtained using the
Soft-Collinear-Effective-Theory (SCET) \cite{SCETBs}.
For the decays $B_s^0 \to PP$ involving an $\eta$- and/or an
$\eta^\prime$-meson, Ref.~\cite{SCETBs} quotes two sets of values, which
differ in the input values of the SCET parameters specific to the iso-singlet
modes, describing gluonic contributions to the $B \to \eta^{(\prime)}$ form
factors and the gluonic parts of the charming penguins. They are not too
different
from each other except for the electroweak-penguin-dominated decay
$\overline{B_s^0}  \to \pi^0\eta^\prime$.
 The errors quoted in the
QCDF case correspond, respectively, to the  assumed variation of the CKM
parameters, variation of the renormalization scales, quark masses, decay
constants, form factors and (whenever applicable) the $\eta$ - $\eta^\prime$
 mixing angle (collectively called ``hadronic 1''), uncertainties in the
expansion of the light-cone distribution amplitudes (called ``hadronic 2''),
and estimates of the power corrections.
The errors shown in the case of the SCET-based results are due to the
estimates of the SU(3)-breaking, $1/m_b$ corrections and  the errors on the
 SCET-parameters. The last
column gives the current experimental data from the CDF
 collaboration~\cite{ExpBs,CDFBsKKrecent}, and the upper limits correspond to
90\% C.L.

A number of remarks on the entries in Table III is in order. We note that
there is general agreement among these methods in the tree and
penguin-dominated $B_s^0 \to PP$ decays, with the variations reflecting
essentially the differences in the input quantities. This agreement is less
marked for the color-suppressed and electroweak-penguin dominated decays.
For the annihilation dominated decays, SCET has no predictions and QCDF has
 essentially no
predictive power as indicated by the estimates for the decays
$\overline{B_s^0} \to \pi^+\pi^-$
and  $\overline{B_s^0} \to \pi^0\pi^0$, which vary over more than an order of
magnitude once the  parametric uncertainties are taken into account. First
measurements of the tree-dominated decay $\overline{B_s^0} \to K^+\pi^-$
and the penguin-dominated decay $\overline{B_s^0} \to K^+K^-$ have been
reported by the CDF collaboration, and are in the right ball-park of the
predictions shown in Table III.  A good number of the decays shown in this
Table will be measured at the LHC and Super B-factories, which would
discriminate among the predictions of the three frameworks.

The decays $B \to \pi \pi$, $B \to K \pi$, $B_s \to K\pi$ and
$B_s \to KK$ have received a lot of theoretical interest, as they can be
related by SU(3)-symmetry, a question of considerable interest is the
amount of SU(3)-breaking in various topologies (diagrams) contributing to
these decays. To that end, we present in Table IV the magnitude
of the decay amplitudes (squared, in units of GeV$^2$) involving the
distinct topologies: $\mathcal{T}$, $\mathcal{P}$, $\mathcal{E}$,
$\mathcal{P_{A}}$ and
 $\mathcal{P_{EW}}$
for the four decays modes of the $B_d^0$ and $B_s^0$ mesons.
The two decays in the upper half of this table are related by U-spin symmetry
$(d \to s)$ (likewise the two decays in the lower half). We note that the
assumption of U-spin symmetry for the (dominant) tree ($\mathcal{T}$) and
penguin ($\mathcal{P}$)
 amplitudes in
the emission diagrams is quite good, it is less so in the other topologies,
including the contributions from the $W$-exchange diagrams, denoted by $\mathcal{E}$
for which there are non-zero contributions for the flavor-diagonal
states $\pi^+\pi^-$ and $K^+K^-$ only. The U-spin breaking is large
 in the electroweak penguin induced amplitudes $\mathcal{P_{EW}}$,
and in the penguin annihilation amplitudes $\mathcal{P_A}$ relating the
decays $B_d \to K^+\pi^-$ and $B_s \to K^+ K^-$.
 In the SM, however,
the amplitudes $\mathcal{P_{EW}}$ are negligibly small.

\begin{table}[htbp]
\caption{Contributions from the various topologies to the decay amplitudes (squared)
for the four indicated decays $B_d^0 \to \pi^+\pi^-; ~K^+\pi^-$
and $B_s^0 \to \pi^+K^-;~ K^+K^- $. Here, $\mathcal{T}$ is the
contribution from the color favored emission diagrams; $\mathcal{P}$
is the penguin contribution from the emission diagrams; $\mathcal{E}$ is
the contribution from the W-exchange diagrams; $\mathcal{P_A}$ is
the contribution from the penguin annihilation amplitudes; and
$\mathcal{P_{EW}}$ is the contribution from the electro-weak penguin
induced amplitude. See text for their definitions.}

\begin{center}
 \begin{tabular}{c|ccccc}
  \hline\hline
 mode ($\mbox{GeV}^2$)& $|\mathcal{T}|^2$&$|\mathcal{P}|^2$&$ |\mathcal{E}|^2$&$|\mathcal{P_A}|^2$&$|\mathcal{P_{EW}}|^2$\\ \hline
   $B_d \to \pi^+\pi^-$   &~~~$0.8$~~~&  $4.8\times 10^{-3}$  &$5.5\times 10^{-3}$ &$1.6\times 10^{-3}$  & $0.6\times 10^{-6}$ \\

   $B_s \to \pi^+ K^- $   &~~~$1.0$~~~&  $5.4\times 10^{-3}$  & $0$                &$3.3\times 10^{-3}$  & $0.8\times 10^{-6}$
   \\ \hline\hline
   $B_d \to K^+\pi^-  $   &~~~$1.2$~~~&  $10.2\times 10^{-3}$   &  0                 &$2.3\times 10^{-3}$  & $2.9\times 10^{-6}$  \\
   $B_s \to K^+K^-    $   &~~~$1.5$~~~&  $11.3\times 10^{-3}$  & $3.5\times 10^{-3}$& $7.3\times 10^{-3}$ & $0.5\times 10^{-6}$  \\
 \hline\hline\end{tabular}\label{Amplitudes}
\end{center}
 \end{table}

The direct $CP$ asymmetry of $\bar B_s\to f$ is defined as
\begin{eqnarray} A_{CP}^{dir}\equiv \frac{BR(\bar B_s^0\to
f)-BR(B_s^0\to \bar f)}{BR(\bar B_s^0\to f)+BR(B_s^0\to \bar f)}
=\frac{|A(\bar B_s\to f)|^2-|A(B_s\to \bar f)|^2}{|A(\bar B_s\to
f)|^2+|A(B_s\to \bar f)|^2}.\label{ACPDef}
\end{eqnarray}
The numerical results for the direct $CP$ asymmetries in $B \to PP$ decays are shown in
Table~\ref{DIRCPPP}. The first error is from the $B_s$ meson
wave function parameter
$\omega_b=0.50^{+0.05}_{-0.05}~\mathrm{GeV}$. As $f_{B_s}$ is an
overall factor, it drops out in the ratio and hence does not give any
 uncertainty in the estimates of the direct CP
asymmetries. The second error is from the hard scale
$t$ varying from $0.75t$ to $1.25t$ (not changing $1/b_i$,
$i=1,2,3$) and $\Lambda^{(5)}_{QCD}=0.25\pm 0.05$ GeV.
 The third error is again from the combined
uncertainty in the $CKM$ matrix elements and the angles of the unitarity
triangle.  The observed CP asymmetry $A_{\rm CP}^{\rm dir}
(\overline{B_s^0} \to K^+\pi^-)$ by the CDF collaboration is also
shown in this table, and is found to be in agreement with the  pQCD
predictions.

Within (large) theoretical errors, the observed CP asymmetry is also
in agreement with the SCET estimate but in stark disagreement with
the QCDF prediction. This deserves a comment. As is well-known,
the decay rates and CP-asymmetries in the decays
 $\overline{B_d^0} \to \pi^+\pi^-$
and $\overline{B_s^0} \to K^+\pi^-$ are related by SU(3) symmetry.
The predicted CP asymmetry $A_{\rm CP}^{\rm dir} (\overline{B_s^0}
\to K^+\pi^-)$ in the QCDF approach shown in Table V is very
similar to the SU(3)-related CP asymmetry  $A_{\rm CP}^{\rm dir}
(\overline{B_d^0} \to \pi^+\pi^-)$. These CP-asymmetries are small
in the QCDF approach and both are consistently in disagreement
with the data, in magnitude and sign. Using SU(3)-symmetry, the
CP-asymmetry $A_{\rm CP}^{\rm dir} (\overline{B_s^0} \to
K^+\pi^-)$ in SCET, given in Table V,  is identical to the values
quoted in Ref.~\cite{SCETBs} for the CP-asymmetry $A_{\rm CP}^{\rm
dir} (\overline{B_d^0} \to \pi^+\pi^-)$. As the dynamical hadronic
quantities in the SCET framework are fitted using the B-factory
data on $B \to \pi \pi$ and $B \to K\pi$~\cite{SCETBs}, it is not
surprising that this phenomenological fit can account for the
SU(3)-related observed CP-asymmetry  in the $B_s^0$ decays, such
as $A_{\rm CP}^{\rm dir} (\overline{B_s^0} \to K^+\pi^-)$. The
fitted quantities, namely the SCET-specific functions $\zeta$,
$\zeta_J$, differ from their corresponding QCDF analogs. But,
crucially, also the contributions of the charming
penguin~\cite{Ciuchini:1997hb}, denoted as $\vert A_{\rm cc}\vert$
and $\arg  A_{\rm cc}$ in~\cite{SCETBs}, which are
power-suppressed in QCDF, are found to be large using data and
SCET. These differences lead to a different phenomenological
profile of the $B \to h_1 h_2$ decays in SCET and QCDF. However,
one has to stress that the charming penguin contribution, which is
not yet shown to factorize in SCET, obviously is not  on the same
theoretical footing as the rest of the decay amplitudes, for which
factorization is proven in the heavy quark limit. Thus,  the
predictions in SCET~\cite{SCETBs} essentially reflect  the
parameterization of the charming penguins from the available data.

In general, one has  also to take into account the uncertainties caused by
the Gegenbauer moments. In recent years, 
the light cone distribution amplitudes have been continually updated
\cite{PseudoscalarUpdate}. In order to check the sensitivity to
the values on the Gegenbauer moments, we calculate the branching
ratio for $\bar B_s\to K^+\pi^-$ with the following values for
the twist-2 LCDAs:
\begin{eqnarray}
a_1^K=0.06\pm0.03,\;\;\; a_2^{\pi,K}=0.25\pm0.15,
\label{eq:recent-gegenb}
\end{eqnarray}
instead of $a_1^K=0.17$ and $a_2^K=0.2$ from eq.~(25) and $a_2^\pi =0.44$ from
eq.~(22).
Using the above values and the asymptotic forms for the twist-3 LCDAs, we obtain
\begin{eqnarray}
{\cal BR}(\bar B_s^0\to
K^+ \pi^-)=(7.2^{+3.0+0.7+0.4}_{-2.2-0.8-0.5})\times 10^{-6},
\end{eqnarray}
where the errors are to be interpreted as before. The agreement between this
result with the corresponding number listed in Table \ref{BRPP} confirms our
expectation that this branching ratio is not changed significantly. In our
calculation, the parameters in the LCDAs are chosen at $\mu=1$ GeV
but as they are scale-dependent quantities, their value at $\mu=2$ GeV,
the typical scale which enters the perturbative calculation in pQCD, is
required.  
 At $\mu=2$ GeV, the values for the
Gegenbauer moments scaled from eq.~(\ref{eq:recent-gegenb}) are:
\begin{eqnarray}
a_1^K=0.05\pm0.02,\;\;\; a_2^{\pi,K}=0.17\pm0.10.
\end{eqnarray}
Using these values, we obtain:
\begin{eqnarray}
{\cal BR}(\bar B_s^0 \to
K^+\pi^-)=(6.9^{+3.0+0.7+0.4}_{-2.1-0.7-0.5})\times 10^{-6}~.
\end{eqnarray}
We remark that the uncertainties caused by the scale dependence
of the Gegenbauer moments are not large, compared with other
uncertainties.

 \begin{table}[tb]
 \caption{The direct $CP$ asymmetries (in \%) in the $B_s\to PP$  decays,
 obtained in the pQCD approach
(This work); the errors for these entries correspond to the
uncertainties in the input hadronic quantities, from the scale-dependence,
and the CKM matrix elements, respectively. The only measured CP asymmetry is
also given~\cite{CDFBsKKrecent}. For comparison, we also
cite the theoretical estimates of the CP asymmetries in the
QCD factorization framework \cite{QCDFBs1},  and in SCET
\cite{SCETBs}.  } \label{DIRCPPP}
\begin{center}
 \begin{tabular}{c|c|c|c|c|c}
 \hline\hline  {Modes}
   & Class &   QCDF &SCET &  This work & EXP\\  \hline
   ${\overline{B}}^{0}_{s}{\to}K^{+}\pi^-$         &  $T$
        & $-6.7^{+2.1+3.1+0.2+15.5}_{-2.2-2.9-0.4-15.2}$
                                                                 & $20\pm17\pm19\pm5$
                                                                 & $24.1^{+3.9+3.3+2.3}_{-3.6-3.0-1.2}$
                                                                & $39\pm 15\pm 8$ \\
   ${\overline{B}}^{0}_{s}{\to}K^{0}{\pi}^{0}$     &$ C$
     & $41.6^{+16.6+14.3+~7.8+40.9}_{-12.0-13.3-14.5-51.0}$
                                                                 & $76\pm26\pm27\pm17$
                                                                 & $59.4^{+1.8+~7.4+2.2}_{-4.0-11.3-3.5}$ \\
   $\overline B^0_s\to K^+ K^-$                      &$P$
       & $4.0^{+1.0+2.0+0.5+10.4}_{-1.0-2.3-0.5-11.3}$
                                                                 & $-6\pm5\pm6\pm2$
                                                                 & $-23.3^{+0.9+4.9+0.8}_{-0.2-4.4-1.1}$  \\
   $\overline B^0_s\to K^0\overline K^0$            &$P$
       & $0.9^{+0.2+0.2+0.1+0.2}_{-0.2-0.2-0.1-0.3}$
                                                                 & $<10$
                                                                 & $0$  \\
   $\overline B^0_s\to\pi^0\eta$                 &$P_{EW}$       & -----
                                                                 & ----
                                                                 & $-0.4^{+0.6+2.2+0.0}_{-0.7-2.2-0.0}$ \\
   $\overline B^0_s\to\pi^0\eta^\prime$           &$P_{EW}$
        & $27.8^{+6.0+9.6+2.0+24.7}_{-7.1+5.7-2.0-27.2} $
                                                                 & ----
                                                                 & $20.6^{+0.0+2.0+2.8}_{-0.7-2.5-1.2}$ \\
   $\overline B^0_s\to K^0\eta$                   & $ C$
      & $46.8^{+18.5+28.6+5.2+34.6}_{-13.2-32.2-12.5-45.6}$
                                                                 & $-56\pm46\pm14\pm6$
                                                                 & $56.4^{+2.9+6.8+3.1}_{-3.4-8.0-3.4}$\\
                                                    &            &
                                                                 & $61\pm59\pm12\pm8$
                                                                 &\\
   $\overline B^0_s\to K^0\eta^\prime$            & $ C$
      & $-36.6^{+8.6+6.0+3.8+19.3}_{-8.2-7.4-2.5-17.3} $
                                                                 & $-14\pm7\pm16\pm2$
                                                                 &$-19.9^{+1.6+5.1+1.4}_{-1.4-5.0-0.9}$ \\
                                                    &            &
                                                                 & $37\pm8\pm14\pm4$
                                                                 & \\
   $\overline B^0_s\to\eta\eta$                    & $P$
      & $-1.6^{+0.5+0.6+0.4+2.2}_{-0.4-0.6-0.7-2.2}$
                                                                 & $7.9\pm4.9\pm2.7\pm1.5$
                                                                 & $-0.6^{+0.2+0.6+0.0}_{-0.2-0.5-0.1}$ \\
                                                    &            &
                                                                 & $-1.1\pm5.0\pm3.9\pm1.0$
                                                                 & \\
   $\overline B^0_s\to\eta\eta^\prime$             & $P$
       & $0.4^{+0.1+0.3+0.1+0.4}_{-0.1-0.3-0.1-0.3}$
                                                                 & $0.04\pm0.14\pm0.39\pm0.43$
                                                                 & $-1.3^{+0.0+0.1+0.1}_{-0.0-0.2-0.1}$\\
                                                    &            &
                                                                 & $2.7\pm0.9\pm0.8\pm7.6$
                                                                 & \\
   $\overline B^0_s\to\eta^\prime \eta^\prime$     & $P$
       & $2.1^{+0.5+0.4+0.2+1.1}_{-0.6-0.4-0.3-1.2}$
                                                                 & $0.9\pm0.4\pm0.6\pm1.9$
                                                                 & $1.9^{+0.2+0.3+0.2}_{-0.2-0.4-0.1}$\\
                                                    &            &
                                                                 & $-3.7\pm1.0\pm1.2\pm5.6$
                                                                 &\\\hline
   $\overline B^0_s\to\pi^+\pi^-$                   &   ann      & -----
                                                                 & ----
                                                                 & $-1.2^{+0.1+1.2+0.1}_{-0.4-1.2-0.1}$ \\
   $\overline B^0_s\to\pi^0\pi^0$                   &   ann      & -----
                                                                 &  ----
                                                                 & $-1.2^{+0.1+1.2+0.1}_{-0.4-1.2-0.1}$\\
 \hline\hline\end{tabular}
\end{center}
 \end{table}

\subsection{The Observables $S_f$ and $H_f$ in time-dependent decays $B_s^0(t)
   \to f$}

Restricting the final state $f$ to have definite CP-parity, the
time-dependent decay width for the $B_s \to f$ decay is
 \cite{BsmixingDun}:
\begin{eqnarray}
\Gamma(B^0_s(t)\to f)= e^{-\Gamma t}\; \overline \Gamma(B_s\to f)
\Big[\cosh \Big(\frac{\Delta \Gamma t}{2}\Big)
+H_f \sinh \Big(\frac{\Delta \Gamma t}{2}\Big)\nonumber\\
-{\cal A}_{\mathrm{CP}}^{dir}\cos( \Delta m t)- S_f \sin (\Delta m
t) \Big],
\end{eqnarray}
 where $\Delta m=m_H-m_L>0$, $\overline{\Gamma}$ is the average decay width,
and  $\Delta \Gamma=\Gamma_H-\Gamma_L$ is the difference of decay
widths for the heavier and lighter $B_s^0$ mass eigenstates. The
time dependent decay width $\Gamma(\bar B^0_s(t)\to f)$ is obtained
from the above expression by flipping the signs of the $\cos(\Delta
m t)$ and $\sin(\Delta m t)$ terms. In the $B_s$ system, we expect a
much larger decay width difference $(\Delta\Gamma/\Gamma)_{B_s}$.
This is estimated within the standard model to have a value 
$(\Delta\Gamma/\Gamma)_{B_s}=-0.12\pm0.05$~\cite{BsmixingLenz}, updated
recently in~\cite{Lenz:2006hd} to $(\Delta\Gamma/\Gamma)_{B_s}=-0.147\pm 0.060$,
 while experimentally $(\Delta
\Gamma/\Gamma)_{B_s}=-0.33^{+0.09}_{-0.11}$ \cite{Bsmixingexp}, so
that both  $S_f$ and $H_f$,   can be extracted from the time
dependent decays of  $B_s$ mesons.
 The definition of the various quantities in the above equation
are as follows:
\begin{eqnarray}
S_f =\frac{2Im[\lambda]}{1+|\lambda|^2}, ~~~ H_f
=\frac{2Re[\lambda]}{1+|\lambda|^2},
\end{eqnarray}
with
 \begin{equation} \lambda=\eta_f e^{2i\epsilon}\frac{A(\bar B_s\to
f)}{A(B_s\to\bar f)},
\end{equation}
 where  $\eta_f$ is $+1(-1)$ for a CP-even (CP-odd) final state $f$ and
$\epsilon=\mbox{arg}[-V_{cb}V_{ts}V^*_{cs}V^*_{tb}]$.
With the convention $\mbox{arg}[V_{cb}]=\mbox{arg}[V_{cs}]=0$, the
parameter can be reduced to $\epsilon=\mbox{arg}[-V_{ts}V_{tb}^*]$.
The results of our calculations for the decays $B_s^0 \to PP$ are listed in Table~\ref{MIXCPBsPP}
and compared with the ones obtained in
 SCET~\cite{SCETBs}.

\begin{table}[tb]
\caption{The mixing-induced CP asymmetries $S_f$ (the first row of
each decay channel) and the observables $H_f$ (the second row) in the $B_s\to PP$
decays calculated in the pQCD approach (This work).
The errors for these entries correspond to the
uncertainties in the input hadronic quantities, from the scale-dependence,
and the CKM matrix elements, respectively.
 Estimates of these quantities in
 SCET~\cite{SCETBs} are also given, quoting two of these for the decays involving an $\eta$
 and/or an $\eta^\prime$-meson. }  \label{MIXCPBsPP}
\begin{center}
{\footnotesize
 \begin{tabular}{c|c|c|c}
 \hline\hline  {Modes}
   &   SCET Theory I  & SCET Theory II & This work  \\  \hline
    $\overline B^0_s\to K_S\pi^0$               & $-0.16\pm0.41\pm0.33\pm0.17$
                                                &
                                                & $-0.61^{+0.08+0.23+0.01}_{-0.06-0.19-0.03}$\\
                                                & $0.80\pm0.27\pm0.25\pm0.11$
                                                &
                                                & $-0.52^{+0.04+0.22+0.03}_{-0.06-0.16-0.02}$\\
  \hline
    $\overline B^0_s\to K^-K^+$                 & $0.19\pm0.04\pm0.04\pm0.01$
                                                &
                                                & $0.28^{+0.03+0.04+0.02}_{-0.03-0.04-0.01}$\\
                                                & $0.979\pm0.008\pm0.007\pm0.002$
                                                &
                                                & $0.93^{+0.01+0.02+0.00}_{-0.01-0.02-0.01}$ \\
  \hline
    $\overline B^0_s\to K^0\overline K^0$       & ---
                                                & ---
                                                & $0.04$\\
                                                & ---
                                                & ---
                                                & $1.00$ \\
  \hline  $\overline B^0_s\to \pi^0\eta$              & $0.45\pm 0.14\pm 0.42\pm 0.30$
                                                & $0.38\pm 0.20\pm 0.42\pm 0.37$
                                                & $0.17^{+0.04+0.10+0.01}_{-0.04-0.12-0.01}$\\
                                                & $-0.89\pm 0.07\pm 0.21\pm 0.15$
                                                & $-0.92\pm 0.08\pm 0.17\pm 0.15$
                                                & $0.99^{+0.00+0.01+0.00}_{-0.01-0.02-0.00}$\\
  \hline  $\overline B^0_s\to \pi^0\eta^\prime$       & $0.45\pm 0.14\pm 0.42\pm 0.30$
                                                & $0.38\pm 0.20\pm 0.42\pm 0.37$
                                                & $-0.17^{+0.00+0.07+0.03}_{-0.01-0.08-0.05}$\\
                                                & $-0.89\pm 0.07\pm 0.21\pm 0.15$
                                                & $-0.92\pm 0.08\pm 0.17\pm 0.15$
                                                & $0.96^{+0.00+0.01+0.01}_{-0.00-0.01-0.01}$\\
  \hline  $\overline B^0_s\to K_S\eta$                & $0.82\pm 0.32\pm 0.11\pm 0.04$
                                                & $0.63\pm 0.61\pm 0.16\pm 0.08$
                                                & $-0.43^{+0.03+0.22+0.02}_{-0.04-0.21-0.03}$\\
                                                & $0.07\pm 0.56\pm 0.17\pm 0.05$
                                                & $0.49\pm 0.68\pm 0.21\pm 0.03$
                                                & $-0.70^{+0.04+0.13+0.01}_{-0.05-0.21-0.01}$\\
  \hline  $\overline B^0_s\to K_S\eta^\prime$         & $0.38\pm 0.08\pm 0.10\pm 0.04$
                                                & $0.24\pm 0.09\pm 0.15\pm 0.05$
                                                & $-0.68^{+0.01+0.06+0.00}_{-0.02-0.05-0.00}$\\
                                                & $-0.92\pm 0.04\pm 0.04\pm 0.02$
                                                & $-0.90\pm 0.05\pm 0.05\pm 0.03$
                                                & $-0.70^{+0.02+0.06+0.00}_{-0.02-0.07-0.00}$\\
  \hline  $\overline B^0_s\to \eta\eta$               & $-0.026\pm 0.040\pm 0.030\pm0.014$
                                                & $-0.077\pm 0.061\pm 0.022\pm 0.026$
                                                & $0.03^{+0.00+0.01+0.00}_{-0.00-0.01-0.00}$\\
                                                & $0.9965\pm0.0041\pm0.0019\pm0.0015$
                                                & $0.9970\pm0.0048\pm0.0017\pm0.0021$
                                                & $1.00^{+0.00+0.00+0.00}_{-0.00-0.00-0.00}$\\
  \hline  $\overline B^0_s\to \eta\eta'$              & $0.041\pm 0.004\pm 0.002\pm0.051$
                                                & $0.015\pm 0.010\pm 0.008\pm 0.069$
                                                & $0.04^{+0.00+0.00+0.00}_{-0.00-0.00-0.00}$\\
                                                & $0.9992\pm0.0002\pm0.0001\pm0.0021$
                                                & $0.9996\pm0.0003\pm0.0003\pm0.0007$
                                                & $1.00^{+0.00+0.00+0.00}_{-0.00-0.00-0.00}$\\
  \hline  $\overline B^0_s\to \eta'\eta'$             & $0.049\pm 0.005\pm 0.005\pm0.031$
                                                & $0.051\pm 0.009\pm 0.017\pm 0.039$
                                                & $0.04^{+0.00+0.01+0.00}_{-0.00-0.01-0.00}$\\
                                                & $0.9988\pm0.0003\pm0.0002\pm0.0017$
                                                & $0.9980\pm0.0007\pm0.0009\pm0.0041$
                                                & $1.00^{+0.00+0.00+0.00}_{-0.00-0.00-0.00}$\\\hline
     $\overline B^0_s\to \pi^+\pi^-$             & ----
                                                & ----
                                                & $0.14^{+0.02+0.08+0.09}_{-0.00-0.02-0.05}$ \\
                                                & ----
                                                & ----
                                                & $0.99^{+0.00+0.00+0.00}_{-0.00-0.01-0.01}$ \\
  \hline  $\overline B^0_s\to \pi^0\pi^0$             & ----
                                                & ----
                                                &  $0.14^{+0.02+0.08+0.09}_{-0.00-0.02-0.05}$  \\
                                                & ----
                                                & ----
                                                & $0.99^{+0.00+0.00+0.00}_{-0.00-0.01-0.01}$ \\
 \hline\hline\end{tabular}
 }
\end{center}
 \end{table}

\subsection{Specific Tests of the pQCD predictions in $\overline{B_d^0} \to PP$ and
   $\overline{B_s^0} \to PP$ Decays}

In this subsection, we confront the predictions of the pQCD
approach to available data in the decay modes $\overline{B_s^0} \to PP$ and
$\overline{B_d^0} \to PP$, in terms of the ratios of the branching ratios and
CP-asymmetries. Restricted by the currently available
data, we shall confine ourselves to the decay modes $\overline{B_s^0} \to
K^+K^-$, $\overline{B_s^0} \to K^+\pi^- $, $\overline{B_d^0} \to \pi^+\pi^-$
and $\overline{B_d^0} \to K^-\pi^+$. The ratio of the branching ratios
defined as $R_1\equiv\frac{BR(\overline{B_s^0}\to
  K^+K^-)}{BR(\overline{B_d^0}\to\pi^+\pi^-)}$  has been studied at some
length in the literature. In $R_1$, the numerator is dominated by the penguin
amplitude,
but the denominator is a mixture of tree and penguin amplitudes, and both
the numerator and denominator have been measured experimentally. It has been
argued that the ratio $R_1$ and
the two $CP$ asymmetries $S_{CP}(\bar B_d^0\to\pi^+\pi^-)$ and
$C_{CP}(\bar B_d^0\to\pi^+\pi^-)=-A_{CP}^{dir}(\bar
B_d^0\to\pi^+\pi^-)$ of the $\bar B_d^0\to\pi^+\pi^-$ channel
depend, in the SU(3) limit,  on only two quantities \cite{KKpipiLM}
which can be determined from data and compared with the various
dynamical models to get a clear picture of two-body non-leptonic
decays. However, not invoking the SU(3) limit, this system of
observables has too many unknowns and a clean test of the
dynamical models is bogged down in the details of the hadronic input.
To enable getting cleaner theoretical handles on the underlying dynamics, we
calculate the ratio
 $R_2\equiv\frac{BR(\bar B_s^0\to K^+K^-)}{BR(\bar B_d^0\to \pi^+K^-)}$,
the ratio $R_3$ and the quantity called $\Delta$,  defined later in this
subsection. The last two (i.e. $R_3$ and $\Delta$) have been advocated
by Lipkin~\cite{KpiLipkin} invoking earlier work by
Gronau~\cite{Gronau:2000zy} as precision tests of the
SM.

The CDF
collaboration has measured the branching ratio of $\overline{ B_s^0} \to
K^+K^-$ in the form \cite{CDFBsKKrecent}:
\begin{eqnarray}
\frac{f_s\cdot BR(\overline{ B_s^0}\to K^+K^-)}{f_d\cdot BR(\overline
{B_d^0}\to\pi^+ K^-)}=0.324\pm 0.019\pm 0.041.
\end{eqnarray}
Using the results \cite{Bsmixingexp}
\begin{eqnarray}
 f_s=(10.4\pm1.4)\%,\;\;\; f_d=(39.8\pm1.0)\%,\\
 BR(\overline{ B_d^0}\to \pi^+\pi^-)=(5.2\pm 0.2)\times 10^{-6},
 \end{eqnarray}
one obtains $R_1$ and $R_2$:
\begin{eqnarray}
R_1=4.69\pm0.94,\;\;\; R_2=1.24\pm0.24.
\end{eqnarray}
We have calculated the branching ratios  for the
related $\overline{B_d^0} \to \pi^+\pi^-, \pi^+K^-$  decays in the pQCD approach, getting:
\begin{equation}
BR(\overline{
B_d^0}\to\pi^+\pi^-)=(5.8^{+3.0+0.5+0.4}_{-2.1-0.4-0.3}) \times
10^{-6} ,\;\;\;
 BR(\overline{
B_d^0}\to\pi^+K^-)=(11.6^{+5.0+5.2+0.7}_{-3.5-2.9-0.3}) \times
10^{-6}.
\end{equation}
Taking into account the correlated errors in the numerator and the
denominator, we get:
\begin{eqnarray}
R_1=2.35^{+0.45+0.99+0.19}_{-0.36-0.59-0.15},\;\;\;
R_2=1.18^{+0.11+0.10+0.01}_{-0.11-0.08-0.01}.
\end{eqnarray}
Adding the theoretical errors in quadrature, we get
$R_1=2.35^{+1.10}_{-0.71}$ and $R_2=1.18^{+0.15}_{-0.14}$. Hence,
in the pQCD approach, 
the ratio $R_1$  is smaller compared to the current data.
 This originates from the fact that
$BR( \overline{B_d^0}\to \pi^+\pi^-)$ is somewhat larger in the
pQCD approach compared to the data, and  $BR( \overline{B_s^0}\to
K^+ K^-)$ is smaller than the currently measured branching ratio
(see Table III).  Furthermore, the value of
$R_1$ also depends on the $s$ quark mass through the chiral scale
parameter $m_0^K=m_K^2/(m_s+m_{u,d})$. If we use
$m_0^K=(1.9\pm0.2)$ GeV, where the errors reflects the uncertainty in the
quark masses, we find that $R_1$ has an additional uncertainty 
$R_1=2.35^{+0.62}_{-0.49}$. Adding this error with the one quoted above in
quadrature yields $R_1=2.35^{+1.26}_{-0.86}$. It would be interesting to investigate how
 the NLO contributions modify the ratio $R_1$ as the tree and penguin
amplitudes are expected to be renormalized differently including
the $O(\alpha_s^2)$ corrections, as shown in~\cite{LMS} in the
context of the $B_d^0 \to PP$ decays.

\begin{figure}
 \centerline{
\psfig{file=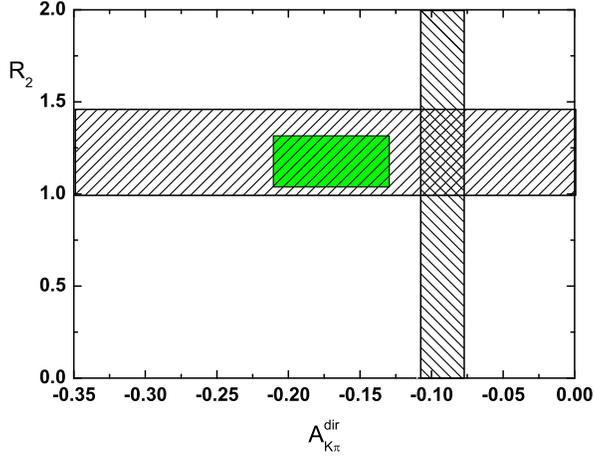,width=9.0cm,angle=0}} \caption{ $R_2$ vs
$A^{dir}_{CP}(\pi K)$. The region inside the solid (red) box is the
pQCD calculation in the LO.
 The experimental results are shown as bands within their $\pm 1\sigma$
errors. }\label{KKKpiLO}
 \end{figure}

\begin{figure}
 \centerline{
\psfig{file=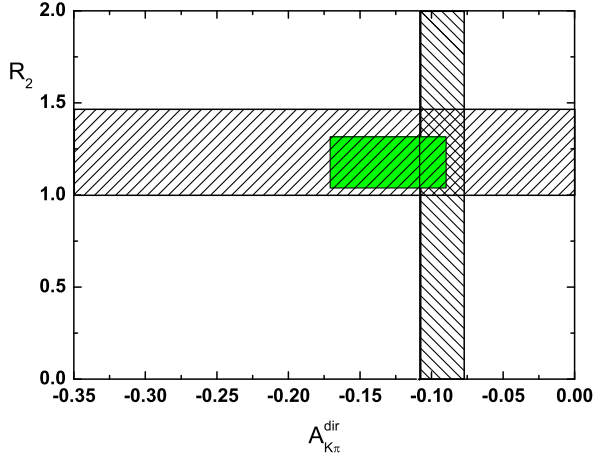,width=9.0cm,angle=0}} \caption{ $R_2$ vs
$A^{dir}_{CP}(\pi K)$. The region inside the solid (green) box is
obtained by estimating $A^{dir}_{CP}(\pi K)$ in NLO, as discussed in the
text.  $R_2$ (pQCD) and the experimental results are the same as in
the previous figure.}
 \label{KKKpiNLO}
 \end{figure}

The ratio $R_2$, on the other hand,  comes out just about right in
pQCD, as in this ratio both the numerator and denominator are
dominated by QCD penguin amplitudes. One also expects that this
ratio is stable under $O(\alpha_s^2)$ corrections, as the
denominator in $R_2$ is itself stable against such corrections
(see Table III in~\cite{LMS}) and very similar arguments apply to
the decay rate for the numerator. The stability of the
color-allowed QCD penguin amplitudes against one-loop corrections
to the hard spectator scattering is also borne out in the QCD
factorization framework~\cite{Beneke:2006wv}. The SU(3)-breaking
effects in $R_2$ are  of the same size as the corresponding
effects in the form factors, typically $\pm~30$\%, and this likely
is the dominant theoretical uncertainty in this ratio. In
Fig.~\ref{KKKpiLO}, we plot the ratio $R_2$ vs. $A_{CP}^{\rm
dir}(K^+\pi^-)$ for the LO pQCD-based calculations worked out by
us and compare them with the current data on these observables.
The experimental value of $R_2$ has already been given earlier,
and we use $A_{CP}^{dir}(B_d^0 \to K^+\pi^-)=-0.093\pm
0.015$~\cite{Bsmixingexp}. We recall that the direct CP asymmetry
$A_{CP}^{dir}(K^+\pi^-)$, calculated by us in the LO pQCD
approach, $A_{CP}^{dir}(\overline
{B_d^0}\to\pi^+K^-)=-0.17^{+0.02+0.03+0.01}_{-0.02-0.02-0.01}$ is
numerically significantly different than the earlier estimates of
the same in this approach (for example, the central value of this
CP asymmetry in LO is quoted as $-0.12$ in~\cite{LMS}). This
mismatch reflects the dependence of $A_{CP}^{dir}(K^+\pi^-)$ on
the input quantities, in particular $V_{ub}$, which have evolved
in the meanwhile. Thus, with our input values, we find good
agreement with data on $R_2$ but not so for
$A_{CP}^{dir}(K^+\pi^-)$, as shown in  Fig.~\ref{KKKpiLO}.
However, taking into account the O$(\alpha_s^2)$ contributions,
this CP-asymmetry is renormalized while $R_2$ remains practically
the same. Taking the central values from Table IV of
Ref.~\cite{LMS}, one gets a K-factor of 0.75 for
$A_{CP}^{dir}(K^+\pi^-)$ (the central value in NLO is $-0.09$ with
the input values used there). Using this K-factor, and our LO
calculations, we estimate $A_{CP}^{dir}(K^+\pi^-)=-0.13 \pm 0.04$
in NLO, making it compatible with the current data. This is
shown in Fig.~\ref{KKKpiNLO}. It should, however, be pointed out
for the sake of clarity that the NLO corrections in~\cite{LMS} are
not complete, as the hard spectator contributions in
O$(\alpha_s^2)$ are not all calculated. Hence a residual
contribution to $A_{CP}^{dir}(K^+\pi^-)$ (and other observables)
can not be logically excluded. However, our discussion here
underscores the dominant source of the uncertainty in
$A_{CP}^{dir}(K^+\pi^-)$, which is of parametric origin.

\begin{figure}
 \centerline{
\psfig{file=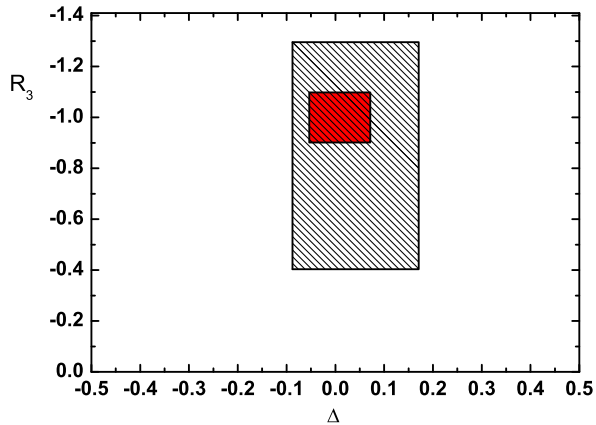,width=9.0cm,angle=0}} \caption{ $R_3$ vs
$\Delta$: The red (smaller) rectangle is the pQCD estimates worked
out in this paper. The
experimental results with their $\pm 1 \sigma$ errors are shown as the larger
 rectangle. }\label{R3}
 \end{figure}

In $\overline{B_d^0}\to K^-\pi^+$ and $\overline{B_s^0}\to K^+\pi^-$, the branching ratios
are very different from each other due to the
differing strong and weak phases entering in the tree and penguin
amplitudes. However, as shown by Gronau~\cite{Gronau:2000zy},
the two relevant products of the CKM matrix elements entering in the
expressions for the direct CP asymmetries in these decays are equal,
and, as stressed by Lipkin~\cite{KpiLipkin} subsequently, the
final states in these decays are charge conjugates, and the strong
interactions
being charge-conjugation invariant, the direct CP asymmetry in
$\overline{B_s^0}\to K^-\pi^+$ can be related to the well-measured
CP asymmetry in the decay $\overline{B_d^0}\to K^+\pi^-$ using U-spin symmetry.
 In this symmetry limit, we have~\cite{Gronau:2000zy,KpiLipkin}
\begin{eqnarray}
&&|A(B_s\to\pi^+K^-)|^2-|A(\bar B_s\to\pi^-K^+)|^2=|A(\bar
B_d\to\pi^+K^-)|^2-|A(B_d\to\pi^-K^+)|^2,\\
&&A^{dir}_{CP}(\bar B_d\to\pi^+K^-)=-A^{dir}_{CP}(\bar
B_s\to\pi^-K^+)\cdot\frac{BR(B_s\to\pi^+K^-)}{BR(\bar
B_d\to\pi^+K^-)}\cdot\frac{\tau(B_d)}{\tau(B_s)}.
\end{eqnarray}
Following the suggestions in the literature, we can test these equations and search for
 possible new physics effects which would likely violate these relations.
To that end, one can
define the following two parameters (using Eq.~(\ref{ACPDef})
for the definition of CP asymmetry):
\begin{eqnarray}
R_3&\equiv&\frac{|A(B_s\to\pi^+K^-)|^2-|A(\bar
B_s\to\pi^-K^+)|^2}{|A(B_d\to\pi^-K^+)|^2-|A(\bar
B_d\to\pi^+K^-)|^2},\\
\Delta&=&\frac{A^{dir}_{CP}(\bar
B_d\to\pi^+K^-)}{A^{dir}_{CP}(\bar
B_s\to\pi^-K^+)}+\frac{BR(B_s\to\pi^+K^-)}{BR(\bar
B_d\to\pi^+K^-)}\cdot\frac{\tau(B_d)}{\tau(B_s)}.
\end{eqnarray}
The standard model predicts $R_3=-1$ and $\Delta=0$ if we assume $U$-spin
symmetry. Since we have a detailed dynamical theory to study the SU(3)
(and U-spin) symmetry violation, we can check how good quantitatively
this symmetry is in the ratios $R_3$ and $\Delta$. We find:
\begin{eqnarray}
 R_3=-1.00^{+0.04+0.03+0.09}_{-0.04-0.04-0.08},\;\;
 \Delta=-0.00^{+0.03+0.03+0.06}_{-0.03-0.02-0.04},
\label{R3delta}
\end{eqnarray}
The differing values of $\omega_B$ and $\omega_{B_s}$, which enter
in the $B_d^0$ and $B_s^0$-meson wave functions,  contribute
dominantly to the first errors, whereas the current uncertainties
on the $CKM$ angle $\gamma$ and $V_{ts}$ are reflected in the
third errors given above. The scale-dependent uncertainties
leading to the second error are relatively small. Adding all the
theoretical errors in quadrature, we get $R_3=-1. 00
^{+0.10}_{-0.10}$ and $\Delta = -0.00^{+0.07}_{-0.05}$. Thus, we
find that these quantities are quite reliably calculable, as
anticipated on theoretical grounds. On the experimental side, the
results for $R_3$ and $\Delta$ are~ \cite{CDFBsKKrecent}:
\begin{equation}
R_3=-0.84\pm0.42\pm0.15,\;\; \Delta=0.04\pm0.11\pm0.08.
\end{equation}
We conclude that the SM is in  good agreement with the data,
as can also be seen in Fig.~\ref{R3} where we plot theoretical predictions for
$R_3$ vs.~$\Delta$ and compare them with the current measurements of the same.
The  measurements of these quantities are rather imprecise at present,
 a situation which we hope will greatly improve at the LHC (and Super-B
factories).

\section{Calculation of $B_s$ to a vector and a pseudo-scalar meson in the PQCD approach}

\subsection{Decay amplitudes}

In the decays of $B_s$ to a vector and a pseudo-scalar meson, only the
longitudinal polarization of the vector meson can contribute, thus
the decay formulas are very similar to those of the $B_s \to PP$ decays.
Since the Lorentz structure of the vector meson wave functions is
different from the pseudo-scalar case, there are two kinds of
emission diagrams in principle. If the emitted meson is a vector meson,
the decay amplitudes  are the same as that of the $B_s\to PP$ case,
since they are both characterized by the  $B_s\to P$ transition form
factors. For the non-factorizable diagrams and also those diagrams
in which a pseudo-scalar meson is emitted, the distribution amplitudes of
the pseudo-scalar meson will be replaced by that of the vector meson as
follows:
\begin{eqnarray}
\phi_2^A(x)\to
\phi_2(x),\;\;\phi_2^P(x)\to\phi_2^s(x),\;\;\phi_2^T(x)\to
\phi_2^t(x),\;\;m_{02}\to M_2,
\end{eqnarray}
where $M_2$ is the mass of the vector meson. The factorizable
emission topology $F_{B\to P}^{SP}$ does not contribute as the
vector meson cannot be generated by the scalar  or the
pseudo-scalar density. Furthermore, we must add a minus sign to
$\phi_3^s(x)$ and $\phi_3^t(x)$ in the annihilation diagram
formulas $F_{ann}$ and $M_{ann}$ if the vector meson is on the
light $\bar s$ quark side.

\subsection{Numerical results for $B_s\to PV$ decays}

The branching ratios of $B_s\to PV$ decays are listed in
Table~\ref{BRPV}. The direct  $CP$ asymmetries of the $B_s \to PV$ decays
 are given in Table~\ref{DIRCPPV}. They are compared with the corresponding
calculations in the QCDF approach~\cite{QCDFBs1}, where the
sources of the errors in the numerical estimates have the same
origin as in the discussion of the $B \to PP$ decays. Comparison
of the entries in Table~\ref{BRPV} shows that the two approaches
give similar results for the tree-dominated decays and the QCD
penguin- and electroweak-penguin dominated decays (except for
those involving an $\eta$ or $\eta^\prime$ in the final states),
and they are radically different for the annihilation-dominated
decays (the last four entries in this table). The branching ratios
for $\bar B_s\to\rho\eta$ and $\bar B_s\to\rho\eta^\prime$ are of
the same order as the QCDF results, but there are large
differences in $\bar B_s\to\omega(\phi)\eta$ and $\bar
B_s\to\rho(\phi)\eta^\prime$. In QCDF, the branching ratios for
$\bar B_s\to\phi\eta$, $\bar B_s\to\phi\eta^\prime$ and $\bar
B_s\to\phi\pi^0$ (electro-weak penguin dominated process) are of
the same magnitude, which implies that the color-suppressed
QCD-penguins have a reduced value similar to the electroweak penguins.
 However, in the pQCD approach, the factorization
scale is low and this dynamically enhances the color-suppressed
QCD penguins sizably.

Predictions for the CP-asymmetries in the two approaches given in
Table~\ref{DIRCPPV} are, however, all quite different. The zeros
shown in some cases for the pQCD approach will be lifted on
including the neglected small subdominant contributions. Also, as
opposed to the QCDF approach, we are able to calculate the CP
asymmetries in the annihilation-dominated topologies. The
mixing-induced $CP$ asymmetries $(S_f)_{B_s}$ and the observable
$(H_f)_{B_s}$, defined in the previous section, are shown in
Table~\ref{MIXBsCPPV}. Currently, there is no data to confront the
estimates given in Tables~\ref{BRPV}, \ref{DIRCPPV} and
\ref{MIXBsCPPV}, and again we hope that this will be remedied at
the LHC (and Super-B factories).

\begin{table}[tb]
\caption{The $CP$-averaged branching ratios ($\times
 10^{-6}$) of $B_s\to PV$ decays obtained in the pQCD approach
(This work); the errors for these entries correspond to the
uncertainties in the input hadronic quantities, from the scale-dependence,
and
the CKM matrix elements, respectively. For comparison, we also
cite the theoretical estimates of the branching ratios in the
QCD factorization framework~\cite{QCDFBs1}. }\label{BRPV}
\begin{center}
 \begin{tabular}{c|c|c|c}
 \hline\hline {Modes}   & Class &   QCDF   & This work   \\  \hline
   ${\overline{B}}^{0}_{s}{\to}\pi^-K^{*+}$       &$T$          &  $8.7^{+4.6+3.5+0.7+0.8}_{-3.7-2.9-1.0-0.7}$
                                                                &  $7.6^{+2.9+0.4+0.5}_{-2.2-0.5-0.3}$\\
   ${\overline{B}}^{0}_{s}{\to}\rho^-K^{+}$       &$T$          & $24.5^{+11.9+9.2+1.8+1.6}_{-9.7-7.8-3.0-1.6}$
                                                                & $17.8^{+7.7+1.3+1.1}_{-5.6-1.6-0.9}$\\
   ${\overline{B}}^{0}_{s}{\to}{\pi}^{0}K^{*0}$   &$ C$        & $0.25^{+0.08+0.10+0.32+0.30}_{-0.08-0.06-0.14-0.14}$
                                                                & $0.07^{+0.02+0.04+0.01}_{-0.01-0.02-0.01}$\\
   ${\overline{B}}^{0}_{s}{\to}{\rho}^{0}K^{0}$   &$ C$        & $0.61^{+0.33+0.21+1.06+0.56}_{-0.26-0.15-0.38-0.36}$
                                                                &  $0.08^{+0.02+0.07+0.01}_{-0.02-0.03-0.00}$\\
   $\overline B^0_s\to K^{*0}\eta$                &$ C$        & $0.26^{+0.15+0.49+0.15+0.57}_{-0.13-0.22-0.05-0.15}$
                                                                & $0.17^{+0.04+0.10+0.03}_{-0.04-0.06-0.01}$\\
   $\overline B^0_s\to K^{*0}\eta^\prime$         &$ C$        & $0.28^{+0.04+0.46+0.23+0.29}_{-0.04-0.24-0.10-0.15} $
                                                                & $0.09^{+0.02+0.03+0.01}_{-0.02-0.02-0.01}$\\
   $\overline B^0_s\to K^{0}\omega$               &$ C$        & $0.51^{+0.20+0.15+0.68+0.40}_{-0.18-0.11-0.23-0.25} $
                                                                & $0.15^{+0.05+0.07+0.02}_{-0.04-0.03-0.01}$\\
   $\overline B^0_s\to K^+ K^{*-}$                &$P$          & $4.1^{+1.7+1.5+1.0+9.2}_{-1.5-1.3-0.9-2.3}$
                                                                & $6.0^{+1.7+1.7+0.7}_{-1.5-1.2-0.3}$\\
   $\overline B^0_s\to K^{*+} K^-$                &$P$          & $5.5^{+1.3+5.0+0.8+14.2}_{-1.4-2.6-0.7-~3.6}$
                                                                & $4.7^{+1.1+2.5+0.0}_{-0.8-1.4-0.0}$\\
   $\overline B^0_s\to K^{0}\overline K^{*0}$     &$P$          & $3.9^{+0.4+1.5+1.3+10.4}_{-0.4-1.4-1.4-~2.8}$
                                                                & $7.3^{+2.5+2.1+0.0}_{-1.7-1.3-0.0}$\\
   $\overline B^0_s\to K^{*0}\overline K^0$       &$P$          & $4.2^{+0.4+4.6+1.1+13.2}_{-0.4-2.2-0.9-~3.2}$
                                                                &  $4.3^{+0.7+2.2+0.0}_{-0.7-1.4-0.0}$\\
   $\overline B^0_s\to\pi^0\phi$                 &$P_{EW}$      & $0.12^{+0.03+0.04+0.01+0.02}_{-0.02-0.04-0.01-0.01} $
                                                                & $0.16^{+0.06+0.02+0.00}_{-0.05-0.02-0.00}$ \\
   $\overline B^0_s\to\rho^0\eta$                &$P_{EW}$      & $0.17^{+0.03+0.07+0.02+0.02}_{-0.03-0.06-0.02-0.01} $
                                                                & $0.06^{+0.03+0.01+0.00}_{-0.02-0.01-0.00}$\\
   $\overline B^0_s\to\rho^0\eta^\prime$         &$P_{EW}$      & $0.25^{+0.06+0.10+0.02+0.02}_{-0.05-0.08-0.02-0.02} $
                                                                & $0.13^{+0.06+0.02+0.00}_{-0.04-0.02-0.01}$\\
   $\overline B^0_s\to\omega\eta$                &$P,C$         & $0.012^{+0.005+0.010+0.028+0.025}_{-0.004-0.003-0.006-0.006} $
                                                                &  $0.04^{+0.03+0.05+0.00}_{-0.01-0.02-0.00}$\\
   $\overline B^0_s\to\omega\eta^\prime$         &$P,C$         & $0.024^{+0.011+0.028+0.077+0.042}_{-0.009-0.006-0.010-0.015}$
                                                                &   $0.44^{+0.18+0.15+0.00}_{-0.13-0.14-0.01}$\\
   $\overline B^0_s\to\phi\eta$                  &$P$           & $0.12^{+0.02+0.95+0.54+0.32}_{-0.02-0.14-0.12-0.13} $
                                                                &  $3.6^{+1.5+0.8+0.0}_{-1.0-0.6-0.0}$\\
   $\overline B^0_s\to\phi\eta^\prime$           &$P$           & $0.05^{+0.01+1.10+0.18+0.40}_{-0.01-0.17-0.08-0.04}$
                                                                &  $0.19^{+0.06+0.19+0.00}_{-0.01-0.13-0.00}$\\
   $\overline B^0_s\to K^0\phi$                  &$P$           & $0.27^{+0.09+0.28+0.09+0.67}_{-0.08-0.14-0.06-0.18} $
                                                                & $0.16^{+0.04+0.09+0.02}_{-0.03-0.04-0.01}$\\\hline
   $\overline B^0_s\to\pi^0 \omega$                    & ann    & $\approx0.0005$
                                                                &   $0.004^{+0.001+0.000+0.000}_{-0.001-0.001-0.000}$\\
   $\overline B^0_s\to\rho^+\pi^-$                     & ann    & $\approx0.003$
                                                                &   $0.22^{+0.05+0.04+0.00}_{-0.05-0.06-0.01}$\\
   $\overline B^0_s\to\pi^+\rho^-$                     & ann    & $\approx0.003$
                                                                &  $0.24^{+0.05+0.05+0.00}_{-0.05-0.06-0.01}$\\
   $\overline B^0_s\to\pi^0\rho^0$                     & ann    & $\approx0.003$
                                                                &  $0.23^{+0.05+0.05+0.00}_{-0.05-0.06-0.01}$\\
 \hline\hline\end{tabular}
\end{center}
 \end{table}

\begin{table}[tb]
\caption{The direct $CP$ asymmetries (in \%) in the $B_s\to PV$  decays,
 obtained in the pQCD approach
(This work); the errors for these entries correspond to the
uncertainties in the input hadronic quantities, from the scale-dependence,
and the CKM matrix elements, respectively.  For comparison, we also
cite the theoretical estimates of the CP asymmetries in the
QCD factorization framework \cite{QCDFBs1}. } \label{DIRCPPV}
\begin{center}
 \begin{tabular}{c|c|c|c}
  \hline\hline {Modes}   & Class &   QCDF   & This work   \\  \hline
   ${\overline{B}}^{0}_{s}{\to}\pi^-K^{*+}$           &$T$        & $0.6^{+0.2+1.4+0.1+19.9}_{-0.1-1.7-0.1-20.1}$
                                                                  & $-19.0^{+2.5+2.7+0.9}_{-2.6-3.4-1.4}$\\
   ${\overline{B}}^{0}_{s}{\to}\rho^-K^{+}$           &$T$        & $-1.5^{+0.4+1.2+0.2+12.1}_{-0.4-1.4-0.3-12.1}$
                                                                  & $14.2^{+2.4+2.3+1.2}_{-2.2-1.6-0.7}$   \\
   ${\overline{B}}^{0}_{s}{\to}{\pi}^{0}K^{*0}$       &$ C$      & $-45.7^{+14.3+13.0+28.4+80.0}_{-16.0-11.6-28.0-59.7}$
                                                                  & $-47.1^{+7.4+35.5+2.9}_{-8.7-29.8-7.0}$ \\
   ${\overline{B}}^{0}_{s}{\to}{\rho}^{0}K^{0}$       &$ C$      & $24.7^{+7.1+14.0+22.8+51.3}_{-5.2-12.4-17.7-52.3}$
                                                                  & $73.4^{+6.4+16.2+2.2}_{-11.7-47.8-3.9}$  \\
   $\overline B^0_s\to K^{*0}\eta$                    &$ C$      & $40.2^{+17.0+24.6+~7.8+65.9}_{-11.5-30.8-14.0-96.3}$
                                                                  & $51.2^{+6.2+14.1+2.0}_{-6.4-12.4-3.3}$ \\
   $\overline B^0_s\to K^{*0}\eta^\prime$             &$ C$      & $-58.6^{+16.9+41.4+19.9+44.9}_{-11.9-11.7-13.9-35.7} $
                                                                  & $-51.1^{+4.6+15.0+3.2}_{-6.6-18.2-4.1}$ \\
   $\overline B^0_s\to K^{0}\omega$                    &$ C$     & $-43.9^{+13.6+18.0+30.6+57.7}_{-13.4-18.2-30.2-49.3} $
                                                                  & $-52.1^{+3.2+22.7+3.2}_{-0.0-15.1-2.0}$  \\
   $\overline B^0_s\to K^+ K^{*-}$                     &$P$       & $2.2^{+0.6+8.4+5.1+68.6}_{-0.7-8.0-5.9-71.0}$
                                                                  & $-36.6^{+2.3+2.8+1.3}_{-2.3-3.5-1.2}$    \\
   $\overline B^0_s\to K^{*+} K^-$                     &$P$       & $-3.1^{+1.0+3.8+1.6+47.5}_{-1.1-2.6-1.3-45.0}$
                                                                  &  $55.3^{+4.4+8.5+5.1}_{-4.9-9.8-2.5}$   \\
   $\overline B^0_s\to K^{0}\overline K^{*0}$          &$P$       & $1.7^{+0.4+0.6+0.5+1.4}_{-0.5-0.5-0.4-0.8}$
                                                                  &  $0$\\
   $\overline B^0_s\to K^{*0}\overline K^0$            &$P$       & $0.2^{+0.0+0.2+0.1+0.2}_{-0.1-0.3-0.1-0.1}$
                                                                  &  $0$\\
   $\overline B^0_s\to\pi^0\phi$                    &$P_{EW}$     & $27.2^{+6.1+9.8+2.7+32.0}_{-6.8-5.6-2.4-37.1} $
                                                                  & $13.3^{+0.3+2.1+1.5}_{-0.4-1.7-0.7}$ \\
   $\overline B^0_s\to\rho^0\eta$                  &$P_{EW}$      & $27.8^{+6.4+9.1+2.6+25.9}_{-6.7-5.7-2.2-28.4} $
                                                                  &  $-9.2^{+1.0+2.8+0.4}_{-0.4-2.7-0.7}$ \\
   $\overline B^0_s\to\rho^0\eta^\prime$            &$P_{EW}$     & $28.9^{+6.1+10.3+1.5+24.8}_{-7.5-~6.3-1.8-27.5} $
                                                                  &  $25.8^{+1.3+2.8+3.4}_{-2.0-3.6-1.5}$\\
   $\overline B^0_s\to\omega\eta$                    &$P,C$       & ---
                                                                  &  $-16.7^{+5.8+15.4+0.8}_{-3.2-19.1-1.7}$ \\
   $\overline B^0_s\to\omega\eta^\prime$             &$P,C$       & ---
                                                                  &  $7.7^{+0.4+4.5+9.4}_{-0.1-4.2-0.4}$\\
   $\overline B^0_s\to\phi\eta$                        &$P$       &  $-8.4^{+2.0+30.1+14.6+36.3}_{-2.1-71.2-44.7-59.7} $
                                                                  &  $-1.8^{+0.0+0.6+0.1}_{-0.1-0.6-0.2}$\\
   $\overline B^0_s\to\phi\eta^\prime$                 &$P$       & $-62.2^{+15.9+132.3+80.8+122.4}_{-10.2-~84.2-46.8-~49.9}$
                                                                  &  $7.8^{+1.5+1.2+0.1}_{-0.5-8.6-0.4}$\\
   $\overline B^0_s\to K^0\phi$                        &$P$       &  $-10.3^{+3.0+4.7+3.7+5.0}_{-2.4-3.0-4.1-7.5} $
                                                                  &  $0$\\\hline
   $\overline B^0_s\to\pi^0 \omega$                    & ann      & ---
                                                                  &  $6.0^{+0.0+0.5+0.8}_{-5.2-3.3-0.4}$\\
   $\overline B^0_s\to\rho^+\pi^-$                     & ann      & ---
                                                                  &   $4.6^{+0.0+2.9+0.6}_{-0.6-3.5-0.3}$  \\
   $\overline B^0_s\to\pi^+\rho^-$                     & ann      & ---
                                                                  &  $-1.3^{+0.9+2.8+0.1}_{-0.4-3.5-0.2}$  \\
   $\overline B^0_s\to\pi^0\rho^0$                     & ann      & ---
                                                                  &  $1.7^{+0.2+2.8+0.2}_{-0.8-3.6-0.1}$\\
 \hline\hline\end{tabular}
\end{center}
 \end{table}

\begin{table}[tb]
\caption{The mixing-induced $CP$ asymmetries $(S_f)_{B_s}$ and $(H_f)_{B_s}$ in
$B_s\to PV$ decays  obtained in the pQCD approach
(This work); the errors for these entries correspond to the
uncertainties in the input hadronic quantities, from the scale-dependence,
and the CKM matrix elements, respectively. } \label{MIXBsCPPV}
\begin{center}
 \begin{tabular}{c|c|c|c}
  \hline\hline {Modes}   & Class &   $(S_f)_{B_s}$   & $(H_f)_{B_s}$   \\  \hline
 ${\overline{B}}^{0}_{s}{\to}{\rho}^{0}K_S$       &$C$        &  $-0.57^{+0.22+0.51+0.02}_{-0.17-0.39-0.05}$
                                                                  &  $-0.36^{+0.10+0.46+0.04}_{-0.13-0.15-0.04}$  \\
 $\overline B^0_s\to K_S\omega$                   &$C$        &  $-0.63^{+0.09+0.28+0.01}_{-0.09-0.11-0.02} $
                                                                  &  $-0.57^{+0.11+0.31+0.02}_{-0.13-0.38-0.02}$  \\
 $\overline B^0_s\to\pi^0\phi$                      &$P_{EW}$     &  $-0.07^{+0.01+0.08+0.02}_{-0.01-0.09-0.03}$
                                                                  &  $0.98^{+0.00+0.01+0.01}_{-0.00-0.03-0.00}$ \\
 $\overline B^0_s\to\rho^0\eta$                     &$P_{EW}$     &  $0.15^{+0.06+0.14+0.01}_{-0.06-0.16-0.01}$
                                                                  &  $0.98^{+0.01+0.01+0.00}_{-0.01-0.03-0.00}$ \\
 $\overline B^0_s\to\rho^0\eta^\prime$              &$P_{EW}$     &  $-0.16^{+0.00+0.10+0.04}_{-0.00-0.12-0.05}$
                                                                  &  $0.95^{+0.01+0.01+0.01}_{-0.00-0.02-0.02}$\\
 $\overline B^0_s\to\omega\eta$                     &$P,C$        &  $-0.02^{+0.01+0.02+0.00}_{-0.03-0.08-0.00}$
                                                                  &  $0.99^{+0.01+0.01+0.00}_{-0.01-0.06-0.00}$ \\
 $\overline B^0_s\to\omega\eta^\prime$              &$P,C$        &  $-0.11^{+0.01+0.04+0.02}_{-0.00-0.04-0.03}$
                                                                  &  $0.99^{+0.00+0.00+0.00}_{-0.00-0.00-0.00}$\\
 $\overline B^0_s\to\phi\eta$                       &$P$          &  $-0.03^{+0.02+0.07+0.01}_{-0.01-0.20-0.02}$
                                                                  &  $1.00^{+0.00+0.00+0.00}_{-0.00-0.01-0.00}$\\
 $\overline B^0_s\to\phi\eta^\prime$                &$P$          &  $0.00^{+0.00+0.02+0.00}_{-0.00-0.02-0.00}$
                                                                  &  $1.00^{+0.00+0.00+0.00}_{-0.00-0.00-0.02}$\\
 $\overline B^0_s\to K_S\phi$                       &$P$          &  $-0.72 $
                                                                  &  $-0.69$\\ \hline
 $\overline B^0_s\to\pi^0 \omega$                  & ann          &  $-0.97^{+0.00+0.00+0.11}_{-0.01-0.00-0.02}$
                                                                  &  $-0.22^{+0.02+0.00+0.12}_{-0.00-0.02-0.29}$\\
 $\overline B^0_s\to\pi^0\rho^0$                   & ann          &  $-0.19^{+0.00+0.02+0.01}_{-0.00-0.02-0.02}$
                                                                  &  $0.99^{+0.00+0.00+0.00}_{-0.00-0.00-0.00}$\\
 \hline\hline\end{tabular}
\end{center}
 \end{table}

\section{$B_s\to VV$ decays in PQCD approach}

\subsection{Decay amplitudes}

There are three kinds of polarizations of a vector meson,
namely longitudinal (L), normal (N) and transverse (T).
The amplitudes for a $B_s$ meson decay to two vector mesons
are also characterized by the polarization states of these
vector mesons. The amplitudes $ A^{(\sigma)}$
for the decay $B_s(P_B) \to V_2(P_2,\epsilon_{2\mu}^{*}) +
V_3(P_3,\epsilon_{3\mu}^{*})$ can be decomposed as follows:
\begin{eqnarray}
A^{(\sigma)}
&=&\epsilon_{2\mu}^{*}(\sigma)\epsilon_{3\nu}^{*}(\sigma) \left[ a
\,\, g^{\mu\nu} + {b \over M_2 M_{3}} P_{B}^\mu P_{B}^\nu + i{c \over
M_2 M_{3}} \epsilon^{\mu\nu\alpha\beta} P_{2\alpha}
P_{3\beta}\right]\;,
\nonumber \\
&\equiv &A_{L}+A_{N}
\epsilon^{*}_{2}(\sigma=T)\cdot\epsilon^{*}_{3}(\sigma=T) +i
\frac{A_{T}}{M_{B_s}^2}\epsilon^{\alpha \beta\gamma \rho}
\epsilon^{*}_{2\alpha}(\sigma)\epsilon^{*}_{3\beta}(\sigma)
P_{2\gamma }P_{3\rho }\;,
\end{eqnarray}
where $M_2$ and $M_3$ are the masses of the vector mesons $V_2$ and $V_3$,
respectively. The definitions of the  amplitudes $ A^{i}$ $(i=L,N,T)$
in terms of the Lorentz-invariant amplitudes $a$, $b$ and $c$ are
\begin{eqnarray}
A_L &=& a \,\, \epsilon_2^{*}(L) \cdot
\epsilon_3^{*}(L) +{b \over M_2 M_{3}} \epsilon_{2}^{*}(L) \cdot
P_3 \,\, \epsilon_{3}^{*}(L) \cdot P_2\;,
\nonumber \\
A_N &=& a ,
\label{id-rel} \\
A_T &=& {c \over r_2 r_{3}}\;. \nonumber
\end{eqnarray}

The longitudinal polarization amplitudes for the $B_s\to VV$
decays can be obtained from those in the $B_s\to PP$ decays with
the following replacement in the distribution amplitudes:
\begin{eqnarray}
\phi_{2(3)}^A(x) \rightarrow \phi_{2(3)}(x), \,\,\,
\phi_{2(3)}^P(x) \rightarrow \phi_{2(3)}^s(x),  \,\,\,
\phi_{2(3)}^T(x) \rightarrow \phi_{2(3)}^t(x),
 \end{eqnarray}
for the emission diagrams, while
\begin{eqnarray}
\phi_{2(3)}^A(x) \rightarrow \phi_{2(3)}(x), \,\,\,
\phi_{2(3)}^P(x) \rightarrow (-)\phi_{2(3)}^s(x),  \,\,\,
\phi_{2(3)}^T(x) \rightarrow (-)\phi_{2(3)}^t(x),
 \end{eqnarray}
 for the annihilation diagrams.
 The factorizable emission  topology contribution
 $F^{SP,i}_{B_{s} \rightarrow V_3}$ ($i=L,N,T$)
 vanish due to the conservation of charge parity.

The normal and transverse polarization amplitudes for $B_s\to
VV$ decays are displayed as follows.
For the factorizable emission diagrams shown in
Fig.\ref{emission}(a) and (b), the formulas are
\begin{eqnarray}
f_{V_2}F^{LL,N}_{B_s\to V_3}(a_i)&=&8\pi
C_FM_{B_s}^4f_{V_2}r_2\int^1_0dx_1dx_3\int^\infty_0b_1db_1b_3db_3\phi_{B_s}(x_1,b_1)
\Big\{h_e(x_1,x_3,b_1,b_3)\nonumber
\\
&&\times E_e(t_a)a_i(t_a)[\phi_3^T(x_3)+2r_3 \phi_3^v(x_3)+r_3 x_3
(\phi_3^v(x_3)-\phi_3^a(x_3))] \non\\ &&
\;\;+r_3[\phi_3^v(x_3)+\phi_3^a(x_3)]E_e(t_a')a_i(t_a')h_e(x_3,x_1,b_3,b_1)
\Big\}, \\
f_{V_2}F^{LL,T}_{B_s \to V_3}( a_i)&=&16\pi
  C_FM_{B_s}^4f_{V_2}r_2\int^1_0dx_1dx_3\int^\infty_0b_1db_1b_3db_3\phi_{B_s}(x_1,b_1)
  \Big\{h_e(x_1,x_3,b_1,b_3)\nonumber\\
  &&\times [\phi_3^T(x_3)+2r_3 \phi_3^v(x_3)-r_3 x_3
  (\phi_3^v(x_3)-\phi_3^a(x_3))]E_e(t_a)a_i(t_a)
 \nonumber\\
 && \;\;+r_3[\phi_3^v(x_3)+\phi_3^a(x_3)]E_e(t_a')a_i(t_a')h_e(x_3,x_1,b_3,b_1)
 \Big\},
\\
  F^{LR,i}_{B_s\to V_3}(a_i)&=&F^{LL,i}_{ B_s\to V_3}(a_i),
\\
  F^{SP,i }_{B_s\to V_3}(a_i)&=&0,
\end{eqnarray}
with $i=L,N,T$.

The non-factorizable emission diagrams are shown in
Fig.\ref{emission}(c) and (d). Their contributions are expressed
in the formulas given below:
\begin{eqnarray}
M_{B_s \to V_3}^{LL,N}(a_i)&=&32\pi
C_FM_{B_s}^4r_2/ \sqrt{6}\int^1_0dx_1dx_2dx_3\int^\infty_0b_1db_1b_2db_2\phi_{B_s}(x_1,b_1) \nonumber\\
 && \times
 \Big\{\left[x_2(\phi_2^v(x_2)+\phi_2^a(x_2))\phi_3^T(x_3)
 -2r_3(x_2+x_3)(\phi_2^v(x_2)\phi_3^v(x_3)+\phi_2^a(x_2)\phi_3^a(x_3))\right]
 \nonumber\\
  &&\;\;h_n(x_1,x_2,x_3,b_1,b_2)E_e'(t_b')a_i(t_b')\nonumber\\
  &&\;\;\;
+(1-x_2)(\phi_2^v(x_2)+\phi_2^a(x_2))\phi_3^T(x_3)
E_e'(t_b)a_i(t_b) h_n(x_1,1-x_2,x_3,b_1,b_2)\Big\},
\\
\ \ \ \ \
M_{B_s\to V_3}^{LL,T}(a_i)&=&64\pi C_FM_{B_s}^4r_2/
\sqrt{6}\int^1_0dx_1dx_2dx_3\int^\infty_0b_1db_1b_2db_2\phi_{B_s}(x_1,b_1)\Big\{
E_e'(t_b')a_i(t_b')\nonumber
\\
 &&\times \big[x_2(\phi_2^v(x_2)+\phi_2^a(x_2))\phi_3^T(x_3)
-2r_3(x_2+x_3)(\phi_2^v(x_2)\phi_3^a(x_3)\nonumber
\\
 && +\phi_2^a(x_2)\phi_3^v(x_3))\big] h_n(x_1,x_2,x_3,b_1,b_2)\nonumber
\\
 && +(1-x_2)[\phi_2^v(x_2)+\phi_2^a(x_2)]\phi_3^T(x_3)
E_e'(t_b)a_i(t_b) h_n(x_1,1-x_2,x_3,b_1,b_2)\Big\},
\end{eqnarray}
\begin{eqnarray}
M_{B_s\to V_3}^{LR,T}(a_i)&=&
 2M_{B_s\to V_3}^{LR,N}(a_i)\nonumber\\
 &=&64\pi
C_FM_{B_s}^4/\sqrt{6}\int^1_0dx_1dx_2dx_3\int^\infty_0b_1db_1b_2db_2\phi_{B_s}(x_1,b_1)\nonumber
\\&&\times
r_3 x_3 \phi_2^T(x_2)(\phi_3^v(x_3)-\phi_3^a(x_3))\non\\
&& \Big\{E_e'(t_b')a_i(t_b')
h_n(x_1,x_2,x_3,b_1,b_2)+E_e'(t_b)a_i(t_b)
h_n(x_1,1-x_2,x_3,b_1,b_2)\Big\},
\end{eqnarray}
\begin{eqnarray} M^{SP,N}_{B_s\to V_3}(a_i) &=&32 \pi C_F
M_{B_s}^4/\sqrt{6}\int^1_0dx_1dx_2dx_3\int^\infty_0b_1db_1b_2db_2
\phi_{B_s}(x_1,b_1)r_2\non
\\
&&\times\Big\{ x_2
(\phi_2^v(x_2)-\phi_2^a(x_2))\phi_3^T(x_3)E_e'(t_b')a_i(t_b')
h_n(x_1,x_2,x_3,b_1,b_2)\non
\\
&&+h_n(x_1,1-x_2,x_3,b_1,b_2)[(1-x_2)(\phi_2^v(x_2)-\phi_2^a(x_2))\phi_3^T(x_3)\non\\
&&\;\;\;-2r_3(1-x_2+x_3)
(\phi_2^v(x_2)\phi_3^v(x_3)-\phi_2^a(x_2)\phi_3^a(x_3))]E_e'(t_b)a_i(t_b)
\Big\} ,
 \end{eqnarray}
\begin{eqnarray}
 M^{SP,T}_{B_s\to V_3}(a_i) &=&64\pi C_F
M_{B_s}^4/\sqrt{6}\int^1_0dx_1dx_2dx_3\int^\infty_0b_1db_1b_2db_2\phi_{B_s}(x_1,b_1)r_2\non
\\
&&\times\Big\{ x_2
(\phi_2^v(x_2)-\phi_2^a(x_2))\phi_3^T(x_3)E_e'(t_b')a_i(t_b')
h_n(x_1,x_2,x_3,b_1,b_2)\non
\\
&&+h_n(x_1,1-x_2,x_3,b_1,b_2)[(1-x_2)(\phi_2^v(x_2)-\phi_2^a(x_2))\phi_3^T(x_3)\non\\
&&\;\;\;-2r_3(1-x_2+x_3)
(\phi_2^v(x_2)\phi_3^a(x_3)-\phi_2^a(x_2)\phi_3^v(x_3))]E_e'(t_b)a_i(t_b)
\Big\}.
\end{eqnarray}

The factorizable annihilation diagrams are shown in
Fig.\ref{annihilation} (a) and (b), and the normal polarization
contributions are:
\begin{eqnarray}
f_{B_s}F_{ann}^{LL,N}(a_i)&=& f_{B_s}F_{ann}^{LR,N}(a_i)\nonumber \\
&=& -8\pi C_FM_{B_s}^4f_{B_s}r_2
r_3\int^1_0dx_2dx_3\int^\infty_0b_2db_2b_3db_3
\Big\{E_a(t_c)a_i(t_c)h_a(x_2,1-x_3,b_2,b_3)
)\non\\
&&\left[(2-x_3)\left(\phi_2^v(x_2)\phi_3^v(x_3)+\phi_2^a(x_2)\phi_3^a(x_3)\right)
+x_3(\phi_2^v(x_2)\phi_3^a(x_3)+\phi_2^a(x_2)\phi_3^v(x_3))\right]
\non\\
&&-h_a(1-x_3,x_2,b_3,b_2)[(1+x_2)
(\phi_2^v(x_2)\phi_3^v(x_3)+\phi_2^a(x_2)\phi_3^a(x_3))
\non\\
&&
-(1-x_2)(\phi_2^v(x_2)\phi_3^a(x_3)+\phi_2^a(x_2)\phi_3^v(x_3))]
E_a(t_c')a_i(t_c')\Big\}.
\end{eqnarray}
Note, that large cancellations between the two diagrams
Fig.\ref{annihilation} (a) and (b) take place, as a result of which
contributions from these diagrams are suppressed.

For the transverse polarization, we have
\begin{eqnarray}
f_{B_s}F_{ann}^{LL,T}(a_i)&=&-f_{B_s}F_{ann}^{LR,T}(a_i)\nonumber\\
&=& -16\pi C_FM_{B_s}^4f_{B_s}r_2
r_3\int^1_0dx_2dx_3\int^\infty_0b_2db_2b_3db_3 \Big\{
[x_3(\phi_2^v(x_2)\phi_3^v(x_3)+\phi_2^a(x_2)\phi_3^a(x_3))\non\\
&&+(2-x_3)(\phi_2^v(x_2)\phi_3^a(x_3)+\phi_2^a(x_2)\phi_3^v(x_3))]
E_a(t_c)a_i(t_c)h_a(x_2,1-x_3,b_2,b_3)\non\\
&&+h_a(1-x_3,x_2,b_3,b_2)[(1-x_2)
(\phi_2^v(x_2)\phi_3^v(x_3)+\phi_2^a(x_2)\phi_3^a(x_3))
\non\\
&&
-(1+x_2)(\phi_2^v(x_2)\phi_3^a(x_3)+\phi_2^a(x_2)\phi_3^v(x_3))]
E_a(t_c')a_i(t_c')\Big\}.
 \end{eqnarray}
We remark that, although the cancellations in this case are not as
severe as for the normal polarization case, the dominant
contributions are still power suppressed by $r_2r_3$,
where $r_{2(3)}=M_{2(3)}/M_{B_s}$.
 For the
$(S-P)(S+P)$ operators, we have
\begin{eqnarray}
f_{B_s}F_{ann}^{SP,T}(a_i)&=&2
f_{B_s}F_{ann}^{SP,N}(a_i)\nonumber\\
&=&-32\pi
C_FM_{B_s}^4f_{B_s}\int^1_0dx_2dx_3\int^\infty_0b_2db_2b_3db_3\Big\{
r_2(\phi_2^v(x_2)+\phi_2^a(x_2))\phi_3^T(x_3)\non\\
&&\times E_a(t_c)a_i(t_c)h_a(x_2,1-x_3,b_2,b_3)
\non\\
&&+r_3
\phi_2^T(x_2)(\phi_3^v(x_3)-\phi_3^a(x_3))E_a(t_c')a_i(t_c')h_a(1-x_3,x_2,b_3,b_2)
\Big\}.
\end{eqnarray}
Again, for this case,  like the $B_s \to PP$ decays, no
cancellations among the contributing diagrams or power suppressions
are involved. The chiral enhancement here
for the transverse and normal polarizations are essential for
the explanation of the large transverse polarization fraction in
the penguin dominant $B$ decays, such as $B\to K^*\phi$ and $B\to
K^*\rho$ decays \cite{kphi,krho}. For the non-factorizable
annihilation diagrams shown in Fig.\ref{annihilation} (c) and (d),
we have
\begin{eqnarray}
M_{ann}^{LL,N}(a_i)&=&M_{ann}^{SP,N}(a_i)\nonumber \\
&=&-64\pi C_FM_{B_s}^4r_2 r_3/\sqrt
{6}\int^1_0dx_1dx_2dx_3\int^\infty_0b_1db_2b_2db_2\phi_{B_s}(x_1,b_1)[\phi_2^v(x_2)\phi_3^v(x_3)\non\\
&&\;\;+\phi_2^a(x_2)\phi_3^a(x_3)]
E_a'(t_d)a_i(t_d)h_{na}(x_1,x_2,x_3,b_1,b_2),
\end{eqnarray}
\begin{eqnarray}
M_{ann}^{LL,T}(a_i)&=& -M_{ann}^{SP,T}(a_i)\nonumber\\
&=&-128 \pi C_FM_{B_s}^4r_2 r_3/\sqrt
{6}\int^1_0dx_1dx_2dx_3\int^\infty_0b_1db_2b_2db_2\phi_{B_s}(x_1,b_1)[\phi_2^v(x_2)\phi_3^a(x_3)\non\\
&&\;\;+\phi_2^a(x_2)\phi_3^v(x_3)]
E_a'(t_d)a_i(t_d)h_{na}(x_1,x_2,x_3,b_1,b_2),
\end{eqnarray}
\begin{eqnarray}
M_{ann}^{LR,T}(a_i)&=&2  M_{ann}^{LR,N}(a_i)\nonumber\\
&=&-64\pi C_FM_{B_s}^4 /\sqrt {6}\int^1_0dx_1dx_2dx_3\int^\infty
b_1db_1b_2db_2\phi_{B_s}(x_1,b_1)
\Big\{h_{na}'(x_1,x_2,x_3,b_1,b_2) \non\\
&&\left[r_2x_2
(\phi_2^v(x_2)+\phi_2^a(x_2))\phi_3^T(x_3)-r_3(1-x_3)\phi_2^T(x_2)
(\phi_3^v(x_3)-\phi_3^a(x_3))\right]
E_a'(t_d')a_i(t_d')\non\\
&&\;\;+\left[r_2(2-x_2)(\phi_2^v(x_2)+\phi_2^a(x_2))\phi_3^T(x_3)
-r_3(1+x_3)\phi_2^T(x_2)(\phi_3^v(x_3)-\phi_3^a(x_3))\right]\non\\
&&\;\;\times E_a'(t_d)a_i(t_d)h_{na}(x_1,x_2,x_3,b_1,b_2)\Big\}.
\end{eqnarray}
These contributions are all  power-suppressed as expected.

\subsection{Numerical results for $B_s\to VV$ decays}

The decay width for $B_s\to V_2V_3$ is given as
\begin{equation}
\Gamma =\frac{P_c}{8\pi M^{2}_{B_s} } \sum_{i=0,\parallel,\perp}
A_{(i)}^\dagger  A_{(i)}\;, \label{dr1}
\end{equation}
where $P_c$ is the momentum of either of the two vector mesons in
the final states. The sum is over the three transversity amplitudes
of the two vector mesons, defined as follows:
\begin{eqnarray}
A_{0}&\equiv&- A_{L}, \nonumber\\
A_{\parallel}&\equiv& \sqrt{2}a, \nonumber \\
A_{\perp}&\equiv& r_{2} r_{3} \sqrt{2(\kappa^{2}-1)} A_{T}\;,
 \label{ase}
\end{eqnarray}
with the ratio $\kappa=P_{2}\cdot P_{3}/(M_{2}M_{3})$.  Note that
the definitions of $A_i (i=0,\parallel,\perp)$ are consistent with
those in~\cite{Beneke:2006hg}, except for an additional minus sign
in $A_0$, so that our definitions of the relative strong phases
$\phi_i(i=\parallel,\perp)$ (see text below)  also differ from the
ones in \cite{Beneke:2006hg} by $\pi$. The polarization fractions
$f_i(i=0,\parallel,\perp)$ are defined as follows:
\begin{eqnarray}
f_i=\frac{|A_i|^2}{|A_0|^2+|A_\parallel|^2+|A_\perp|^2}\;.
\end{eqnarray}

We first give the numerical results of the form factors at maximal
recoil:
\begin{equation}
\begin{array}{lll}
 V^{B \rightarrow K^{\ast}}=0.25^{+0.05+0.00}_{-0.05-0.00}, &  A_0^{B \rightarrow
K^{\ast}}=0.30^{+0.06+0.00}_{-0.05-0.01}, & A_1^{B \rightarrow K^{\ast}}=0.19^{+0.04+0.00}_{-0.03-0.00}, \\
 V^{B
\rightarrow \rho}=0.21^{+0.05+0.00}_{-0.04-0.00}, &
 A_0^{B \rightarrow \rho}=0.25^{+0.05+0.00}_{-0.04-0.01}, & A_1^{B \rightarrow
\rho}=0.17^{+0.04+0.00}_{-0.03-0.00},
 \\
 V^{B \rightarrow \omega}=0.19^{+0.04+0.00}_{-0.04-0.00}, &  A_0^{B \rightarrow
\omega}=0.23^{+0.05+0.00}_{-0.04-0.01}, & A_1^{B \rightarrow
\omega}=0.15^{+0.03+0.00}_{-0.03-0.00}.\\
V^{B_s \rightarrow K^{\ast}}=0.21^{+0.04+0.00}_{-0.03-0.01},&
A_0^{B_s \rightarrow
K^{\ast}}=0.25^{+0.05+0.00}_{-0.05-0.01}, & A_1^{B_s \rightarrow K^{\ast}}=0.16^{+0.03+0.00}_{-0.03-0.01}, \\
 V^{B_s
\rightarrow \phi}=0.25^{+0.05+0.00}_{-0.04-0.01},&  A_0^{B_s
\rightarrow \phi}=0.30^{+0.05+0.00}_{-0.05-0.01}, &
 A_1^{B_s \rightarrow \phi}=0.19^{+0.03+0.00}_{-0.03-0.01}.
\end{array}
\end{equation}
where the first error in the above entries is due to the input
hadronic parameters $f_{B(B_s)}$, $\omega_b$  and the second one
is from the hard scale and $\Lambda_{QCD}$. The entries in the
first three lines involve the decays of the $B^\pm$ and $B_d^0$
mesons to two vector mesons, and as such are not required for the
$B_s^0 \to VV$ decays being worked out in this paper. We list them
here to see the SU(3)-breaking effects in the form factors. Within
errors, these form factors are in agreement with the ones in the
light cone QCD sum rules~\cite{Ball:2004ye} and with the slightly
different estimates of the same in~\cite{Beneke:2006hg}. The
branching ratios of $B_s\to VV$ decays are listed in
Table~\ref{BRVV}. They are compared with the corresponding results
in the QCDF approach~\cite{Beneke:2006hg}. A comparison shows that
the tree-, penguin- and electroweak  dominated decays are
comparable in the two approaches, with the numerical differences
reflecting the input parameters. The color-suppressed decays, but
more markedly the annihilation-dominated decays, differ in these
approaches, a feature which is well appreciated in the literature.
The experimental upper bounds on some of the decays are also
listed. Except for the decay $\overline{B_s} \to \phi \phi$, where
an experimental measurement may be just around the corner, all
other decay modes remain essentially unexplored.

\begin{table}[tb]
\caption{The CP-averaged branching ratios in $B_s\to VV$ decays
($\times 10^{-6}$)  obtained in the pQCD approach (This work); the
errors for these entries correspond to the uncertainties in the
input hadronic quantities, from the scale-dependence, and the CKM
matrix elements, respectively.
 For comparison, we also
cite the updated theoretical estimates in the QCD factorization
framework~\cite{Beneke:2006hg}. The experimental upper limits (at
90\% C.L.) are from the Particle Data Group~\cite{PDG}.}
\label{BRVV}
\begin{center}
\begin{tabular}{c|c|c|c|c}
\hline \hline Channel & Class & QCDF~\cite{Beneke:2006hg} & This work& Exp~\cite{PDG}\\
\hline \hline $\bar{B}_s \to \rho^0 K^{\ast0}$ & $C$&
$1.5^{+1.0+3.1}_{-0.5-1.5}$
& $0.33^{+0.09+0.14+0.00}_{-0.07-0.09-0.01}$ & $<767$\\
$\bar{B}_s \to \omega K^{\ast0}$ & $C$&
$1.2^{+0.7+2.3}_{-0.3-1.1}$
& $0.31^{+0.10+0.12+0.07}_{-0.07-0.06-0.02}$&  \\
$\bar{B}_s\to \rho^-K^{\ast+}$ & $T$& $25.2^{+1.5+4.7}_{-1.7-3.1}$
& $20.9^{+8.2+1.4+1.2}_{-6.2-1.4-1.1}$ &  \\
$\bar{B}_s \to K^{*-} K^{*+}$&$P$& $9.1^{+2.5+10.2}_{-2.2-5.9}$
& $6.7^{+1.5+3.4+0.5}_{-1.2-1.4-0.2}$& \\
$\bar{B}_s \to K^{*0} \overline{K}^{*0}$&$P$&
$9.1^{+0.5+11.3}_{-0.4-6.8}$
& $7.8^{+1.9+3.8+0.0}_{-1.5-2.2-0.0}$& $<1681$\\
$\bar{B}_s \to \phi K^{*0}$&$P$& $0.4^{+0.1+0.5}_{-0.1-0.3}$
& $0.65^{+0.16+0.27+0.10}_{-0.13-0.18-0.04}$&$<1013$\\
$\bar{B}_s \to \phi\phi$&$P$& $21.8^{+1.1+30.4}_{-1.1-17.0}$
& $35.3^{+8.3+16.7+0.0}_{-6.9-10.2-0.0}$& $14 \pm 8$ \\
$\bar{B}_s \to \rho^+\rho^-$&ann& $0.34^{+0.03+0.60}_{-0.03-0.38}$
& $1.0^{+0.2+0.3+0.0}_{-0.2-0.2-0.0}$& \\
$\bar{B}_s \to \rho^0\rho^0$&ann& $0.17^{+0.01+0.30}_{-0.01-0.19}$
& $0.51^{+0.12+0.17+0.01}_{-0.11-0.10-0.01}$& $<320$\\
$\bar{B}_s \to \rho^0\omega$&ann& $<0.01  $
& $0.007^{+0.002+0.001+0.000}_{-0.001-0.001-0.000}$&\\
$\bar{B}_s\to \omega\omega$&ann& $0.11^{+0.01+0.20}_{-0.01-0.12}$
& $0.39^{+0.09+0.13+0.01}_{-0.08-0.07-0.00}$&\\
$\bar{B}_s\to\phi\rho^0$&$P_{EW}$&$0.40^{+0.12+0.25}_{-0.10-0.04}$
& $0.23^{+0.09+0.03+0.00}_{-0.07-0.01-0.01}$&$<617$\\
$\bar{B}_s \to \phi\omega$&$P$& $0.10^{+0.05+0.48}_{-0.03-0.12}$
& $0.16^{+0.09+0.10+0.01}_{-0.05-0.04-0.00}$&\\
\hline\hline
\end{tabular}
\end{center}
\end{table}

The results for the longitudinal polarization fraction $f_0$,
parallel polarization fraction $f_\parallel$ and perpendicular
polarization fraction $f_\perp$,  their relative phases
$\phi_{\parallel}\equiv Arg(A_{\parallel}/A_{0})$ and
$\phi_{\perp}\equiv Arg(A_{\perp}/A_{0})$ (both in radians), and
the direct CP-asymmetry in the $\overline{B_s^0} \to VV$ decays
are displayed in Table~\ref{HAVV}. In calculating the
CP-asymmetries, we have used the definition given earlier in
Eq.~\ref{ACPDef}. Again, there are no data available to confront
the entries in Table~\ref{HAVV}.

Two remarks on the $B_s \to \phi \phi$ decays presented above are in order.

First, $B_s\to\phi\phi$ is a $b\to s$ penguin dominated process.
The CDF collaboration \cite{Bs phi phi ex} has reported the decay
branching fraction of this channel as $(14 \pm 8) \times 10^{-6}$,
but a thorough angular analysis is still lacking. This channel is
very similar to the decay $B \to \phi K^*$, which is well measured
in the experiment. In the $B \to \phi K^*$ decays, the data show
that the fraction of the left-handed polarization reaches about
50\%. This result is quite different from the expectation in the
factorization assumption that the longitudinal polarization should
dominate due to the quark helicity analysis. There exist lots of
theoretical attempts to solve this contradiction. To distinguish,
which one is the most appropriate scheme, we must investigate 
the decay $B_s \to \phi\phi$ and other similar channels.

In the pQCD approach,  the weak annihilation diagram induced by the
operator $O_6$ can enhance the transverse polarization sizably. In
 Ref.~\cite{Bs phi phi th 1}, the longitudinal fraction of $B
\to \phi K^*$ is about 75 \%, which is much smaller than that
obtained based on  the factorization assumption, but still larger
than the data.  Moreover, the branching fraction is overestimated.
Li~\cite{Bs phi phi th 2} has suggested a strategy to solve these two
problems together, by invoking a smaller value for the form factor $A_0$. 
A smaller $A_0$ is also
consistent with the predictions in other approaches, such as the
covariant light front quark model \cite{haiyang cheng}. In the
pQCD approach, a smaller $A_0$ requires smaller Gegenbauer
moments of the longitudinal polarized distribution amplitudes of the
recoiling vector meson. Following this lead, and  performing a calculation using the
asymptotic distribution amplitude for $K^{\ast}$, one indeed finds
that the longitudinal polarization fraction can be reduced; Ref.~\cite{Bs phi phi th 2}
finds that this fraction  can be reduced to 59\%.

Based on the above discussion, we also analyzed the $B_s \to V V$ decays
adopting the asymptotic forms of the twist-3 distribution amplitudes
in the pQCD approach, while keeping the leading-twist distribution
amplitudes up to the second Gegenbauer moments, as done for the
pseudo-scalar meson case discussed earlier. In addition, we also test the sensitivity
of the branching ratio of the $B_s\to\phi \phi$ decay on the Gengenbauer
moments of the twist-2 distribution amplitudes. The result of the branching ratio
 for this channel with asymptotic twist-2 distribution
amplitudes is $22.2 \times 10^{-6}$, which is much smaller than
that for the case with higher Gegenbauer moments and is closer to
the experiment. However, we find that the polarization
fractions do not change too much using the asymptotic forms of
distribution amplitudes. The reason is that the contributions from
the annihilation topology, which enhance the transverse
polarization fraction, also decrease with a decreased value of 
the form factor $A_0$. This is different from  the $B \to \phi K^{\ast}$ decay.

Second, it is to be noted that the longitudinal polarization of $B
\to \phi K^{\ast}$ and $\bar B_s\to\phi\phi$ can be related to
each other in the SU(3) limit. The discrepancy between them
represents SU(3) symmetry breaking effects, which can be reflected
in the following aspects in the pQCD approach specifically: In the
first place, the shape parameter of $B$ and $B_s$ wavefunction in
the initial state can give  rise to differences in these
two U-spin related process. Then, the longitudinal
and transverse decay constants as well as the Gegenbauer moments
in the two vector mesons in the final state can also
contribute to the SU(3) symmetry breaking effects significantly.
In addition,  these two decay
processes also differ due to the absence of the time-like penguin
annihilation in the $B \to \phi K^{\ast}$ decay.

\begin{table}[tb]
\caption{The CP-averaged polarization fractions, relative phases
and direct CP asymmetries in the $B_s\to VV$ decays
  obtained in the pQCD approach;
 the errors for these entries correspond to the
uncertainties in the input hadronic quantities, from the
scale-dependence, and the CKM matrix elements,
respectively.}\label{HAVV}
\begin{center}
{\footnotesize
\begin{tabular}{c||ccccccc}
\hline \hline Channel & $f_0(\%)$ & $f_\parallel(\%)$ &
$f_{\perp}(\%)$
& $\phi_{\|}(rad)$ & $\phi_{\perp}(rad)$ & CP(\%)\\
\hline \hline $\bar{B}_s \to \rho^0 K^{\ast0}$
& $45.5^{+0.4+6.9+0.6}_{-0.3-4.3-0.9}$ & $27.6^{+0.1+2.1+0.4}_{-0.2-3.4-0.3}$ & $26.9^{+0.2+2.3+0.4}_{-0.3-3.5-0.3}$ & $2.7^{+0.2+0.2+0.3}_{-0.3-0.3-0.1}$ & $2.8^{+0.3+0.3+0.4}_{-0.2-0.2-0.1}$ &$61.8^{+3.2+17.1+4.4}_{-4.7-22.8-2.3}$\\
$\bar{B}_s \to \omega K^{\ast0}$
& $53.2^{+0.3+3.5+2.3}_{-0.2-2.9-1.3}$ & $23.6^{+0.2+1.5+0.4}_{-0.1-1.7-1.0}$& $23.1^{+0.1+1.4+1.0}_{-0.2-1.7-0.6}$ & $1.4^{+0.1+0.2+0.0}_{-0.1-0.2-0.1}$ & $1.4^{+0.1+0.2+0.0}_{-0.1-0.2-0.0}$ &$-62.1^{+4.8+19.7+5.5}_{-3.9-12.6-1.9}$\\
$\bar{B}_s\to \rho^-K^{\ast+}$
& $93.7^{+0.1+0.2+0.0}_{-0.2-0.3-0.2}$ & $3.4^{+0.1+0.2+0.1}_{-0.0-0.1-0.0}$ & $2.9^{+0.1+0.1+0.1}_{-0.1-0.1-0.0}$ & $3.0^{+0.1+0.1+0.1}_{-0.0-0.0-0.0}$ & $3.1^{+0.0+0.0+0.0}_{-0.1-0.1-0.1}$ &$-8.2^{+1.0+1.2+0.4}_{-1.2-1.7-1.1}$\\
$\bar{B}_s \to K^{*-} K^{*+}$
&$43.8^{+5.1+2.1+3.7}_{-4.0-2.3-1.5}$&$30.1^{+2.1+0.9+0.8}_{-2.7-1.0-1.9}$&$26.1^{+1.8+1.4+0.7}_{-2.4-1.0-1.8}$&$1.7^{+0.2+0.1+0.1}_{-0.2-0.1-0.0}$&$1.7^{+0.2+0.1+0.1}_{-0.2-0.1-0.0}$&
$9.3^{+0.4+3.3+0.3}_{-0.7-3.6-0.2}$\\
$\bar{B}_s \to K^{*0} \overline{K}^{*0}$ &$49.7^{+5.7+0.6+0.0}_{-4.8-3.8-0.0}$&$26.8^{+2.6+2.1+0.0}_{-3.0-0.3-0.0}$&$23.5^{+2.2+1.7+0.0}_{-2.7-0.3-0.0}$&$1.4^{+0.2+0.0+0.0}_{-0.1-0.0-0.0}$&$1.4^{+0.1+0.0+0.0}_{-0.1-0.0-0.0}$ &0\\
$\bar{B}_s \to \phi K^{*0}$ &$71.2^{+3.2+2.7+0.0}_{-3.0-3.7-0.0}$&$15.5^{+1.6+2.1+0.0}_{-1.7-1.5-0.0}$&$13.3^{+1.4+1.7+0.0}_{-1.5-1.3-0.0}$&$1.4^{+0.1+0.0+0.0}_{-0.1-0.1-0.0}$&$1.4^{+0.1+0.0+0.0}_{-0.1-0.1-0.0}$ &0\\
$\bar{B}_s \to \phi\phi$& $61.9^{+3.6+2.5+0.0}_{-3.2-3.3-0.0}$ &$20.7^{+1.7+1.8+0.0}_{-2.0-1.4-0.0}$&$17.4^{+1.5+1.5+0.0}_{-1.6-1.1-0.0}$&$1.3^{+0.2+0.1+0.0}_{-0.1-0.0-0.0}$&$1.3^{+0.2+0.0+0.0}_{-0.1-0.0-0.0}$ &0\\
$\bar{B}_s \to \rho^+\rho^-$ & $\sim$ 100 &$\sim$ 0&$\sim$ 0&$4.3^{+0.1+0.1+0.0}_{-0.0-0.1-0.0}$&$4.7^{+0.0+0.3+0.1}_{-0.0-0.5-0.0}$&$-2.1^{+0.2+1.7+0.1}_{-0.1-1.3-0.1}$\\
$\bar{B}_s \to \rho^0\rho^0$ &$\sim$ 100 &$\sim$ 0&$\sim$ 0&$4.3^{+0.1+0.1+0.0}_{-0.0-0.1-0.0}$&$4.7^{+0.0+0.3+0.1}_{-0.0-0.5-0.0}$&$-2.1^{+0.2+1.7+0.1}_{-0.1-1.3-0.1}$\\
$\bar{B}_s \to \rho^0\omega$ &$\sim$ 100&$\sim$ 0&$\sim$ 0&$4.5^{+0.0+0.0+0.0}_{-0.0-0.0-0.0}$&$3.2^{+0.0+0.1+0.0}_{-0.1-0.0-0.2}$&$6.0^{+0.7+2.7+1.0}_{-0.5-3.9-0.4}$\\
$\bar{B}_s\to \omega\omega$ &$\sim$ 100&$\sim$ 0&$\sim$ 0&$4.3^{+0.0+0.1+0.0}_{-0.0-0.1-0.0}$&$4.7^{+0.0+0.2+0.0}_{-0.0-0.5-0.0}$&$-2.0^{+0.1+1.7+0.1}_{-0.1-1.3-0.1}$\\
$\bar{B}_s \to \phi\rho^0$ &$87.0^{+0.2+0.9+0.9}_{-0.2-0.3-0.4}$&$6.8^{+0.1+0.2+0.3}_{-0.1-0.5-0.4}$&$6.2^{+0.1+0.1+0.1}_{-0.1-0.5-0.4}$&$3.5^{+0.0+0.0+0.0}_{-0.0-0.1-0.1}$&$3.5^{+0.0+0.1+0.1}_{-0.0-0.0-0.0}$&$10.1^{+0.9+1.6+1.3}_{-0.9-1.8-0.5}$\\
$\bar{B}_s \to \phi\omega$ &$44.3^{+0.0+5.4+0.9}_{-7.5-6.1-0.4}$&$28.5^{+3.8+3.1+0.1}_{-0.0-2.8-0.5}$&$27.2^{+3.7+3.0+0.2}_{-0.0-2.6-0.4}$&$3.0^{+0.1+0.2+0.0}_{-0.1-0.2-0.0}$&$3.0^{+0.1+0.2+0.1}_{-0.0-0.2-0.0}$&$3.6^{+0.6+2.4+0.6}_{-0.6-2.4-0.2}$\\
\hline\hline
\end{tabular}\
}
\end{center}
 \end{table}

\section{summary}

With the LHC era almost upon us, where apart from the decisive searches for
the Higgs boson(s) and supersymmetry (or, alternatives thereof),
also dedicated studies of the $B_s^0$-meson physics (as well as that of the
heavier $b$-hadrons) will be carried out. First results on the decay
characteristics of the $B_s^0$-mesons are already available from the
Tevatron, in particular the
$B_s^0$ - $\overline{B_s^0}$-mixing induced mass difference $\Delta M_s$,
the branching ratios for the decays $\overline{B_s^0} \to K^+\pi^-$,
$\overline{B_s^0} \to K^+ K^-$, $\overline{B_s^0} \to \phi \phi$,
and the first direct CP asymmetry $A_{\rm CP}^{\rm dir}(
\overline{B_s^0} \to K^+\pi^-)$. These measurements owe themselves to
the two  experiments CDF and D0; they are impressive and prove that
cutting-edge B physics can also be carried out by general purpose
hadron collider experiments.
 At the LHC, we
will have a dedicated $b$-physics experiment, LHCb, but also the two
main general purpose detectors ATLAS and CMS will be able to contribute
handsomely to the ongoing research in  $b$-physics.
 Conceivably, also at a Super-B factory
with dedicated running at the $\Upsilon (5S)$-resonance, the current
experimental databank on the $B_s$-meson will be greatly enlarged.

Anticipating these developments, we have presented in this work the results
of a comprehensive study of the decays $\overline{B_s^0} \to h_1 h_2$,
where $h_1$ and $h_2$ are light (i.e., charmless) pseudo-scalar and vector
mesons. This study has been carried out in the context of the pQCD
approach, taking into account the most recent information on the CKM
matrix elements and weak phases and updating the input hadronic parameters.
The decay amplitudes contain the relevant emission and annihilation
diagrams of the tree and penguin nature. Explicit formulae for all these
amplitudes, including the various pQCD-related functions,  are provided
in the Appendices. Numerical results for the charge-conjugation averaged
branching ratios, the direct CP asymmetries and the mixing-induced
CP-asymmetries $S_f$ for the CP-eigenstates $f$, and the time-dependent
observable $H_f$ are presented in the form of tables. In addition, for the
$\overline{B_s^0} \to V V$ decays, we also calculate the
 charge- conjugation-averaged transversity amplitudes, their relative strong
 phases and magnitudes. The results for the altogether 49
$\overline{B_s^0} \to h_1 h_2$ decays include also theoretical errors
coming from the input hadronic parameters, variation of the
hard scattering scale together with the uncertainty on $\Lambda_{\rm QCD}$,
and the combined uncertainty in the CKM matrix elements and the angles of
the unitarity triangle. For the last mentioned error, we use the recently
updated results from the CKMfitter~\cite{CKMfitter:2006}.

Our results are compared with the available data on the decays
$\overline{B_s^0} \to h_1 h_2$ from the Tevatron, and some
selected $\overline{B_d^0} \to h_1 h_2$ decays from the B-factory
experiments, updating the theoretical calculations in the pQCD
approach with our input parameters. In particular, we revisited
the well-measured direct CP asymmetry $A_{K\pi}^{dir}(B_d^0 \to
K^+\pi^-)$,  working out the parametric sensitivity of this
observable, and argued that data can be accommodated in the pQCD
approach by invoking $O(\alpha_s^2)$ contributions, which we
estimated from the existing literature. The successful predictions
of this direct CP-asymmetry in  the pQCD approach is set forth
with the first observed CP-asymmetry in the decay
$\overline{B_s^0} \to K^+\pi^-$. This CP-asymmetry is
experimentally large and is in good agreement with our numerical
results, within the stated errors. We have also analyzed a number
of ratios of branching ratio involving the $B_s^0 \to PP$ and
$B_d^0 \to PP$ decays. They include the ratios $R_1\equiv
\frac{BR(\overline{B_s^0} \to K^+K^-)}{BR(\overline{B_d^0} \to
\pi^+\pi^-)}$, $R_2\equiv \frac{BR(\overline{B_s^0} \to K^+K^-)}
{BR(\overline{B_d^0} \to K^-\pi^+)}$, and two more, called $R_3$
and $\Delta$, defined in Eq. (72) and (73), respectively, which
involve the decays $\overline{B_d^0} \to K^-\pi^+$ and
$\overline{B_s^0} \to K^+\pi^-$ and their charge conjugates.
Except for the ratio $R_1$, for which the pQCD calculations
presented here are on the lower side of the data (within large errors), the others are
well accounted for. We also compare our results with the
corresponding ones in the QCDF and SCET approaches. What concerns
the branching ratios, we find reasonable agreement for some
topological amplitudes, but also major disagreement, in particular
for those decays which are dominated by the color-suppressed tree
and annihilation amplitudes. There are striking differences in the
CP asymmetries, in particular with the QCDF-based estimates, which
will be precisely tested in the future.

Experimental precision on $B_s$ decays will
improve enormously in the coming years, thanks to dedicated
experiments at the Tevatron and LHC. Many of the decay rates and CP
asymmetries worked out here will be put to experimental scrutiny.
They can be combined with the B-factory data on the corresponding
 $B \to PP, PV, VV$ decays,
eliminating some of the large parametric uncertainties in theoretically
well-motivated ratios to test the SM precisely in exclusive hadronic decays.
 In principle, theoretical
predictions presented here can be systematically improved by
including higher order perturbative  corrections (in $\alpha_s$)
and sub-leading power corrections in $1/m_b$.

\section*{Acknowledgment}

  This work is partly supported by National Science Foundation of
China under the Grant Numbers 10475085 and 10625525. One of us (C.-D.L.) would
like to acknowledge the financial support of the Sino-German Center for
Science Promotion (Grant No.~GZ 369),
 the Alexander-von-Humboldt-Foundation and DESY. We would like
to thank Vladimir Braun for a discussion on the SU(3)-breaking effects in the
light-cone-distribution amplitudes of light mesons.

\newpage

\appendix

\section{pQCD functions}\label{PQCDfunctions}

In this section, we group the functions which appear in the
factorization formulae.

The hard scales are chosen as \begin{eqnarray}
t_a&=&\mbox{max}\{{\sqrt
{x_3}M_{B_s},1/b_1,1/b_3}\},\\
t_a^\prime&=&\mbox{max}\{{\sqrt
{x_1}M_{B_s},1/b_1,1/b_3}\},\\
t_b&=&\mbox{max}\{\sqrt
{x_1x_3}M_{B_s},\sqrt{|1-x_1-x_2|x_3}M_{B_s},1/b_1,1/b_2\},\\
t_b^\prime&=&\mbox{max}\{\sqrt{x_1x_3}M_{B_s},\sqrt
{|x_1-x_2|x_3}M_{B_s},1/b_1,1/b_2\},\\
t_c&=&\mbox{max}\{\sqrt{1-x_3}M_{B_s},1/b_2,1/b_3\},\\
t_c^\prime
&=&\mbox{max}\{\sqrt {x_2}M_{B_s},1/b_2,1/b_3\},\\
t_d&=&\mbox{max}\{\sqrt {x_2(1-x_3)}M_{B_s},
\sqrt{1-(1-x_1-x_2)x_3}M_{B_s},1/b_1,1/b_2\},\\
t_d^\prime&=&\mbox{max}\{\sqrt{x_2(1-x_3)}M_{B_s},\sqrt{|x_1-x_2|(1-x_3)}M_{B_s},1/b_1,1/b_2\}.
\end{eqnarray}

The functions $h$  in the decay amplitudes consist of two parts: one
is the jet function $S_t(x_i)$ derived by the threshold
re-summation\cite{L3}, the other is the propagator of virtual quark
and gluon. They are defined by
\begin{eqnarray}
h_e(x_1,x_3,b_1,b_3)&=&\left[\theta(b_1-b_3)I_0(\sqrt
x_3M_{B_s}b_3)K_0(\sqrt
x_3 M_{B_s}b_1)\right.\\
&& \left.+\theta(b_3-b_1)I_0(\sqrt x_3M_{B_s}b_1)K_0(\sqrt
x_3M_{B_s}b_3)\right]K_0(\sqrt {x_1x_3}M_{B_s}b_1)S_t(x_3),\nonumber\\
h_n(x_1,x_2,x_3,b_1,b_2)&=&\left[\theta(b_2-b_1)K_0(\sqrt
{x_1x_3}M_{B_s}b_2)I_0(\sqrt
{x_1x_3}M_{B_s}b_1)\right. \nonumber\\
&&\;\;\;\left. +\theta(b_1-b_2)K_0(\sqrt
{x_1x_3}M_{B_s}b_1)I_0(\sqrt{x_1x_3}M_{B_s}b_2)\right]\nonumber\\
&&\times
\left\{\begin{array}{ll}\frac{i\pi}{2}H_0^{(1)}(\sqrt{(x_2-x_1)x_3}
M_{B_s}b_2),& x_1-x_2<0\\
K_0(\sqrt{(x_1-x_2)x_3}M_{B_s}b_2),& x_1-x_2>0
\end{array}
\right. ,
\end{eqnarray}
\begin{eqnarray}
h_a(x_2,x_3,b_2,b_3)&=&(\frac{i\pi}{2})^2
S_t(x_3)\Big[\theta(b_2-b_3)H_0^{(1)}(\sqrt{x_3}M_{B_s}b_2)J_0(\sqrt
{x_3}M_{B_s}b_3)\nonumber\\
&&\;\;+\theta(b_3-b_2)H_0^{(1)}(\sqrt {x_3}M_{B_s}b_3)J_0(\sqrt
{x_3}M_{B_s}b_2)\Big]H_0^{(1)}(\sqrt{x_2x_3}M_{B_s}b_2),
\\
h_{na}(x_1,x_2,x_3,b_1,b_2)&=&\frac{i\pi}{2}\left[\theta(b_1-b_2)H^{(1)}_0(\sqrt
{x_2(1-x_3)}M_{B_s}b_1)J_0(\sqrt {x_2(1-x_3)}M_{B_s}b_2)\right. \nonumber\\
&&\;\;\left.
+\theta(b_2-b_1)H^{(1)}_0(\sqrt{x_2(1-x_3)}M_{B_s}b_2)J_0(\sqrt
{x_2(1-x_3)}M_{B_s}b_1)\right]\nonumber\\
&&\;\;\;\times K_0(\sqrt{1-(1-x_1-x_2)x_3}M_{B_s}b_1),
\\
h_{na}^\prime(x_1,x_2,x_3,b_1,b_2)&=&\frac{i\pi}{2}\left[\theta(b_1-b_2)H^{(1)}_0(\sqrt
{x_2(1-x_3)}M_{B_s}b_1)J_0(\sqrt{x_2(1-x_3)}M_{B_s}b_2)\right. \nonumber\\
&&\;\;\;\left. +\theta(b_2-b_1)H^{(1)}_0(\sqrt
{x_2(1-x_3)}M_{B_s}b_2)J_0(\sqrt{x_2(1-x_3)}M_{B_s}b_1)\right]\nonumber\\
&&\;\;\;\times
\left\{\begin{array}{ll}\frac{i\pi}{2}H^{(1)}_0(\sqrt{(x_2-x_1)(1-x_3)}M_{B_s}b_1),&
x_1-x_2<0\\
K_0(\sqrt {(x_1-x_2)(1-x_3)}M_{B_s}b_1),&
x_1-x_2>0\end{array}\right. ,
\end{eqnarray}
where $H_0^{(1)}(z) = \mathrm{J}_0(z) + i\, \mathrm{Y}_0(z)$.

The $S_t$ re-sums the threshold logarithms $\ln^2x$ appearing in the
hard kernels to all orders and it has been parameterized as
  \begin{eqnarray}
S_t(x)=\frac{2^{1+2c}\Gamma(3/2+c)}{\sqrt \pi
\Gamma(1+c)}[x(1-x)]^c,
\end{eqnarray}
with $c=0.4$. In the nonfactorizable contributions, $S_t(x)$ gives
a very small numerical effect to the amplitude~\cite{L4}.
Therefore, we drop $S_t(x)$ in $h_n$ and $h_{na}$.

The evolution factors $E^{(\prime)}_e$ and $E^{(\prime)}_a$ 
entering in the expressions for the matrix elements (see section 3) are
given by
\begin{eqnarray}
E_e(t)&=&\alpha_s(t) \exp[-S_B(t)-S_3(t)],
 \ \ \ \
 E'_e(t)=\alpha_s(t)
 \exp[-S_B(t)-S_2(t)-S_3(t)]|_{b_1=b_3},\\
E_a(t)&=&\alpha_s(t)
 \exp[-S_2(t)-S_3(t)],\
 \ \ \
E'_a(t)=\alpha_s(t) \exp[-S_B(t)-S_2(t)-S_3(t)]|_{b_2=b_3},
\end{eqnarray}
in which the Sudakov exponents are defined as
\begin{eqnarray}
S_B(t)&=&s\left(x_1\frac{M_{B_s}}{\sqrt
2},b_1\right)+\frac{5}{3}\int^t_{1/b_1}\frac{d\bar \mu}{\bar
\mu}\gamma_q(\alpha_s(\bar \mu)),\\
S_2(t)&=&s\left(x_2\frac{M_{B_s}}{\sqrt
2},b_2\right)+s\left((1-x_2)\frac{M_{B_s}}{\sqrt
2},b_2\right)+2\int^t_{1/b_2}\frac{d\bar \mu}{\bar
\mu}\gamma_q(\alpha_s(\bar \mu)),
\end{eqnarray}
 with the quark
anomalous dimension $\gamma_q=-\alpha_s/\pi$. Replacing the
kinematic variables of $M_2$ to $M_3$ in $S_2$, we can get the
expression for $S_3$. The explicit form for the  function
$s(Q,b)$ is:
\begin{eqnarray}
s(Q,b)&=&~~\frac{A^{(1)}}{2\beta_{1}}\hat{q}\ln\left(\frac{\hat{q}}
{\hat{b}}\right)-
\frac{A^{(1)}}{2\beta_{1}}\left(\hat{q}-\hat{b}\right)+
\frac{A^{(2)}}{4\beta_{1}^{2}}\left(\frac{\hat{q}}{\hat{b}}-1\right)
-\left[\frac{A^{(2)}}{4\beta_{1}^{2}}-\frac{A^{(1)}}{4\beta_{1}}
\ln\left(\frac{e^{2\gamma_E-1}}{2}\right)\right]
\ln\left(\frac{\hat{q}}{\hat{b}}\right)
\nonumber \\
&&+\frac{A^{(1)}\beta_{2}}{4\beta_{1}^{3}}\hat{q}\left[
\frac{\ln(2\hat{q})+1}{\hat{q}}-\frac{\ln(2\hat{b})+1}{\hat{b}}\right]
+\frac{A^{(1)}\beta_{2}}{8\beta_{1}^{3}}\left[
\ln^{2}(2\hat{q})-\ln^{2}(2\hat{b})\right],
\end{eqnarray} where the variables are defined by
\begin{eqnarray}
\hat q\equiv \mbox{ln}[Q/(\sqrt 2\Lambda)],~~~ \hat b\equiv
\mbox{ln}[1/(b\Lambda)], \end{eqnarray} and the coefficients
$A^{(i)}$ and $\beta_i$ are \begin{eqnarray}
\beta_1=\frac{33-2n_f}{12},~~\beta_2=\frac{153-19n_f}{24},\nonumber\\
A^{(1)}=\frac{4}{3},~~A^{(2)}=\frac{67}{9}
-\frac{\pi^2}{3}-\frac{10}{27}n_f+\frac{8}{3}\beta_1\mbox{ln}(\frac{1}{2}e^{\gamma_E}),
\end{eqnarray}
$n_f$ is the number of the quark flavors and $\gamma_E$ is the
Euler constant. We will use the one-loop running coupling
constant, i.e. we pick up the four terms in the first line of the
expression for the function $s(Q,b)$.

\section{Analytic Formulae for the $B_s \to PP$  decay
amplitudes} \label{bspp}

Before we give the analytic formulae for $B_s\to PP$ decays, we
analyze the amplitudes in some special cases which can simplify
the formulae.
\begin{itemize}
\item In the non-factorizable emission diagrams, the formulae can
be simplified if the emission meson $M_2$ is $\pi$, $\eta$ or
$\eta^\prime$. Their distribution amplitudes $\phi^A(x)$ and
$\phi^P(x)$ are symmetric and $\phi^T(x)$ is antisymmetric under
the change $x\to 1-x$. From Eq. (\ref{ppenlr}) we can see that the
amplitude is identically zero for the $(V-A)(V+A)$ operators; From
Eq. (\ref{ppenll}) and Eq. (\ref{ppensp}) we can see that the
$(V-A)(V-A)$ contribution is the same as the $(S-P)(S+P)$
contribution which can be simplified as \begin{eqnarray} M_{B_s\to
M_3}^{LL}(a_i)&=&32\pi
C_FM_{B_s}^4/\sqrt{6}\int^1_0dx_1dx_2dx_3\int^\infty_0
b_1db_1b_2db_2\phi_B(x_1,b_1) \phi_2^A(x_2)
\nonumber\\
&&~\times
h_n(x_1,x_2,x_3,b_1,b_2)\Big[-x_3\phi_3^A(x_3)+2r_3x_3\phi_3^T(x_3)\Big]
 a_i(t_b^\prime) E_e^\prime(t_b^\prime)\nonumber\\
 &=&M_{B_s\to
M_3}^{SP}(a_i).
 \end{eqnarray}

\item In the annihilation factorizable diagrams, the $(V-A)(V-A)$
and $(V-A)(V+A)$ operators give the same contribution, both  from
the vector current. If the final state mesons are charge conjugate
with each other, the $(V-A)(V-A)$ and $(V-A)(V+A)$ operators give
identically zero contributions due to the conservation of the
vector current. Such operators do not contribute to the $B_s\to
K\pi$ decay either, in the $SU(3)$ limit. The $SU(3)$ symmetry breaking,
i.e. the difference in the distribution amplitudes of $\pi$ and
$K$ meson, can induce small deviations from zero. We expect this
kind of operators can not give important contributions to the
branching ratios.

\item In the nonfactorizable annihilation amplitudes of $\bar
B_s\to \pi\pi$, $B_s\to \eta(\bar nn)\eta(\bar nn)$ and $B_s\to
\eta(\bar ss)\eta(\bar ss)$, the $(V-A)(V-A)$ operators and
$(S-P)(S+P)$ operators give equal contributions as can be seen
from the factorization formulae $M_{ann}$ by making
$x_2\leftrightarrow 1-x_3$.
\end{itemize}

\subsection{The case without $\eta^{(\prime)}$}

Tree operator dominant decays:
\begin{eqnarray}
 A(\bar B_{s}^0\to\pi^{-}K^{+})
&=&  \frac{G_F}{\sqrt{2}}  V_{ub}V_{ud}^{*} \Big\{ f_{\pi} F_{B_s\to
K}^{LL}\left[a_1\right] + M_{B_s\to K}^{LL} [C_{1}] \Big \}\nonumber
   \\
  &&- \frac{G_F}{\sqrt{2}} V_{tb}V_{td}^{*} \bigg\{ f_{\pi}F_{B_s\to K}^{LL}\left[
   a_{4} +a_{10}\right]
   +f_{\pi} F_{B_s\to K}^{SP} \left[a_{6} +a_{8}\right]
 \nonumber
\\
 && +  M_{B_s\to K}^{LL} \left[C_{3}+C_{9}\right]
+ f_{B_s}
 F_{ann}^{LL}\left[a_{4}-\frac{1}{2}a_{10}\right] + f_{B_s} F_{ann}^{SP}\left[a_{6}-
\frac{1}{2}a_{8}\right]
  \nonumber  \\
&&  + M_{ann}^{LL}\left[C_{3}-\frac{1}{2}C_{9}\right]
 + M_{ann}^{LR}\left[C_{5}-\frac{1}{2}C_{7}\right]
\bigg \},
\end{eqnarray}
\begin{eqnarray}
 \sqrt{2}A(\bar B_{s}^0\to\pi^{0} K^{0})
&=& \frac{G_F}{\sqrt{2}} V_{ub}V_{ud}^{*} \Big \{f_{\pi} F_{B_s\to
K}^{LL} \left[a_{2}\right]+ M_{B_s\to K}^{LL} [C_{2}]\Big \}
 \nonumber
 \\
  &&- \frac{G_F}{\sqrt{2}} V_{tb}V_{td}^{*} \bigg \{f_{\pi} F_{B_s\to K}^{LL} \left[
 -a_{4}-\frac{3}{2}a_7+\frac{3}{2}a_9+\frac{1}{2}a_{10}\right]
 \nonumber\\
  &&  +f_{\pi} F_{B_s\to K}^{SP} \left[-a_{6}+\frac{1}{2}a_{8}\right]+  M_{B_s\to K}^{LL}
  \left[-C_{3}+\frac{3}{2}C_{8}+\frac{1}{2}C_{9}
  +\frac{3}{2}C_{10}\right]\nonumber\\
  &&
+  f_{B_s} F_{ann}^{SP}\left[-a_{6}+\frac{1}{2}a_{8}\right]
  +  f_{B_s}  F_{ann}^{LL}\left[-a_{4}+ \frac{1}{2}a_{10}\right] \nonumber
  \\
  &&
+  M_{ann}^{LL}\left[-C_{3}+\frac{1}{2}C_{9}\right]
   +  M_{ann}^{LR}\left[-C_{5}+\frac{1}{2}C_{7}\right]\bigg \}.
\end{eqnarray}

Pure annihilation type decays:
\begin{eqnarray}
A(\bar B_{s}^0\to\pi^{+}\pi^{-}) &=&\sqrt{2}A(\bar
B_{s}^0\to\pi^{0}\pi^{0})
\nonumber\\
 &=& \frac{G_F}{\sqrt{2}} V_{ub}V_{us}^{*} M_{ann}^{LL}\left[C_2\right] -
\frac{G_F}{\sqrt{2}} V_{tb}V_{ts}^{*}
M_{ann}^{LL}\left[2C_{4}+2C_6+\frac{1}{2}C_{8}+\frac{1}{2}C_{10}\right]
.
\end{eqnarray}

QCD penguin operator dominant decays:
\begin{eqnarray}
 A(\bar B_{s}^0\to K^{-}K^{+})
&=&  \frac{G_F}{\sqrt{2}} V_{ub}V_{us}^{*} \Big \{ f_{K} F_{B_s\to
K}^{LL} \left[a_{1} \right] + M_{B_s\to K}^{LL} [C_{1}]+
M_{ann}^{LL}[C_2]\Big\} \nonumber
  \\
 &&- \frac{G_F}{\sqrt{2}}
 V_{tb}V_{ts}^{*} \bigg \{ f_{K} F_{B_s\to K}^{LL} \left[
  a_{4}+a_{10}\right]
 + f_{K} F_{B_s\to K}^{SP} \left[a_{6}+a_{8}\right] \nonumber\\
&& + M_{B_s\to K}^{LL} \left[C_{3}+C_{9}\right]
  + M_{B_s\to K}^{LR} \left[C_{5}+C_{7}\right]
  + f_{B_s} F_{ann}^{SP}\left[a_{6}- \frac{1}{2}a_{8}\right]  \nonumber\\
&&  +
M_{ann}^{LL}\left[C_{3}-\frac{1}{2}C_9+C_{4}-\frac{1}{2}C_{10}\right]
+ M_{ann}^{LR}\left[C_{5}-\frac{1}{2}C_{7}\right]
   + M_{ann}^{SP}\left[C_{6}-\frac{1}{2}C_{8}\right]\nonumber\\
  &&+ \left( M_{ann}^{LL}\left[C_{4}+C_{10}\right]
   + M_{ann}^{SP}\left[C_{6}+C_{8}\right]\right)_{K^-\leftrightarrow K^+}\bigg\},
\end{eqnarray}
\begin{eqnarray}
 A(\bar B_{s}^0\to\bar K^0  K^{0})
&=&   - \frac{G_F}{\sqrt{2}}
 V_{tb}V_{ts}^{*} \bigg \{ f_{K} F_{B_s\to K}^{LL} \left[
  a_{4}-\frac{1}{2}a_{10}\right]
 + f_{K} F_{B_s\to K}^{SP} \left[a_{6}-\frac{1}{2}a_{8}\right]\nonumber\\
 && + M_{B_s\to
K}^{LL} \left[C_{3}-\frac{1}{2}C_{9}\right]
  + M_{B_s\to K}^{LR} \left[C_{5}-\frac{1}{2}C_{7}\right]
  + f_{B_s} F_{ann}^{SP}\left[a_{6}- \frac{1}{2}a_{8}\right] \nonumber\\
&&
 + M_{ann}^{LL}\left[C_{3}-\frac{1}{2}C_9+C_{4}-\frac{1}{2}C_{10}\right]
  + M_{ann}^{LR}\left[C_{5}-\frac{1}{2}C_{7}\right]
   \\\nonumber
  &&+  M_{ann}^{LL}\left[C_{4}-\frac{1}{2}C_{10}\right]_{K^0\leftrightarrow \bar K^0}
   + \left(M_{ann}^{SP}\left[C_{6}-\frac{1}{2}C_{8}\right]+ [K^0\leftrightarrow \bar K^0]\right)\bigg\},
\end{eqnarray}

\subsection{The case with $\eta^{(\prime)}$}

As discussed in the last section, we use the quark flavor basis for
the mixing of $\eta$ and $\eta\prime$. So we divide the amplitudes
into the $\eta_n=(\bar uu+\bar dd)/\sqrt2$ and $\bar ss$ component.
\begin{eqnarray}
\sqrt2A(\bar B_{s}^0\to  \eta_n K^{0}) &=& \frac{G_F}{\sqrt{2}}
V_{ub}V_{ud}^{*} \Big\{ f_n F_{B_s\to K}^{LL} \left[a_{2}\right] +
M_{B_s\to K}^{LL} [C_{2}] \Big \}\nonumber
\\
 &&- \frac{G_F}{\sqrt{2}}
V_{tb}V_{td}^{*} \Big\{f_n F_{B_s\to K}^{SP}\left[a_{6}
   -\frac{1}{2}a_{8}\right]+ f_{B_s}
F_{ann}^{LL}\left[a_{4} -\frac{1}{2}a_{10}\right]
\nonumber \\
 && +f_n F_{B_s\to K}^{LL} \left[
 2a_{3}+a_{4}-2a_{5}
-\frac{1}{2}a_{7}+\frac{1}{2}a_{9}-\frac{1}{2}a_{10}\right]\nonumber
   \\
 &&
 + M_{B_s\to K}^{LL}
   \left[C_{3}+2C_{4}+2C_{6}+\frac{1}{2}C_{8}-\frac{1}{2}C_9
  +\frac{1}{2}C_{10}\right]
 \nonumber
  \\
 &&
  +f_{B_s} F_{ann}^{SP}\left[a_{6} -\frac{1}{2}a_{8}\right]
 + M_{ann}^{LL}\left[C_{3}-\frac{1}{2}C_{9}\right]
 + M_{ann}^{LR}\left[C_{5}-\frac{1}{2}C_{7}\right] \Big\},
\end{eqnarray}
\begin{eqnarray}
A(\bar B_{s}^0\to   K^{0}\eta_s) &=& - \frac{G_F}{\sqrt{2}}
V_{tb}V_{td}^{*} \bigg\{ f_s F_{B_s\to K}^{LL}\left[
 a_{3}-a_{5}
+\frac{1}{2}a_{7} -\frac{1}{2}a_{9} \right]
 +   f_KF_{B_s\to \eta_s}^{LL}\left[
 a_{4} -\frac{1}{2}a_{10}\right] \nonumber
  \\
 &&
   +  f_K  F_{B_s\to \eta_s}^{SP}\left[a_{6} -\frac{1}{2}a_{8}\right]
 +  M_{B_s\to K}^{LL}\left[C_{4}+C_{6}-\frac{1}{2}C_{8}-\frac{1}{2}C_{10}\right]\nonumber
  \\
 &&
 +  M_{B_s\to \eta_s}^{LL}\left[C_{3}-\frac{1}{2}C_9\right]
  +  M_{B_s\to \eta_s}^{LR}\left[C_{5}-\frac{1}{2}C_{7}\right]
  +  f_{B_s}   F_{ann}^{LL}\left[a_{4} -\frac{1}{2}a_{10}\right]\nonumber
  \\
 &&
  +  f_{B_s}  F_{ann}^{SP}\left[a_{6} -\frac{1}{2}a_{8}\right] +  M_{ann}^{LL}
  \left[C_{3}-\frac{1}{2}C_{9}\right]
 +  M_{ann}^{LR}\left[C_{5}-\frac{1}{2}C_{7}\right]\bigg\}.
\end{eqnarray}
The  decay amplitudes for the physical states are then
\begin{eqnarray}
A(\bar B_s \to \eta K^0) = A(\bar B_{s}^0\to  \eta_n K^{0})
\cos \phi -A(\bar B_{s}^0\to   K^{0}\eta_s) \sin \phi, \\
A(\bar B_s \to \eta' K^0) = A(\bar B_{s}^0\to \eta_nK^{0}) \sin \phi
+ A(\bar B_{s}^0\to   K^{0}\eta_s) \cos \phi.
\end{eqnarray}
For $\bar B_s \to \eta^{(\prime)} \pi^0$ the decay amplitudes are
defined similarly,
\begin{eqnarray}
A(\bar B_s \to\pi^0 \eta ) = A(\bar B_{s}^0\to  \pi^{0}\eta_n)
\cos \phi -A(\bar B_{s}^0\to \pi^{0} \eta_s) \sin \phi, \\
A(\bar B_s \to\pi^0 \eta' ) = A(\bar B_{s}^0\to \pi^{0}\eta_n) \sin
\phi + A(\bar B_{s}^0\to   \pi^{0}\eta_s) \cos \phi,
\end{eqnarray}
where
\begin{eqnarray}
A(\bar B_{s}^0\to\pi^{0} \eta_n) &=& \frac{G_F}{\sqrt{2}}
V_{ub}V_{us}^{*}
 \bigg\{ f_{B_s}F_{ann}^{LL}\left[a_{2}\right] +  M_{ann}^{LL}[C_{2}] \bigg\}
 \nonumber
  \\
  &&- \frac{G_F}{\sqrt{2}}  V_{tb}V_{ts}^{*}\bigg\{f_{B_s}
F_{ann}^{LL}\left[\frac{3}{2}a_9 +\frac{3}{2}a_7
\right]+M_{ann}^{SP}\left[\frac{3}{2}C_8\right]
 +M_{ann}^{LL}\left[\frac{3}{2}C_{10}\right]
  \bigg\},
\end{eqnarray}
and
\begin{eqnarray}
 \sqrt{2}A(\bar B_{s}^0\to\pi^{0}\eta_s)
&=& \frac{G_F}{\sqrt{2}} V_{ub}V_{us}^{*} \bigg \{ f_{\pi} F_{
B_s\to \eta_s}^{LL} \left[a_{2}\right]+ M_{ B_s\to
\eta_s}^{LL}\left[C_{2}\right]\bigg \}\nonumber
\\
&& - \frac{G_F}{\sqrt{2}} V_{tb}V_{ts}^{*}  \bigg \{
f_{\pi}F_{B_s\to
\eta_s}^{LL}\left[\frac{3}{2}a_{9}-\frac{3}{2}a_{7}\right] +M_{
B_s\to \eta_s}^{LL} \left[\frac{3}{2}C_{8}+\frac{3}{2}C_{10}\right]
\bigg\}.
\end{eqnarray}

For $\bar B_s \to \eta^{(\prime)} \eta^{(\prime)} $, we have
\begin{eqnarray}
A(\bar B_{s}^0\to \eta_n\eta_n) &=&
 \frac{G_F}{\sqrt{2}} V_{ub}V_{us}^{*}
M_{ann}^{LL}[C_{2}]  - \frac{G_F}{\sqrt{2}} V_{tb}V_{ts}^{*}
M_{ann}^{LL}\left[2C_4+2C_6+\frac{1}{2}C_8+\frac{1}{2}C_{10}\right],
\\
 \sqrt 2A(\bar B_{s}^0\to \eta_n\eta_s)
&=& \frac{G_F}{\sqrt{2}} V_{ub}V_{us}^{*} \Big\{f_n F_{B_s\to
\eta_s}^{LL}\left[a_{2}\right] + M_{B_s\to
\eta_s}^{LL}\left[C_{2}\right]\Big \}\nonumber
\\
 &&- \frac{G_F}{\sqrt{2}} V_{tb}V_{ts}^{*} \Big\{ f_n F_{B_s\to \eta_s}^{LL}\left[2a_3-2a_5
-\frac{1}{2}a_7 +\frac{1}{2}a_9  \right]  \nonumber
   \\
&&+ M_{B_s\to \eta_s}^{LL}
\left[2C_4+2C_6+\frac{1}{2}C_8+\frac{1}{2}C_{10}\right]\Big \},
\\
 A(\bar B_{s}^0\to \eta_s\eta_s)
&=& - \sqrt{2}{G_F}{} V_{tb}V_{ts}^{*} \Big\{ f_sF_{B_s\to
\eta_s}^{LL}\left[a_3+a_{4}-a_5
 +\frac{1}{2}a_7 -\frac{1}{2}a_9-\frac{1}{2}a_{10}\right]
  \nonumber
 \\
 &&+ f_s F_{B_s\to \eta_s}^{SP}\left[a_{6}
 -\frac{1}{2}a_{8}\right]
  + M_{B_s\to \eta_s}^{LL} \left[C_3+C_{4}+C_6-\frac{1}{2}C_{8}-\frac{1}{2}C_9-\frac{1}{2}C_{10}\right]\nonumber
 \\
 &&
+ f_{B_s} F_{ann}^{SP}\left[a_{6} -\frac{1}{2}a_{8}\right]+
M_{ann}^{LL}\left[C_3+C_{4}+C_6-\frac{1}{2}C_{8}-\frac{1}{2}C_9
-\frac{1}{2}C_{10}\right] \Bigg\} ,
\end{eqnarray}
The  decay amplitudes for the physical states are then
\begin{eqnarray}
{\sqrt 2}A(\bar B_s \to\eta \eta ) &=&  A(\bar B_{s}^0\to
\eta_n\eta_n) \cos^2 \phi + A(\bar B_{s}^0\to \eta_s \eta_s)
\sin^2 \phi-\sin(2\phi) A(\bar B_{s}^0\to \eta_n \eta_s) ,
\\
 A(\bar B_s \to\eta \eta' ) &=& \left[A(\bar B_{s}^0\to
\eta_n\eta_n)-A(\bar B_{s}^0\to \eta_s\eta_s)\right] \cos \phi
\sin\phi+ A(\bar B_{s}^0\to \eta_n \eta_s)\cos(2 \phi),
 \\
 {\sqrt 2}A(\bar B_s \to\eta' \eta' ) &=&  A(\bar
B_{s}^0\to \eta_n\eta_n) \sin^2 \phi + A(\bar B_{s}^0\to \eta_s
\eta_s)
 \cos^2 \phi+\sin(2\phi) A(\bar B_{s}^0\to \eta_n \eta_s).
\end{eqnarray}

\section{Analytic Formulae for the $B_s \to PV$  decay amplitudes}

\subsection{The tree  dominant decays}

\begin{eqnarray}
 A(\bar B_{s}^0\to\pi^{-}K^{*+})
&=&  \frac{G_F}{\sqrt{2}}  V_{ub}V_{ud}^{*} \Bigg\{ f_{\pi}
F_{B_s\to K^*}^{LL}\left[a_1\right] + M_{B_s\to K^*}^{LL} [C_{1}]
\Bigg \}\nonumber
   \\
  &&- \frac{G_F}{\sqrt{2}} V_{tb}V_{td}^{*} \Bigg\{
f_{\pi}F_{B_s\to K^*}^{LL}\left[
   a_{4} +a_{10}\right]
   -f_{\pi} F_{B_s\to K^*}^{SP} \left[a_{6} +a_{8}\right]
 \nonumber  \\
&&+  M_{B_s\to K^*}^{LL} \left[C_{3}+C_{9}\right]  + f_{B_s}
 F_{ann}^{LL}\left[a_{4}-\frac{1}{2}a_{10}\right]
 - f_{B_s} F_{ann}^{SP}\left[a_{6}- \frac{1}{2}a_{8}\right] \nonumber  \\
&&   + M_{ann}^{LL}\left[C_{3}-\frac{1}{2}C_{9}\right]
 - M_{ann}^{LR}\left[C_{5}-\frac{1}{2}C_{7}\right]
\Bigg \},
\end{eqnarray}
\begin{eqnarray}
 A(\bar B_{s}^0\to\rho^{-}K^{+})
&=&  \frac{G_F}{\sqrt{2}}  V_{ub}V_{ud}^{*} \Bigg\{ f_{\rho}
F_{B_s\to K}^{LL}\left[a_1\right] + M_{B_s\to K}^{LL} [C_{1}] \Bigg
\}\nonumber
   \\
  &&- \frac{G_F}{\sqrt{2}} V_{tb}V_{td}^{*} \Bigg\{ f_{\rho}F_{B_s\to K}^{LL}\left[
   a_{4} +a_{10}\right]+  M_{B_s\to K}^{LL} \left[C_{3}+C_{9}\right]
 \nonumber
\\
 &&  + M_{B_s\to
K}^{LR}\left[C_{5}+C_{7}\right]
  + f_{B_s}
 F_{ann}^{LL}\left[a_{4}-\frac{1}{2}a_{10}\right] + f_{B_s} F_{ann}^{SP}\left[a_{6}-
\frac{1}{2}a_{8}\right]
  \nonumber  \\
&&  + M_{ann}^{LL}\left[C_{3}-\frac{1}{2}C_{9}\right]
 + M_{ann}^{LR}\left[C_{5}-\frac{1}{2}C_{7}\right]
\Bigg \},
\end{eqnarray}
\begin{eqnarray}
 \sqrt{2}A(\bar B_{s}^0\to\pi^{0} K^{*0})
&=& \frac{G_F}{\sqrt{2}} V_{ub}V_{ud}^{*} \Bigg \{f_{\pi} F_{B_s\to
K^*}^{LL} \left[a_{2}\right]+ M_{B_s\to K^*}^{LL} [C_{2}]\Bigg \}
 \nonumber
 \\
  &&- \frac{G_F}{\sqrt{2}} V_{tb}V_{td}^{*} \Bigg \{f_{\pi} F_{B_s\to K^*}^{LL} \left[
 -a_{4}-\frac{3}{2}a_7+\frac{1}{2}a_{10}+\frac{3}{2}a_9\right]
\nonumber\\
  &&  - f_{\pi} F_{B_s\to K^*}^{SP}
  \left[-a_{6}+\frac{1}{2}a_{8}\right]  +  M_{B_s\to K^*}^{LL}
  \left[-C_{3}+\frac{3}{2}C_{8}+\frac{1}{2}C_{9}
  +\frac{3}{2}C_{10}\right]
  \nonumber
  \\
  && +  f_{B_s}  F_{ann}^{LL}\left[-a_{4}+ \frac{1}{2}a_{10}\right]
   -  f_{B_s} F_{ann}^{SP}\left[-a_{6}+\frac{1}{2}a_{8}\right]
\nonumber
  \\
  && +  M_{ann}^{LL}\left[-C_{3}+\frac{1}{2}C_{9}\right]
   -  M_{ann}^{LR}\left[-C_{5}+\frac{1}{2}C_{7}\right]\Bigg \},
\end{eqnarray}
\begin{eqnarray}
 \sqrt{2}A(\bar B_{s}^0\to\rho^{0} K^{0})
&=& \frac{G_F}{\sqrt{2}} V_{ub}V_{ud}^{*} \Bigg \{f_{\rho} F_{B_s\to
K}^{LL} \left[a_{2}\right]+ M_{B_s\to K}^{LL} [C_{2}]\Bigg \}
 \nonumber
 \\
  &&- \frac{G_F}{\sqrt{2}} V_{tb}V_{td}^{*} \Bigg \{f_{\rho} F_{B_s\to K}^{LL} \left[
 -a_{4}+\frac{3}{2}a_7+\frac{1}{2}a_{10}+\frac{3}{2}a_9\right]
 +  M_{B_s\to K}^{LR} \left[-C_{5}+\frac{1}{2}C_{7}\right]
  \nonumber\\
  &&
   +  M_{B_s\to K}^{LL}
  \left[-C_{3}+\frac{1}{2}C_{9}
  +\frac{3}{2}C_{10}\right]
  -M_{B_s\to K}^{SP} \left[\frac{3}{2}C_{8}\right]
  +  f_{B_s}  F_{ann}^{LL}\left[-a_{4}+ \frac{1}{2}a_{10}\right]
  \nonumber
  \\
  &&
  +  f_{B_s} F_{ann}^{SP}\left[-a_{6}+\frac{1}{2}a_{8}\right]
 +  M_{ann}^{LL}\left[-C_{3}+\frac{1}{2}C_{9}\right]
   +  M_{ann}^{LR}\left[-C_{5}+\frac{1}{2}C_{7}\right]\Bigg \},
\end{eqnarray}
\begin{eqnarray}
\sqrt2A(\bar B_{s}^0\to  \omega K^{0}) &=& \frac{G_F}{\sqrt{2}}
V_{ub}V_{ud}^{*} \Bigg\{ f_\omega F_{B_s\to K}^{LL}
\left[a_{2}\right] + M_{B_s\to K}^{LL} [C_{2}] \Big\}
\nonumber \\
 &&-  \frac{G_F}{\sqrt{2}} V_{tb}V_{td}^{*} \bigg\{ f_\omega F_{B_s\to K}^{LL}  \left[
 2a_{3}+a_{4}+2a_{5}
+\frac{1}{2}a_{7}+\frac{1}{2}a_{9}-\frac{1}{2}a_{10}\right]
\nonumber
   \\
 &&  + M_{B_s\to K}^{LL}\left[C_{3}+2C_{4}-\frac{1}{2}C_9
  +\frac{1}{2}C_{10}\right] +
M_{B_s\to K}^{LR}\left[C_{5}-\frac{1}{2}C_{7}\right]
 \nonumber
  \\
 &&
    - M_{B_s\to K}^{SP}\left[2C_{6}+\frac{1}{2}C_{8}\right]
 + f_{B_s}  F_{ann}^{LL}\left[a_{4} -\frac{1}{2}a_{10}\right]
  +f_{B_s} F_{ann}^{SP}\left[a_{6} -\frac{1}{2}a_{8}\right]
\nonumber
  \\
 &&  + M_{ann}^{LL}\left[C_{3}-\frac{1}{2}C_{9}\right]
 + M_{ann}^{LR}\left[C_{5}-\frac{1}{2}C_{7}\right] \Bigg\},
\end{eqnarray}

\subsection{The pure annihilation type decays}

\begin{eqnarray}
A(\bar B_{s}^0\to\pi^{+}\rho^{-}) &=&  \frac{G_F}{\sqrt{2}}
V_{ub}V_{us}^{*} \Bigg\{ f_{B_s}F_{ann}^{LL}\left[a_{2}\right] +
M_{ann}^{LL}[C_2] \Bigg\} - \frac{G_F}{\sqrt{2}}
V_{tb}V_{ts}^{*}\Bigg \{
 f_{B_s} F_{ann}^{LL}\left[a_{3}
 +a_{9}\right]\nonumber \\
&&
  -f_{B_s} F_{ann}^{LR}\left[a_{5} +a_{7} \right]
   +M_{ann}^{LL}\left[C_{4}+C_{10}\right]
  - M_{ann}^{SP}\left[C_{6}+C_{8}\right]
  +[\pi^+\leftrightarrow\rho^-]\Bigg \},
\\
A(\bar B_{s}^0\to\rho^{+}\pi^{-})  &=&  \frac{G_F}{\sqrt{2}}
V_{ub}V_{us}^{*} \Bigg\{ f_{B_s}F_{ann}^{LL}\left[a_{2}\right] +
M_{ann}^{LL}[C_2] \Bigg\} - \frac{G_F}{\sqrt{2}}
V_{tb}V_{ts}^{*}\Bigg \{
 f_{B_s} F_{ann}^{LL}\left[a_{3}
 +a_{9}\right]\nonumber \\
&&
  -f_{B_s} F_{ann}^{LR}\left[a_{5} +a_{7} \right]
   +M_{ann}^{LL}\left[C_{4}+C_{10}\right]
  - M_{ann}^{SP}\left[C_{6}+C_{8}\right]
  +[\rho^+\leftrightarrow\pi^-]\Bigg \},
\\
2A(\bar B_{s}^0\to\pi^{0}\rho^{0}) &=& A(\bar
B_{s}^0\to\pi^{+}\rho^{-})+A(\bar B_{s}^0\to\rho^{+}\pi^{-}),
\\
2A(\bar B_{s}^0\to\pi^{0}\omega) &=& \frac{G_F}{\sqrt{2}}
V_{ub}V_{us}^{*} \Bigg \{ M_{ann}^{LL}[C_2]\Bigg\}-
\frac{G_F}{\sqrt{2}} V_{tb}V_{ts}^{*}\Bigg \{
M_{ann}^{LL}\left[\frac{3}{2}C_{10}\right]
-M_{ann}^{SP}\left[\frac{3}{2}C_{8}\right]+[\pi^0\leftrightarrow\omega]\Bigg
\}.
\end{eqnarray}

\subsection{The QCD penguin  dominant decays}

\begin{eqnarray}
 A(\bar B_{s}^0\to K^{-}K^{*+})
&=&  \frac{G_F}{\sqrt{2}} V_{ub}V_{us}^{*} \Big \{ f_{K} F_{B_s\to
K^*}^{LL} \left[a_{1} \right] + M_{B_s\to K^*}^{LL} [C_{1}]+ f_{B_s}
 F_{ann}^{LL}\left[a_{2}\right]+ M_{ann}^{LL}[C_2] \Big\}
\nonumber\\
&&   - \frac{G_F}{\sqrt{2}}
 V_{tb}V_{ts}^{*} \Bigg \{ f_{K} F_{B_s\to K^*}^{LL} \left[
  a_{4}+a_{10}\right]-f_{K} F_{B_s\to K^*}^{SP} \left[a_{6}+a_{8}\right]
  \nonumber\\
&&+  M_{B_s\to K^*}^{LL} \left[C_{3}+C_{9}\right]
   +
  f_BF_{ann}^{LL}\left[a_{3}+a_{4}-a_{5} +\frac{1}{2}a_{7}-\frac{1}{2}a_{9}
-\frac{1}{2}a_{10}\right]\nonumber\\
 &&
  - M_{ann}^{LR}\left[C_{5}-\frac{1}{2}C_{7}\right]
   +
    f_{B_s} F_{ann}^{LL}\left[a_{3}-a_{5}-a_{7} +a_{9}
\right]_{K^* \leftrightarrow K}\nonumber
\\
&&- M_{B_s\to K^*}^{LR} \left[C_{5}+C_{7}\right]
   - f_{B_s} F_{ann}^{SP}\left[a_{6}- \frac{1}{2}a_{8}\right]\nonumber\\
 && +
M_{ann}^{LL}\left[C_{3}-\frac{1}{2}C_9+C_{4}-\frac{1}{2}C_{10}\right]
- M_{ann}^{SP}\left[C_{6}-\frac{1}{2}C_{8}\right]\nonumber\\
  &&
  + M_{ann}^{LL}\left[C_{4}+C_{10}\right]_{K^* \leftrightarrow K}
   - M_{ann}^{SP}\left[C_{6}+C_{8}\right]_{K^* \leftrightarrow K} \Bigg \},
\end{eqnarray}
\begin{eqnarray}
 A(\bar B_{s}^0\to K^{*-}K^{+})
&=&  \frac{G_F}{\sqrt{2}} V_{ub}V_{us}^{*} \Big \{ f_{K^*} F_{B_s\to
K}^{LL} \left[a_{1} \right] + M_{B_s\to K^*}^{LL} [C_{1}] + f_{B_s}
 F_{ann}^{LL}\left[a_{2}\right] + M_{ann}^{LL}[C_2] \Big\} \nonumber
  \\
 &&- \frac{G_F}{\sqrt{2}}
 V_{tb}V_{ts}^{*} \bigg \{ f_{K^*} F_{B_s\to K}^{LL} \left[
  a_{4}+a_{10}\right]
   + M_{B_s\to K}^{LL} \left[C_{3}+C_{9}\right]
  + M_{B_s\to K}^{LR} \left[C_{5}+C_{7}\right]
    \nonumber\\
&&+
  f_BF_{ann}^{LL}\left[a_{3}+a_{4}-a_{5} +\frac{1}{2}a_{7}-\frac{1}{2}a_{9}
-\frac{1}{2}a_{10}\right]+ f_{B_s} F_{ann}^{SP}\left[a_{6}-
\frac{1}{2}a_{8}\right]
 \nonumber\\
&&+  f_{B_s} F_{ann}^{LL}\left[a_{3}- a_{5}-a_{7}+a_{9} \right]_{K^*
\leftrightarrow K} + M_{ann}^{LR}\left[C_{5}-\frac{1}{2}C_{7}\right]
  \nonumber\\
&&+
M_{ann}^{LL}\left[C_{3}-\frac{1}{2}C_9+C_{4}-\frac{1}{2}C_{10}\right]
   - M_{ann}^{SP}\left[C_{6}-\frac{1}{2}C_{8}\right]\nonumber\\
  &&
   +
M_{ann}^{LL}\left[C_{4}+C_{10}\right]_{K^* \leftrightarrow K}
   - M_{ann}^{SP}\left[C_{6}+C_{8}\right]_{K^* \leftrightarrow K} \Bigg \},
\end{eqnarray}
\begin{eqnarray}
 A(\bar B_{s}^0\to\bar K^0  K^{*0})
&=&   - \frac{G_F}{\sqrt{2}}
 V_{tb}V_{ts}^{*} \Bigg \{ f_{K} F_{B_s\to K^*}^{LL} \left[
  a_{4}-\frac{1}{2}a_{10}\right]
 - f_{K} F_{B_s\to K^*}^{SP} \left[a_{6}-\frac{1}{2}a_{8}\right]\nonumber
 \\
 && + M_{B_s\to
K^*}^{LL} \left[C_{3}-\frac{1}{2}C_{9}\right]
  +
 f_{B_s} F_{ann}^{LL}\left[a_{3}+a_{4}-a_{5} +\frac{1}{2}a_{7} -\frac{1}{2}a_{9}
-\frac{1}{2}a_{10}\right]\nonumber
\\
&&- M_{B_s\to K^*}^{LR} \left[C_{5}-\frac{1}{2}C_{7}\right]
   - f_{B_s} F_{ann}^{SP}\left[a_{6}- \frac{1}{2}a_{8}\right]\nonumber\\
&&
 + M_{ann}^{LL}\left[C_{3}-\frac{1}{2}C_9+C_{4}-\frac{1}{2}C_{10}\right] +
M_{ann}^{LL}\left[C_{4}-\frac{1}{2}C_{10}\right]_{K^*
\leftrightarrow K}\nonumber\\
&&   - M_{ann}^{LR}\left[C_{5}-\frac{1}{2}C_{7}\right]
   -\left( M_{ann}^{SP}\left[C_{6}-\frac{1}{2}C_{8}\right]+[K^* \leftrightarrow K]\right)
   \nonumber\\
 &&
  +
    f_{B_s} F_{ann}^{LL}\left[a_{3}-a_{5}+\frac{1}{2}a_{7}-\frac{1}{2}a_{9}
\right]_{K^* \leftrightarrow K}
   \bigg \},
\end{eqnarray}
\begin{eqnarray}
 A(\bar B_{s}^0\to\bar K^{*0}  K^{0})
&=&   - \frac{G_F}{\sqrt{2}}
 V_{tb}V_{ts}^{*} \Bigg \{ f_{K^*} F_{B_s\to K}^{LL} \left[
  a_{4}-\frac{1}{2}a_{10}\right]+ M_{B_s\to
K}^{LL} \left[C_{3}-\frac{1}{2}C_{9}\right]
 \nonumber\\
 &&
  + M_{B_s\to K}^{LR} \left[C_{5}-\frac{1}{2}C_{7}\right]  +
 f_{B_s} F_{ann}^{LL}\left[a_{3}+a_{4}-a_{5} +\frac{1}{2}a_{7}-\frac{1}{2}a_{9}
-\frac{1}{2}a_{10}\right]
  \nonumber\\
&& + f_{B_s} F_{ann}^{SP}\left[a_{6}- \frac{1}{2}a_{8}\right]
 + M_{ann}^{LL}\left[C_{3}-\frac{1}{2}C_9+C_{4}-\frac{1}{2}C_{10}\right] \nonumber\\
 &&
  + M_{ann}^{LR}\left[C_{5}-\frac{1}{2}C_{7}\right]
   - \left(M_{ann}^{SP}\left[C_{6}-\frac{1}{2}C_{8}\right]+ [K^*
\leftrightarrow K]\right)\\
  &&+  f_{B_s} F_{ann}^{LL}\left[a_{3} -a_{5}+\frac{1}{2}a_{7}-\frac{1}{2}a_{9}
\right]_{K^* \leftrightarrow K}
 + M_{ann}^{LL}\left[C_{4}-\frac{1}{2}C_{10}\right]_{K^*
\leftrightarrow K}
    \Bigg \}.\nonumber
\end{eqnarray}

\begin{eqnarray}
A(\bar B_{s}^0\to    K^{0} \phi) &=& - \frac{G_F}{\sqrt{2}}
V_{tb}V_{td}^{*} \Bigg\{ f_\phi F_{B_s\to K}^{LL}\left[
 a_{3} +a_{5}
 -\frac{1}{2}a_{7}-\frac{1}{2}a_{9} \right]
 + f_{K}  F_{B_s\to \phi}^{LL}\left[
 a_{4} -\frac{1}{2}a_{10}\right] \nonumber
  \\
 &&
   -  f_{K}  F_{B_s\to \phi}^{SP}\left[a_{6} -\frac{1}{2}a_{8}\right]
 +  M_{B_s\to K}^{LL}\left[C_{4}-\frac{1}{2}C_{10}\right]
 +  M_{B_s\to \phi}^{LL}\left[C_{3}-\frac{1}{2}C_9\right]
 \nonumber
  \\
 &&
  -  M_{B_s\to K}^{SP}\left[C_{6}-\frac{1}{2}C_{8}\right]
  -  M_{B_s\to \phi}^{LR}\left[C_{5}-\frac{1}{2}C_{7}\right]
  +  f_{B_s}   F_{ann}^{LL}\left[a_{4} -\frac{1}{2}a_{10}\right]\nonumber
  \\
 &&
  - f_{B_s}  F_{ann}^{SP}\left[a_{6} -\frac{1}{2}a_{8}\right]
+  M_{ann}^{LL}\left[C_{3}-\frac{1}{2}C_{9}\right]
 -  M_{ann}^{LR}\left[C_{5}-\frac{1}{2}C_{7}\right]\bigg\}.
\end{eqnarray}
\subsection{The Electroweak penguin dominant decays}

\begin{eqnarray}
\sqrt{2}A(\bar B_{s}^0\to\pi^{0}\phi) &=& \frac{G_F}{\sqrt{2}}
V_{ub}V_{us}^{*} \Bigg \{f_\pi F^{LL}_{B_s\to\phi}[a_2]+
M_{B_s\to\phi}^{LL}[C_2]\Bigg\}\nonumber \\
&&
 - \frac{G_F}{\sqrt{2}}
V_{tb}V_{ts}^{*}\Bigg \{f_\pi
F^{LL}_{B_s\to\phi}\left[\frac{3}{2}a_9-\frac{3}{2}a_7\right]+
M_{B_s\to\phi}^{LL}\left[\frac{3}{2}C_{8}+\frac{3}{2}C_{10}\right]\Bigg
\}.
\end{eqnarray}

\subsection{The decays involving $\eta$ and $\eta'$}

\begin{eqnarray}
\sqrt2A(\bar B_{s}^0\to  \eta_nK^{*0}) &=& \frac{G_F}{\sqrt{2}}
V_{ub}V_{ud}^{*} \Big\{ f_n F_{B_s\to K^*}^{LL} \left[a_{2}\right] +
M_{B_s\to K^*}^{LL} [C_{2}] \Big\}
\nonumber \\
 &&-  \frac{G_F}{\sqrt{2}} V_{tb}V_{td}^{*} \bigg\{ f_n F_{B_s\to K^*}^{LL}  \left[
 2a_{3}+a_{4}-2a_{5}
-\frac{1}{2}a_{7}+\frac{1}{2}a_{9}-\frac{1}{2}a_{10}\right]
\nonumber
   \\
 &&
  -f_n F_{B_s\to K^*}^{SP}\left[a_{6}
   -\frac{1}{2}a_{8}\right]+ M_{B_s\to K^*}^{LL}\left[C_{3}+2C_{4}-\frac{1}{2}C_9
  +\frac{1}{2}C_{10}\right]
  \nonumber
 \\
 &&
    + M_{B_s\to K^*}^{SP}\left[2C_{6}+\frac{1}{2}C_{8}\right]
  + f_{B_s}  F_{ann}^{LL}\left[a_{4} -\frac{1}{2}a_{10}\right]
  +f_{B_s} F_{ann}^{SP}\left[a_{6} -\frac{1}{2}a_{8}\right]
\nonumber
  \\
 &&  + M_{ann}^{LL}\left[C_{3}-\frac{1}{2}C_{9}\right]
 + M_{ann}^{LR}\left[C_{5}-\frac{1}{2}C_{7}\right] \Bigg\},
\\
\
\
\
A(\bar B_{s}^0\to   K^{*0}\eta_s) &=& - \frac{G_F}{\sqrt{2}}
V_{tb}V_{td}^{*} \Bigg\{ f_s F_{B_s\to K^*}^{LL}\left[
 a_{3} -a_{5}
 +\frac{1}{2}a_{7} -\frac{1}{2}a_{9} \right]
 + f_{K^*}  F_{B_s\to \eta_s}^{LL}\left[
 a_{4} -\frac{1}{2}a_{10}\right] \nonumber
  \\
 &&
 +  M_{B_s\to K^*}^{LL}\left[C_{4}-\frac{1}{2}C_{10}\right]
 +  M_{B_s\to \eta_s}^{LL}\left[C_{3}-\frac{1}{2}C_9\right]
 \nonumber
  \\
 &&
  +  M_{B_s\to K^*}^{SP}\left[C_{6}-\frac{1}{2}C_{8}\right]
  +  M_{B_s\to \eta_s}^{LR}\left[C_{5}-\frac{1}{2}C_{7}\right]
  +  f_{B_s}   F_{ann}^{LL}\left[a_{4} -\frac{1}{2}a_{10}\right]    \nonumber
  \\
 &&
 +  f_{B_s}  F_{ann}^{SP}\left[a_{6} -\frac{1}{2}a_{8}\right]
+  M_{ann}^{LL}\left[C_{3}-\frac{1}{2}C_{9}\right]
 +  M_{ann}^{LR}\left[C_{5}-\frac{1}{2}C_{7}\right]\bigg\}.
\end{eqnarray}
The  decay amplitudes for the physical states are then
\begin{eqnarray}
A(\bar B_s \to \eta K^{*0}) &=& A(\bar B_{s}^0\to  \eta_nK^{*0})
\cos \phi -A(\bar B_{s}^0\to \eta_s  K^{*0}) \sin \phi, \\
A(\bar B_s \to \eta' K^{*0}) &=& A(\bar B_{s}^0\to  \eta_nK^{*0})
\sin \phi  + A(\bar B_{s}^0\to  \eta_sK^{*0}) \cos \phi.
\end{eqnarray}
For $\bar B_s \to \eta^{(\prime)} \rho^0$ the decay amplitudes are
defined similarly,
\begin{eqnarray}
A(\bar B_s \to\rho^0 \eta ) &=& A(\bar B_{s}^0\to \rho^{0}\eta_n)
\cos \phi -A(\bar B_{s}^0\to \rho^{0} \eta_s) \sin \phi, \\
A(\bar B_s \to\rho^0 \eta' ) &=& A(\bar B_{s}^0\to \rho^{0}\eta_n)
\sin \phi + A(\bar B_{s}^0\to \rho^{0}\eta_s) \cos \phi,
\end{eqnarray}
where
\begin{eqnarray}
2A(\bar B_{s}^0\to\rho^{0} \eta_n) &=& - \frac{G_F}{\sqrt{2}}
V_{tb}V_{ts}^{*}\Bigg\{ f_{B_s}
F_{ann}^{LL}\left[\frac{3}{2}a_9-\frac{3}{2}a_7 \right]
 +M_{ann}^{LL}\left[\frac{3}{2}C_{10}\right]
  - M_{ann}^{SP}\left[\frac{3}{2}C_8\right]\Bigg\}
  \nonumber
  \\
 &&+\frac{G_F}{\sqrt{2}}
V_{ub}V_{us}^{*}
 \Bigg\{ f_{B_s}F_{ann}^{LL}\left[a_{2}\right] +  M_{ann}^{LL}[C_{2}] \Bigg\}
  +\big [\rho^0 \leftrightarrow \eta_n\big ],
\end{eqnarray}
and
\begin{eqnarray}
 \sqrt{2}A(\bar B_{s}^0\to\rho^{0}\eta_s)
&=& \frac{G_F}{\sqrt{2}} V_{ub}V_{us}^{*} \Bigg \{ f_{\rho} F_{
B_s\to \eta_s}^{LL} \left[a_{2}\right]+ M_{ B_s\to
\eta_s}^{LL}\left[C_{2}\right]\Bigg \}
\\
&&
 - \frac{G_F}{\sqrt{2}} V_{tb}V_{ts}^{*}\Bigg \{
f_{\rho}F_{B_s\to\eta_s}^{LL}\left[\frac{3}{2}a_{7}+\frac{3}{2}a_{9}\right]
+M_{ B_s\to \eta_s}^{LL} \left[\frac{3}{2}C_{10}\right]
  - M_{B_s\to \eta_s}^{SP}\left[\frac{3}{2}C_{8}\right] \Bigg\}.\nonumber
\end{eqnarray}

For decays $\bar B_s \to \eta^{(\prime)} \omega$, we have
\begin{eqnarray}
2A(\bar B_{s}^0\to \eta_n\omega) &=&
 \frac{G_F}{\sqrt{2}} V_{ub}V_{us}^{*}\Bigg \{ f_{B_s} F_{ann}^{LL}\left[a_{2}\right] +
M_{ann}^{LL}[C_{2}] \Bigg\}\nonumber
\\
 &&- \frac{G_F}{\sqrt{2}}
V_{tb}V_{ts}^{*}\Bigg\{
M_{ann}^{LL}\left[2C_4+\frac{1}{2}C_{10}\right] -
M_{ann}^{SP}\left[2C_6+\frac{1}{2}C_8\right]\nonumber
\\
 && +f_{B_s} F_{ann}^{LL}\left[2a_3-2a_5
-\frac{1}{2}a_7 +\frac{1}{2}a_9 \right]
  \Bigg\}+[\eta_n\leftrightarrow \omega],
\end{eqnarray}
\begin{eqnarray}
 \sqrt 2A(\bar B_{s}^0\to \omega \eta_s)
&=& \frac{G_F}{\sqrt{2}} V_{ub}V_{us}^{*} \Bigg\{f_\omega F_{B_s\to
\eta_s}^{LL}\left[a_{2}\right] + M_{B_s\to
\eta_s}^{LL}\left[C_{2}\right]\Bigg \}\nonumber
\\
 &&- \frac{G_F}{\sqrt{2}} V_{tb}V_{ts}^{*} \Bigg\{
  f_\omega F_{B_s\to\eta_s}^{LL}\left[2a_3+2a_5
+\frac{1}{2}a_7 +\frac{1}{2}a_9  \right]  \nonumber
   \\
&&+ M_{B_s\to \eta_s}^{LL} \left[2C_4+\frac{1}{2}C_{10}\right]
   -M_{B_s\to
\eta_s}^{SP}\left[2C_6+\frac{1}{2}C_8\right]\Bigg \}.
\end{eqnarray}
And the decay amplitudes for physical states are
\begin{eqnarray}
A(\bar B_s \to \eta\omega ) &=& A(\bar B_{s}^0\to \eta_n \omega)
\cos \phi -A(\bar B_{s}^0\to \omega \eta_s) \sin \phi, \\
A(\bar B_s \to \eta'\omega ) &=& A(\bar B_{s}^0\to \eta_n\omega)
\sin \phi + A(\bar B_{s}^0\to \omega\eta_s) \cos \phi.
\end{eqnarray}
For decays  $\bar B_s \to \eta\phi$, we have
\begin{eqnarray}
 \sqrt 2A(\bar B_{s}^0\to \eta_n\phi)
&=& \frac{G_F}{\sqrt{2}} V_{ub}V_{us}^{*} \Bigg\{f_n F_{B_s\to
\phi}^{LL}\left[a_{2}\right] + M_{B_s\to
\phi}^{LL}\left[C_{2}\right]\Bigg \}\nonumber
\\
 &&- \frac{G_F}{\sqrt{2}} V_{tb}V_{ts}^{*} \Bigg\{
  f_n F_{B_s\to \phi}^{LL}\left[2a_3-2a_5
-\frac{1}{2}a_7 +\frac{1}{2}a_9  \right]  \nonumber
   \\
&&+ M_{B_s\to \phi}^{LL} \left[2C_4+\frac{1}{2}C_{10}\right]
   +M_{B_s\to
\phi}^{SP}\left[2C_6+\frac{1}{2}C_8\right]\Bigg \},
\\
 A(\bar B_{s}^0\to \eta_s\phi)
&=& - \frac{G_F}{\sqrt{2}} V_{tb}V_{ts}^{*} \Bigg\{ f_sF_{B_s\to
\phi}^{LL}\left[a_3+a_{4}-a_5
 +\frac{1}{2}a_7 -\frac{1}{2}a_9-\frac{1}{2}a_{10}\right]
 \nonumber
 \\
 &&
 - f_s F_{B_s\to \phi}^{SP}\left[a_{6}
 -\frac{1}{2}a_{8}\right]
 + M_{B_s\to \phi}^{LL} \left[C_3+C_{4}-\frac{1}{2}C_9-\frac{1}{2}C_{10}\right]
 \nonumber
 \\
 &&+ M_{B_s\to \phi}^{SP}
 \left[C_6-\frac{1}{2}C_{8}\right]
 + f_{B_s} F_{ann}^{LL}\left[a_3+a_{4}-a_5  +\frac{1}{2}a_7 -\frac{1}{2}a_9-\frac{1}{2}a_{10}\right]
\nonumber
 \\
 &&
   + M_{ann}^{LL}\left[C_3+C_{4}-\frac{1}{2}C_9
-\frac{1}{2}C_{10}\right] - f_{B_s} F_{ann}^{SP}\left[a_{6}
-\frac{1}{2}a_{8}\right]
  \nonumber
 \\
 &&
  - M_{ann}^{LR}\left[C_5-\frac{1}{2}C_{7}\right]
 - M_{ann}^{SP}
  \left[C_6-\frac{1}{2}C_{8}\right]\Bigg\} +[\eta_s\leftrightarrow
  \phi].
\end{eqnarray}
And the decay amplitudes for physical states are
\begin{eqnarray}
A(\bar B_s \to \eta\phi ) &=& A(\bar B_{s}^0\to \eta_n \phi)
\cos \phi -A(\bar B_{s}^0\to \eta_s \phi) \sin \phi, \\
A(\bar B_s \to \eta'\phi ) &=& A(\bar B_{s}^0\to \eta_n\phi) \sin
\phi + A(\bar B_{s}^0\to \eta_s\phi) \cos \phi.
\end{eqnarray}

\section{Analytic Formulae for the $B_s \to VV$  decay
amplitudes}

There are three kinds of polarizations in the $B_s$ meson decays to
two vector final states, namely: Longitudinal (L), parallel (N) and
transverse (T). The decay amplitudes are classified accordingly,
with $i=L,N,T$.

\subsection{Tree  dominant decays}

\begin{eqnarray}
 A^i(\bar B_{s}^0\to\rho^{-}K^{*+})
&=&  \frac{G_F}{\sqrt{2}}  V_{ub}V_{ud}^{*} \Big\{ f_{\rho}
F_{B_s\to K^*}^{LL,i}\left[a_1\right] + M_{B_s\to K^*}^{LL,i}
[C_{1}] \Big \}\nonumber
   \\
  &&- \frac{G_F}{\sqrt{2}} V_{tb}V_{td}^{*} \bigg\{ f_{\rho}
  F_{B_s\to K^*}^{LL,i}\left[
   a_{4} +a_{10}\right]
 \nonumber
\\
 && +  M_{B_s\to K^*}^{LL,i} \left[C_{3}+C_{9}\right] - M_{B_s\to
K^*}^{LR,i}\left[C_{5}+C_{7}\right]
 \nonumber  \\
&& + f_{B_s}
 F_{ann}^{LL,i}\left[a_{4}-\frac{1}{2}a_{10}\right] - f_{B_s} F_{ann}^{SP,i}\left[a_{6}-
\frac{1}{2}a_{8}\right]
  \nonumber  \\
&&  + M_{ann}^{LL,i}\left[C_{3}-\frac{1}{2}C_{9}\right]
 - M_{ann}^{LR,i}\left[C_{5}-\frac{1}{2}C_{7}\right]
\bigg \},
\end{eqnarray}

\begin{eqnarray}
 \sqrt{2}A^i(\bar B_{s}^0\to\rho^{0} K^{*0})
&=& \frac{G_F}{\sqrt{2}} V_{ub}V_{ud}^{*} \Big \{f_{\rho} F_{B_s\to
K^*}^{LL,i} \left[a_{2}\right]+ M_{B_s\to K^*}^{LL,i} [C_{2}]\Big \}
 \nonumber\\
 && +\frac{G_F}{\sqrt{2}} V_{tb}V_{td}^{*} \Big \{f_{\rho} F_{B_s\to K^*}^{LL,i} \left[
 -a_{4}+\frac{3}{2}a_7+\frac{3}{2}a_9+\frac{1}{2}a_{10}\right]
  -  M_{B_s\to K^*}^{LR,i} \left[-C_{5}+\frac{1}{2}C_{7}\right] \nonumber
  \\
  && +  M_{B_s\to K^*}^{LL,i}
  \left[-C_{3}+\frac{1}{2}C_{9}
  +\frac{3}{2}C_{10}\right]
   -M_{B_s\to K^*}^{SP,i} \left[\frac{3}{2}C_{8}\right]
  \nonumber
  \\
  && +  f_{B_s}  F_{ann}^{LL,i}\left[-a_{4}+ \frac{1}{2}a_{10}\right]
   -  f_{B_s} F_{ann}^{SP,i}\left[-a_{6}+\frac{1}{2}a_{8}\right]
\nonumber
  \\
  && +  M_{ann}^{LL,i}\left[-C_{3}+\frac{1}{2}C_{9}\right]
   -  M_{ann}^{LR,i}\left[-C_{5}+\frac{1}{2}C_{7}\right]\bigg \},
\end{eqnarray}

\begin{eqnarray}
\sqrt2A^i(\bar B_{s}^0\to  \omega K^{*0}) &=& \frac{G_F}{\sqrt{2}}
V_{ub}V_{ud}^{*} \Big\{ f_{\omega} F_{B_s\to K^*}^{LL,i}
\left[a_{2}\right] + M_{B_s\to K^*}^{LL,i} [C_{2}] \Big\}
\nonumber \\
 &&-  \frac{G_F}{\sqrt{2}} V_{tb}V_{td}^{*} \bigg\{ f_{\omega} F_{B_s\to K^*}^{LL,i}  \left[
 2a_{3}+a_{4}+2a_{5}
+\frac{1}{2}a_{7}+\frac{1}{2}a_{9}-\frac{1}{2}a_{10}\right]
\nonumber
   \\
 &&   + M_{B_s\to K^*}^{LL,i}\left[C_{3}+2C_{4}-\frac{1}{2}C_9
  +\frac{1}{2}C_{10}\right]
 \nonumber
  \\
 &&   - M_{B_s\to K^*}^{LR,i}\left[C_{5}-\frac{1}{2}C_{7}\right]
    - M_{B_s\to K^*}^{SP,i}\left[2C_{6}+\frac{1}{2}C_{8}\right]
   \nonumber
  \\
 &&
 + f_{B_s}  F_{ann}^{LL,i}\left[a_{4} -\frac{1}{2}a_{10}\right]
  -f_{B_s} F_{ann}^{SP,i}\left[a_{6} -\frac{1}{2}a_{8}\right]
\nonumber
  \\
 &&  + M_{ann}^{LL,i}\left[C_{3}-\frac{1}{2}C_{9}\right]
 - M_{ann}^{LR,i}\left[C_{5}-\frac{1}{2}C_{7}\right] \bigg\}.
\end{eqnarray}

\subsection{Pure annihilation type decays}

\begin{eqnarray}
A^i(\bar B_{s}^0\to\rho^{+}\rho^{-}) &=&  \frac{G_F}{\sqrt{2}}
V_{ub}V_{us}^{*} \Big\{ f_{B_s}F_{ann}^{LL,i}\left[a_{2}\right] +
M_{ann}^{LL,i}[C_2] \Big\}
\nonumber \\
&&- \frac{G_F}{\sqrt{2}} V_{tb}V_{ts}^{*}\bigg \{ f_{B_s}
F_{ann}^{LL,i}\left[2a_{3}
 +\frac{1}{2}a_{9}\right]
  +f_{B_s} F_{ann}^{LR,i}\left[2a_{5} +\frac{1}{2}a_{7} \right]
  \nonumber \\
&& +M_{ann}^{LL,i}\left[2C_{4}+\frac{1}{2}C_{10}\right]
+M_{ann}^{SP,i}\left[2C_{6}+\frac{1}{2}C_{8}\right]\bigg \},
\end{eqnarray}
\begin{eqnarray}
\sqrt{2}A^i(\bar B_{s}^0\to\rho^{0}\rho^{0}) &=&
\frac{G_F}{\sqrt{2}} V_{ub}V_{us}^{*} \Big \{ f_{B_s}
F_{ann}^{LL,i}\left[a_{2}\right]
+ M_{ann}^{LL,i}[C_2]\Big\} \nonumber \\
&&- \frac{G_F}{\sqrt{2}} V_{tb}V_{ts}^{*}\bigg \{ f_{B_s}
F_{ann}^{LL,i}\left[2a_{3} +\frac{1}{2}a_{9} \right] +f_{B_s}
F_{ann}^{LR,i}\left[2a_{5} +\frac{1}{2}a_{7} \right]
  \nonumber\\
&&+M_{ann}^{LL,i}\left[2C_{4}+\frac{1}{2}C_{10}\right]
+M_{ann}^{SP,i}\left[2C_{6}+\frac{1}{2}C_{8}\right]\bigg \},
\end{eqnarray}
\begin{eqnarray}
\sqrt 2 A^i(\bar B_{s}^0\to \omega\omega) &=&
 \frac{G_F}{\sqrt{2}} V_{ub}V_{us}^{*}\Bigg \{ f_{B_s} F_{ann}^{LL,i}\left[a_{2}\right] +
M_{ann}^{LL,i}[C_{2}] \Bigg\}\nonumber
\\
 &&- \frac{G_F}{\sqrt{2}}
V_{tb}V_{ts}^{*}\Bigg\{ f_{B_s} F_{ann}^{LL,i}\left[2a_3
+\frac{1}{2}a_9 \right] +f_{B_s} F_{ann}^{LR,i}\left[2a_5
+\frac{1}{2}a_7 \right]
  \nonumber\\
&&+ M_{ann}^{LL,i}\left[2C_4+\frac{1}{2}C_{10}\right]
+M_{ann}^{SP,i}\left[2C_6+\frac{1}{2}C_8\right]\Bigg\},
\end{eqnarray}
\begin{eqnarray}
2A^i(\bar B_{s}^0\to\rho^{0}\omega) &=& \frac{G_F}{\sqrt{2}}
V_{ub}V_{us}^{*}
 \Big\{ f_{B_s}F_{ann}^{LL,i}\left[a_{2}\right] +  M_{ann}^{LL,i}[C_{2}] \Big\}
 - \frac{G_F}{\sqrt{2}}  V_{tb}V_{ts}^{*}\Big\{ f_{B_s}
F_{ann}^{LL,i}\left[\frac{3}{2}a_9 \right]  \nonumber
  \\
 && + f_{B_s}
F_{ann}^{LR,i}\left[\frac{3}{2}a_7 \right]
+M_{ann}^{LL,i}\left[\frac{3}{2}C_{10}\right]
+M_{ann}^{SP,i}\left[\frac{3}{2}C_8\right]\Big\} +\left [\rho^0
\leftrightarrow \omega\right ].
\end{eqnarray}

\subsection{QCD penguin  dominant decays}

\begin{eqnarray}
 A^i(\bar B_{s}^0\to K^{*-}K^{*+})
&=&  \frac{G_F}{\sqrt{2}} V_{ub}V_{us}^{*} \Big \{ f_{K^*} F_{B_s\to
K^*}^{LL,i} \left[a_{1} \right] + M_{B_s\to K^*}^{LL,i} [C_{1}] +
f_{B_s}
 F_{ann}^{LL,i}\left[a_{2}\right]\nonumber\\
&&  + M_{ann}^{LL,i}[C_2] \Big\}  - \frac{G_F}{\sqrt{2}}
 V_{tb}V_{ts}^{*} \bigg \{ f_{K^*} F_{B_s\to K^*}^{LL,i} \left[
  a_{4}+a_{10}\right]\nonumber\\
&& + M_{B_s\to K^*}^{LL,i} \left[C_{3}+C_{9}\right]
-M_{B_s\to K}^{LR,i} \left[C_{5}+C_{7}\right]  \nonumber\\
&&+
  f_{B_s} F_{ann}^{LL,i}\left[a_{3}+a_{4}-\frac{1}{2}a_{9}
-\frac{1}{2}a_{10}\right]
  + f_{B_s} F_{ann}^{LR,i}\left[a_{5} -\frac{1}{2}a_{7} \right]\nonumber\\
&& -f_{B_s} F_{ann}^{SP,i}\left[a_{6}- \frac{1}{2}a_{8}\right] +
M_{ann}^{LL,i}\left[C_{3}-\frac{1}{2}C_9+C_{4}-\frac{1}{2}C_{10}\right] \nonumber\\
 &&
-M_{ann}^{LR,i}\left[C_{5}-\frac{1}{2}C_{7}\right]
   + M_{ann}^{SP,i}\left[C_{6}-\frac{1}{2}C_{8}\right]\nonumber\\
  &&+ \left( f_{B_s} F_{ann}^{LL,i}\left[a_{3} +a_{9}
\right]+ f_{B_s} F_{ann}^{LR,i}\left[a_{5}+a_{7}\right]\right.\nonumber\\
&& \left.+ M_{ann}^{LL,i}\left[C_{4}+C_{10}\right]
   + M_{ann}^{SP,i}\left[C_{6}+C_{8}\right]\right)_{K^{*-}\leftrightarrow K^{*+}}
    \bigg \},
\end{eqnarray}
\begin{eqnarray}
 A^i(\bar B_{s}^0\to\bar K^{*0}  K^{*0})
&=&   - \frac{G_F}{\sqrt{2}}
 V_{tb}V_{ts}^{*} \bigg \{ f_{K^*} F_{B_s\to K^*}^{LL,i} \left[
  a_{4}-\frac{1}{2}a_{10}\right] + M_{B_s\to
K^*}^{LL,i} \left[C_{3}-\frac{1}{2}C_{9}\right]\nonumber\\
 &&
-M_{B_s\to K^*}^{LR,i} \left[C_{5}-\frac{1}{2}C_{7}\right]   +
 f_{B_s} F_{ann}^{LL,i}\left[a_{3}+a_{4}-\frac{1}{2}a_{9}
-\frac{1}{2}a_{10}\right]
  \nonumber\\
&& -f_{B_s} F_{ann}^{SP,i}\left[a_{6}- \frac{1}{2}a_{8}\right]
 + M_{ann}^{LL,i}\left[C_{3}-\frac{1}{2}C_9+C_{4}-\frac{1}{2}C_{10}\right] \nonumber\\
 &&
- M_{ann}^{LR,i}\left[C_{5}-\frac{1}{2}C_{7}\right]
   +
   M_{ann}^{SP,i}\left[C_{6}-\frac{1}{2}C_{8}\right]
    \nonumber \\ && +\left(f_{B_s} F_{ann}^{LL,i}\left[a_{3}-\frac{1}{2}a_{9}
\right]  +f_{B_s} F_{ann}^{LR,i}\left[a_{5}-\frac{1}{2}a_{7}
\right]\right)_{K^{*0}\leftrightarrow \bar K^{*0}} \nonumber \\
&& +\left( M_{ann}^{LL,i}\left[C_{4}-\frac{1}{2}C_{10}\right]
+M_{ann}^{SP,i}\left[C_{6}-\frac{1}{2}C_{8}\right]
\right)_{K^{*0}\leftrightarrow \bar K^{*0}}\bigg \},
\\
A^i(\bar B_{s}^0\to   K^{*0}\phi) &=& - \frac{G_F}{\sqrt{2}}
V_{tb}V_{td}^{*} \bigg\{ f_\phi F_{B_s\to K^*}^{LL,i}\left[
 a_{3}+a_{5}
 -\frac{1}{2}a_{7}-\frac{1}{2}a_{9}\right]
 +   f_{K^*}F_{B_s\to \phi}^{LL,i}\left[
 a_{4} -\frac{1}{2}a_{10}\right] \nonumber
\nonumber
  \\
 &&
 +  M_{B_s\to K^*}^{LL,i}\left[C_{4}-\frac{1}{2}C_{10}\right]
 +  M_{B_s\to \phi}^{LL,i}\left[C_{3}-\frac{1}{2}C_9\right]
 - M_{B_s\to K^*}^{SP,i}\left[C_{6}-\frac{1}{2}C_{8}\right]
 \nonumber
  \\
 &&
-M_{B_s\to \phi}^{LR,i}\left[C_{5}-\frac{1}{2}C_{7}\right]+  f_{B_s}
F_{ann}^{LL,i}\left[a_{4} -\frac{1}{2}a_{10}\right] -f_{B_s}
F_{ann}^{SP,i}\left[a_{6} -\frac{1}{2}a_{8}\right] \nonumber
  \\
 &&
+  M_{ann}^{LL,i}\left[C_{3}-\frac{1}{2}C_{9}\right]
-M_{ann}^{LR,i}\left[C_{5}-\frac{1}{2}C_{7}\right]\bigg \}.
\\
 \sqrt 2A^i(\bar B_{s}^0\to \omega\phi)
&=& \frac{G_F}{\sqrt{2}} V_{ub}V_{us}^{*} \Bigg\{f_\omega F_{B_s\to
\phi}^{LL,i}\left[a_{2}\right] + M_{B_s\to
\phi}^{LL,i}\left[C_{2}\right]\Bigg \}\nonumber
\\
 &&- \frac{G_F}{\sqrt{2}} V_{tb}V_{ts}^{*} \Bigg\{ f_\omega F_{B_s\to \phi}^{LL,i}\left[2a_3
+ 2a_5 +\frac{1}{2}a_7+\frac{1}{2}a_9
   \right]\nonumber
   \\
&&+ M_{B_s\to \phi}^{LL,i} \left[2C_4+\frac{1}{2}C_{10}\right]
-M_{B_s\to \phi}^{SP,i}\left[2C_6+\frac{1}{2}C_8\right]\Bigg \},
\\
\sqrt 2 A^i(\bar B_{s}^0\to \phi\phi) &=& - \frac{2G_F}{\sqrt{2}}
V_{tb}V_{ts}^{*} \Bigg\{ f_\phi F_{B_s\to
\phi}^{LL,i}\left[a_3+a_{4}+a_5
 -\frac{1}{2}a_7 -\frac{1}{2}a_9-\frac{1}{2}a_{10}
\right]
  \nonumber
 \\
 &&
 + M_{B_s\to \phi}^{LL,i} \left[C_3+C_{4}-\frac{1}{2}C_9-\frac{1}{2}C_{10}\right]
 -M_{B_s\to \phi}^{LR,i} \left[C_5-\frac{1}{2}C_{7}\right]
  \nonumber
 \\
 &&
 - M_{B_s\to \phi}^{SP,i}
 \left[C_6-\frac{1}{2}C_{8}\right]
 + f_{B_s} F_{ann}^{LL,i}\left[a_3+a_{4}-\frac{1}{2}a_9-\frac{1}{2}a_{10}\right]
   \nonumber
 \\
 &&
  + f_{B_s} F_{ann}^{LR,i}\left[a_5  -\frac{1}{2}a_7 \right]
 -f_{B_s} F_{ann}^{SP,i}\left[a_{6} -\frac{1}{2}a_{8}\right]
 - M_{ann}^{LR,i}\left[C_5-\frac{1}{2}C_{7}\right]
 \nonumber
 \\
 &&
+ M_{ann}^{LL,i}\left[C_3+C_{4}-\frac{1}{2}C_9
-\frac{1}{2}C_{10}\right] +M_{ann}^{SP,i}
  \left[C_6-\frac{1}{2}C_{8}\right]\Bigg\} .
\end{eqnarray}

\subsection{Electroweak penguin dominant decays}

\begin{eqnarray}
 \sqrt{2}A^i(\bar B_{s}^0\to\rho^{0}\phi)
&=& \frac{G_F}{\sqrt{2}} V_{ub}V_{us}^{*} \Bigg \{ f_{\rho} F_{
B_s\to \phi}^{LL,i} \left[a_{2}\right]+ M_{ B_s\to
\phi}^{LL,i}\left[C_{2}\right]\Bigg \}\nonumber
\\
&& - \frac{G_F}{\sqrt{2}} V_{tb}V_{ts}^{*}\Bigg \{
f_{\rho}F_{B_s\to \phi}^{LL,i}\left[\frac{3}{2}(a_{9}+a_{7})
\right] +M_{ B_s\to \phi}^{LL,i} \left[\frac{3}{2}C_{10}\right]
-M_{B_s\to \phi}^{SP,i}\left[\frac{3}{2}C_{8}\right] \Bigg\}~.
\nonumber\\
\end{eqnarray}


\begin{thebibliography}{11}
\bibitem{Chargeconj}
The use of charge conjugate modes in all the decays and branching ratios are
implied in this paper.
\end{thebibliography}
\end{document}